
%
%

%
%
\font\twelvrm=cmr12
\font\ninerm=cmr9
\font\twelvi=cmmi12
\font\ninei=cmmi9
\font\twelvex=cmex10 scaled\magstep1
\font\twelvbf=cmbx12
\font\ninebf=cmbx9
\font\twelvit=cmti12
\font\twelvsy=cmsy10 scaled\magstep1
\font\ninesy=cmsy9
\font\twelvtt=cmtt12

\font\twelvsl=cmsl12

\font\abstractfont=cmr10
\font\abstractitalfont=cmti10

\def\twelvepoint{\def\rm{\fam0\twelvrm}
	\textfont0=\twelvrm \scriptfont0=\ninerm \scriptscriptfont0=\sevenrm
	\textfont1=\twelvi \scriptfont1=\ninei \scriptscriptfont1=\seveni
	\textfont2=\twelvsy \scriptfont2=\ninesy \scriptscriptfont2=\sevensy
	\textfont3=\twelvex \scriptfont3=\tenex \scriptscriptfont3=\tenex
	\textfont\itfam=\twelvit \def\it{\fam\itfam\twelvit}
	\textfont\slfam=\twelvsl \def\sl{\fam\slfam\twelvsl}
	\textfont\ttfam=\twelvtt \def\tt{\fam\ttfam\twelvtt}
	\textfont\bffam=\twelvbf \def\bf{\fam\bffam\twelvbf}
	\scriptfont\bffam=\ninebf  \scriptscriptfont\bffam=\sevenbf
        \skewchar\ninei='177
        \skewchar\twelvi='177
        \skewchar\seveni='177
}

\newdimen\normalwidth
\newdimen\double
\newdimen\single
\newdimen\indentlength		\indentlength=.5in

\newif\ifdrafton

\def\galley{
 \draftonfalse
 \twelvepoint
 \rm
 \font\chapterfont=cmbx10 scaled\magstep1
 \font\sectionfont=cmbx12
 \font\subsectionfont=cmbx12
 \font\headingfont=cmr10 scaled\magstep2
 \font\titlefont=cmbx10 scaled\magstep2
 \normalwidth=5.7in
 \double=.34in
 \single=.17in
 \hsize=\normalwidth
 \vsize=8.7in
 \hoffset=0.48in
 \voffset=0.1in
 \hfuzz=0.5pt
 \baselineskip=\double plus 2pt minus 2pt }

\parindent=\indentlength
\clubpenalty=10000
\widowpenalty=10000
\displaywidowpenalty=500
\overfullrule=2pt
\tolerance=100

\newcount\chapterno	\chapterno=0
\newcount\sectionno	\sectionno=0
\newcount\appno		\appno=0
\newcount\subsectionno	\subsectionno=0
\newcount\eqnum	\eqnum=0
\newcount\refno \refno=0
\newcount\chap
\newcount\figno \figno=0
\newcount\tableno \tableno=0
\newcount\lettno \lettno=0

\def\bodypaging{
 \headline={\ifodd\chap \hfil \else \tenrm\hfil\twelvrm\folio \fi}
 \footline={\rm \ifodd\chap \global\chap=0 \tenrm\hfil\twelvrm\folio\hfil
 \else \hfil \fi}}

%
\def\preq{\mainid.\the\eqnum}
\def\Elabel#1{\xdef#1{\mainid.\the\eqnum}}
\def\eq{\global\advance\eqnum by1 \eqno(\mainid.\the\eqnum)}
\def\eqlabel#1{\eq {\xdef#1{\mainid.\the\eqnum}}}
\def\quieteqlabel#1{\advance\eqnum by1 {\xdef#1{\mainid.\the\eqnum}}}
\def\newlett{
  \global\lettno=1\global\advance\eqnum by1 \eqno(\mainid.\the\eqnum a)}
\def\lett{\global\advance\lettno by 1
  \eqno(\mainid.\the\eqnum{\ifcase\lettno\or a\or b\or c\or d\or e\or
    f\or g\or h\or \fi})}
\def\newlettlabel#1{\newlett {\xdef#1{\mainid.\the\eqnum}}}
\def\lettlabel#1{\lett {\xdef#1{\mainid.\the\eqnum}}}
%
\newwrite\refs
\def\startrefs#1{\immediate\openout\refs=#1
\immediate\write\refs{\global\chap=1}
\immediate\write\refs{\vfill\noexpand\eject\noexpand\vglue.2in}
\immediate\write\refs{\noexpand\centerline{\headingfont REFERENCES}}
\immediate\write\refs{\noexpand\vglue.5in\baselineskip=\single}
\immediate\write\refs{\parindent=16pt \parskip=\single}}

\def\ref#1{\advance\refno by1 \the\refno \immediate\write\refs
{\noexpand\item{\the\refno.}#1 \hfill\par}}
\def\andref#1{\advance\refno by1\kern-.4em[\the\refno]\immediate\write\refs
{\noexpand\item{\the\refno.}#1 \hfill\par}}
\def\cite#1{[#1]}

\def\refname#1{ \xdef#1{\the\refno}}

\newif\iftoc \tocfalse
\newif\ifpart \partfalse
\newwrite\conts
\def\startcontents{\iftoc\immediate\openout\conts=contents
\immediate\write\conts{\noexpand\centerline{\headingfont CONTENTS}}
\immediate\write\conts{\noexpand\vglue.5in\baselineskip=\single}
\immediate\write\conts{\parskip=0pt \parindent=0pt}\else\relax\fi}

\def\chaptercont#1{\iftoc\immediate\write\conts{
  \vskip\single\the\chapterno.\ #1\hfill\folio\vskip0cm}\else\relax\fi}
\def\andchaptercont#1{\iftoc\immediate\write\conts{
  \hskip .5in \ #1\hfill}\else\relax\fi}
\def\sectioncont#1{\iftoc\immediate\write\conts{\hskip.5cm
  \the\chapterno.\the\sectionno\ #1\hfill\folio\vskip0cm}\else\relax\fi}
\def\subsectioncont#1{\iftoc\immediate\write\conts{\hskip1cm
  \the\chapterno.\the\sectionno.\the\subsectionno\
  #1\hfill\folio\vskip0cm}\else\relax\fi}
\def\appendixcont#1{\iftoc\immediate\write\conts{\vskip\single
  Appendix \mainid.\ #1\hfill\folio\vskip0cm}\else\relax\fi}

\def\partcont#1{\iftoc
 \ifpart\immediate\write\conts{\noexpand\vfill\noexpand\eject}\fi
 \immediate\write\conts{\vskip\single
 \noexpand\centerline{- PART #1 -}}
 \parttrue
 \else\relax\fi}


\newif\ifpskip

\def\chapter#1{
  \ifpskip\vfill\eject \else \bigskip\bigskip\bigskip \fi
 \global\advance\chapterno by1 \sectionno=0 \subsectionno=0 \eqnum=0
 \def\mainid{\the\chapterno}
 \vglue.8cm\centerline{\chapterfont
 \uppercase\expandafter{\romannumeral\the\chapterno.\ #1}}
 \nobreak\vskip.2cm\nobreak
 \ifpskip\global\chap=1\fi
 \chaptercont{#1}}

\def\andchapter#1{\centerline{\chapterfont\raise.1cm
 \hbox{\uppercase{#1}}}\vskip.2cm\nobreak\andchaptercont{#1}}

\def\section#1{
 \global\advance\sectionno by1 \subsectionno=0
 \vskip.8cm\centerline{\sectionfont \the\chapterno.\the\sectionno\ #1}
 \nobreak\vskip.2cm\nobreak
 \sectioncont{#1}}

\def\andsection#1{\centerline{\sectionfont\raise.1cm\hbox{#1}}
 \nobreak\vskip.2cm\nobreak}
\def\subsection#1{
 \global\advance\subsectionno by1
 \vskip.2cm\leftline{\subsectionfont
 \the\chapterno.\the\sectionno.\the\subsectionno\ #1}
 \subsectioncont{#1}}
\def\appendix#1{
 \vfil\eject
 \global\advance\appno by1 \subsectionno=0 \eqnum=0
 \def\mainid{\ifcase\appno\or A\or B\or C\or D\or E\or F\or G\or H\or \fi}
 \vglue.8cm\centerline{\sectionfont
 \uppercase{\mainid .\ \ #1}}
 \nobreak\vskip.2cm\nobreak
 \global\chap=1
 \appendixcont{#1}}

\def\loneappendix#1{
 \vfil\eject
 \global\advance\appno by1 \subsectionno=0 \eqnum=0
 \def\mainid{\ifcase\appno\or A\or B\or C\or D\or E\or F\or G\or H\or \fi}
 \vglue.8cm\centerline{\sectionfont
 \uppercase{APPENDIX: #1}}
 \nobreak\vskip.2cm\nobreak
 \global\chap=1
 \appendixcont{#1}}
\def\Slabel#1{\xdef#1{\mainid
              \ifnum\sectionno>0
                    { {.\the\sectionno}
                        \ifnum\subsectionno>0
                             { .\the\subsectionno}
                         \fi}
                \fi}}
%
\newif\iftable \tablefalse
\newwrite\tablelist
\def\starttablelist{\iftable
 \immediate\openout\tablelist=tablelist
 \immediate\write\tablelist{\noexpand\centerline{\headingfont LIST OF TABLES}}
 \immediate\write\tablelist{\vskip\double\vskip\single\baselineskip=\single}
 \immediate\write\tablelist{\parskip=0pt \parindent=0pt}\else\relax\fi}
\def\inserttable#1#2{\global\advance\tableno by 1 \vfill\vskip\double
 \centerline{Table \the\tableno:\ #2}\nobreak\vskip\double\nobreak
 \begingroup #1\endgroup \vskip\double
 \iftable\immediate\write\tablelist{\hskip1cm
 \the\tableno. #2 \hfill\folio\vskip0cm}\else\relax\fi}
\def\andinserttable#1#2{\vfill\vskip\double
 \centerline{Table \the\tableno\ (continued):\ #2}
 \nobreak\vskip\double\nobreak
 \begingroup #1\endgroup \vskip\double}

\def\noadvancetable#1#2{\vfill\vskip\double
 \centerline{Table \the\tableno:\ #2}\nobreak\vskip\double\nobreak
 #1 \vskip\double
 \iftable\immediate\write\tablelist{\hskip1cm
 \the\tableno. #2 \hfill\folio\vskip0cm}\else\relax\fi}
\def\continuetable#1#2{\vfill\vskip\double
 \centerline{Table \the\tableno{ }(continued):\ #2}\nobreak
 \vskip\double\nobreak
 \begingroup #1\endgroup \vskip\double}
\def\Tlabel#1{\global\advance\tableno by 1 \xdef#1{\the\tableno} \the\tableno}
%
\newif\iffig \figfalse
\newwrite\figlist
\newwrite\figs
\def\startfiglist{\iffig\immediate\openout\figlist=figlist
 \immediate\openout\figs=figs
 \immediate\write\figlist{\noexpand\centerline{\headingfont LIST OF FIGURES}}
 \immediate\write\figlist{\vskip\double\vskip\double\baselineskip=\single}
 \immediate\write\figlist{\parskip=0pt \parindent=0pt}\else\relax\fi}
\def\insertpagefig#1{\iffig\global\advance\figno by 1
 \immediate\write\figs{\vfil\noexpand\eject\bodypaging\pageno=\the\pageno
 \noexpand\vglue 20cm}
 \immediate\write\figs{\noexpand\centerline{Figure \the\figno. #1}}
 \immediate\write\figlist{\hskip1cm
 \the\figno. #1 \hfill\folio\vskip0cm}\advance\pageno by1 \else\relax\fi}
\def\insertfig#1#2{\iffig\global\advance\figno by 1
 \vglue #2
 \centerline{Figure \the\figno. #1}
 \immediate\write\figlist{\hskip1cm
 \the\figno. #1 \hfill\folio\vskip0cm} \else\relax\fi}
\def\insertfig#1#2{\iffig\global\advance\figno by 1
 \vglue #2
 \centerline{Figure \the\figno. #1}
 \immediate\write\figlist{\hskip1cm
 \the\figno. #1 \hfill\folio\vskip0cm} \else\relax\fi}

\def\Flabel#1{\global\advance\figno by 1 \xdef#1{\the\figno} \the\figno}
\def\noadvancefig#1#2{\iffig
 \vglue #2
 \centerline{Figure \the\figno. #1}
 \immediate\write\figlist{\hskip1cm
 \the\figno. #1 \hfill\folio\vskip0cm} \else\relax\fi}


\input tables  

\galley  

%
%
%
%
\def \arglt#1#2#3#4{(#1;#2,#3,\ldots,#4)}
\def \argrt#1#2#3#4{(#1,#2,\ldots,#3;#4)}
\def \argltpos#1#2#3#4{(#1^{+};#2,#3,\ldots,#4)}
\def \argltneg#1#2#3#4{(#1^{-};#2,#3,\ldots,#4)}
\def \argrtpos#1#2#3#4{(#1,#2,\ldots,#3;#4^{+})}
\def \argrtneg#1#2#3#4{(#1,#2,\ldots,#3;#4^{-})}
\def \argltbreak#1#2#3#4#5{(#1;#2,\ldots,#3,#4,\ldots,#5)}
\def \argrtbreak#1#2#3#4#5{(#1,\ldots,#2,#3,\ldots,#4;#5)}
\def \argltdblbreak#1#2#3#4#5#6#7{(#1;#2,\ldots,#3,#4,\ldots,
     #5,#6,\ldots,#7)}
\def \argrtdblbreak#1#2#3#4#5#6#7{(#1,\ldots,#2,#3,\ldots,
     #4,#5,\ldots,#6;#7)}
\def \argltposbreak#1#2#3#4#5{(#1^{+};#2,\ldots,#3,#4,\ldots,#5)}
\def \argltnegbreak#1#2#3#4#5{(#1^{-};#2,\ldots,#3,#4,\ldots,#5)}

%
\def \positronon{\bar\psi}
\def \electronon{\psi}

\def \Wnorm{W}
\def \Wms{{\cal{W}}}
\def \Wbar{{\overline{\Wms}}}

\def \PHI{{\mit\Phi}}

\def \pole{{\mit\Pi}}

\def \gauge{{\cal G}}

\def \dblscalar{{\cal G}}

\def \amp{{\cal M}}

%
%
\def \permsum#1#2{\sum_{{\cal{P}}(#1\ldots #2)}}

\def \braket#1#2{ \langle #1 \thinspace\thinspace #2 \rangle }
\def \bra#1{ \langle #1 | }
\def \ket#1{ | #1 \rangle }
\def \num{{\cal{N}}}

\def \jay{{j}}

\def \eps{\epsilon}
\def \vareps{\varepsilon}
\def \down{{}_\downarrow}
\def \up{{}_\uparrow}
\def \ts{\thinspace}


\def \dyadic#1{\vbox{\ialign{##\crcr
     $\hfil
{\thinspace\scriptstyle\leftrightarrow}
\hfil$\crcr\noalign{\kern-.01pt\nointerlineskip}
     $\hfil\displaystyle{#1}\hfil$\crcr}}}

\def\centeronto#1#2{{\setbox0=\hbox{#1}\setbox1=\hbox{#2}\ifdim
\wd1>\wd0\kern.5\wd1\kern-.5\wd0\fi
\copy0\kern-.5\wd0\kern-.5\wd1\copy1\ifdim\wd0>\wd1
\kern.5\wd0\kern-.5\wd1\fi}}
\def\slash#1{\centeronto{$#1$}{$/$}}

\def \varsp{\thinspace}


\hyphenation{ap-pen-dix  spin-or  spin-ors}

\def \msreplacefermion{A.20}

\def \msconvent{A}

\def \cc{B}

\def \polepropappendix{C}

\def \prtosm{D}

\def \slashsqr{A.6}

\def \fierz{A.14}

\def \linkidnosum{A.15}
\def \linkidsummed{A.16}

\def \splitid{C.6}


\def \spinnorm{A.11}

\def \antisym{A.13a}

\def \reverseid{C.4}

\def \Dstart{D.11}

\def \Dend{D.26}

\def \RELABELstart{D.22}

\def \RELABELend{D.23}

%
%
{\nopagenumbers
\hfill\hbox{
CLNS 91/1119}

\hfill\hbox{
September 1992}
\vfill
\baselineskip=\double
{\titlefont
        \centerline{MULTIPHOTON PRODUCTION AT HIGH}
        \centerline{ENERGIES IN THE  STANDARD MODEL I}
}
\bigskip
\bigskip
\bigskip
\centerline{Gregory MAHLON$^{1}$ and Tung--Mow YAN$^{2}$}
\medskip

{\baselineskip=.20in plus 2pt minus 2pt
\centerline{
{\it Newman Laboratory of Nuclear Studies,}
}
\centerline{
{\it Cornell University,
Ithaca, NY 14853, USA}
}
}

\bigskip
\bigskip
\bigskip
\centerline{ABSTRACT}
{\narrower\medskip\baselineskip=.20in plus 2pt minus 2pt
{\abstractfont
We examine multiphoton production in the electroweak
sector of the Standard Model  in the high energy
limit using the equivalence theorem in combination with
spinor helicity techniques.  We obtain recursion relations
for currents consisting of a charged scalar, spinor, or
vector line that radiates {\abstractitalfont n}\ photons.
Closed form solutions  to these recursion relations
for arbitrary {\abstractitalfont n}\ are
presented for the cases of like-helicity and one
unlike-helicity photon production.  We apply the currents
singly and in pairs to obtain amplitudes for processes
involving  the production of {\abstractitalfont n}\ photons
with up to two unlike helicities from a pair
of charged particles.  The replacement of one or more photons
by transversely polarized {\abstractitalfont Z}-bosons is
also discussed.
}\smallskip}

\vfill

\noindent
\hrule
\smallskip

\noindent
$^1$ e-mail:  gdm@beauty.tn.cornell.edu

\noindent
$^2$ e-mail:  yan@lnssun5.tn.cornell.edu

\eject}

\pageno=2  \bodypaging
\startrefs{refs1}

\chapter{INTRODUCTION}

In this paper we will continue a study of the high energy
scatterings involving many vector bosons and Higgs bosons
in a spontaneously symmetry-broken gauge theory begun in
reference
[\ref{C. Dunn and T.--M. Yan, Nucl. Phys. {\bf B352}, 402 (1991).}
{\kern-.354em}{].}\refname\DY
In particular, we will consider the Weinberg-Salam-Glashow
model
[\ref{S. L. Glashow, Nucl. Phys. {\bf 22}, 579 (1961);
S. Weinberg, Phys. Rev. Lett. {\bf 19}, 1264 (1967);
A. Salam in {\it Proc. 8th Nobel Symposium,
Aspen\"asgarden,} edited by N. Svartholm, (Almqvist and
Wiksell, Stockholm, 1968) p. 367.}],
\refname\WSGmodel{\kern-.4em}
with a focus upon processes involving
an arbitrary number of photons, plus a pair of  charged
particles ($W^{+}W^{-}$, $\ell^{\pm}W^{\mp}$, or $\ell^{+}\ell^{-}$),
and possibly a $Z$, Higgs boson, or neutrino.
This work is a generalization of the analysis of Berends and Giele
[\ref{F. A. Berends and W. T. Giele, Nucl. Phys. {\bf B306},
759 (1988).}],
\refname\BG  who discussed the case of
electron-positron annihilation to $n$ photons, with one
photon of opposite helicity.  Kleiss and
Stirling~[\ref{R. Kleiss and W.J. Stirling,
Phys. Lett. {\bf 179B}, 159 (1986).}~{\kern-.4em}~]\refname{\KS}
have examined this process for an arbitrary helicity configuration.
Their result is presented in a form
which is convenient for
numerical evaluation.  The special
cases for $n<4$ have been studied in detail by the CALCUL
collaboration~[\ref{
P. De Causmaecker, R. Gastmans, W. Troost, and T.T. Wu,
Phys. Lett. {\bf 105B}, 215 (1981);
P. De Causmaecker, R. Gastmans, W. Troost, and T.T. Wu,
Nucl. Phys. {\bf B206}, 53 (1982);
F. A. Berends, R. Kleiss, P. De Causmaecker, R. Gastmans,
W. Troost, and T.T. Wu, Nucl. Phys. {\bf B206}, 61 (1982);
F.A. Berends, P. De Causmaecker, R. Gastmans, R. Kleiss,
W. Troost, and T.T. Wu, Nucl. Phys. {\bf B239}, 382 (1984);
{\bf B239}, 395 (1984); {\bf B264}, 243 (1986); {\bf B264}, 265 (1986).
}].\refname\CALCUL

The work here and in reference \cite\DY\ is based upon
three main ideas.  First, we employ
the equivalence theorem
[\ref{J. M. Cornwall, D. N. Levin, and G. Tiktopoulous,
Phys. Rev. {\bf D10}, 1145 (1974);
B. W. Lee, C. Quigg,
and H. Thacker, Phys. Rev. {\bf D16}, 1519 (1977);
M. S. Chanowitz and M. K. Gaillard, Nucl. Phys. {\bf B261},
379 (1985);
G. J. Gounaris, R. Kogerler and H. Neufeld,
Phys. Rev. {\bf D34}, 3257 (1986).}],
\refname\ET
which allows us to identify
the longitudinal degrees of freedom of the $W^{\pm}$ and $Z$ bosons
with the corresponding would-be Goldstone bosons  $\phi^{\pm}$ and
$\phi_2$, up to corrections of the order of the vector boson mass
divided by the center-of-mass energy.
Second, the multispinor representation of a vector field
[\ref{J. Schwinger, {\it Particles, Sources and Fields,}\
(Addison-Wesley, Redwood City, 1970), Vol. I;
Ann. Phys. {\bf 119}, 192 (1979).},
\refname\SCHWINGER  {\kern-.75em}
\ref{
The spinor technique was first introduced by the
CALCUL collaboration, in the context
of massless Abelian gauge theory (ref. \cite\CALCUL).
By now, many papers have been published on the subject.  A
partial list of references follows.
\smallskip
\noexpand\item{ }
P. De Causmaecker, thesis, Leuven University, 1983;
R. Farrar and F. Neri, Phys. Lett. {\bf 130B}, 109 (1983);
R. Kleiss, Nucl. Phys. {\bf B241}, 61 (1984);
Z. Xu, D.H. Zhang, and Z. Chang, Tsingua University
preprint TUTP-84/3, 84/4, and 84/5a (1984), and
Nucl. Phys. {\bf B291}, 392 (1984);
J.F. Gunion and Z. Kunszt, Phys. Lett. {\bf 161B}, 333 (1985);
F.A. Berends, P.H. Davereldt, and R. Kleiss, Nucl. Phys. {\bf B253},
441 (1985);
R. Kleiss and W.J. Stirling, Nucl. Phys. {\bf B262}, 235 (1985);
J.F. Gunion and Z. Kunszt, Phys. Lett. {\bf 159B}, 167 (1985);
{\bf 161B}, 333 (1985);
S.J. Parke and T.R. Taylor, Phys. Rev. Lett. {\bf 56}, 2459 (1986);
Z. Kunszt, Nucl. Phys. {\bf B271}, 333 (1986);
J.F. Gunion and J. Kalinowski, Phys. Rev. {\bf D34}, 2119 (1986);
R. Kleiss and W.J. Stirling, Phys. Lett. {\bf 179B}, 159 (1986);
M. Mangano and S.J. Parke, Nucl. Phys. {\bf B299}, 673 (1988);
M. Mangano, S.J. Parke, and Z. Xu, Nucl. Phys. {\bf B298},
653 (1988);
D.A. Kosower, B.--H. Lee, and V.P. Nair, Phys. Lett. {\bf 201B},
85 (1988);
M. Mangano and S.J. Parke, Nucl. Phys. {\bf B299}, 673 (1988);
F.A. Berends and W.T. Giele, Nucl. Phys. {\bf B313}, 595 (1989);
M. Mangano, Nucl. Phys. {\bf B315}, 391 (1989);
D.A. Kosower, Nucl. Phys. {\bf B335}, 23 (1990);
Phys. Lett. {\bf B254}, 439 (1991);
Z. Bern and D.A. Kosower, Nucl. Phys. {\bf B379}, 451 (1992);
C.S. Lam, McGill preprint McGill/92-32, 1992.},
\refname\REVIEW {\kern-.75em}
\ref{
Many of the results for processes containing six or fewer particles are
collected in
R. Gastmans and T.T. Wu, {\it The Ubiquitous Photon:  Helicity Method
for QED and QCD} (Oxford University Press, New York, 1990).},
\refname\UBIQUITOUS {\kern-.75em}
\ref{
The excellent
review by Mangano and Parke provides
a guide to the
various approaches to  and extensive literature on the subject:
M. Mangano and S.J. Parke,  Phys. Reports {\bf 200}, 301 (1991).}]
\refname\REVIEWx\
allows us to treat
fermions and vector bosons on an equal footing by replacing the
conventional Lorentz 4-vector with  a second rank spinor which may be
thought of as a combination of two spin-$1\over2$ objects.
We will use Weyl-van der Waerden spinors in this work
(see Appendix \msconvent{ }for a summary of our conventions).
The final essential tool in our analysis is the use of recursion
relations \cite{\DY,\BG}.

Whereas reference \cite\DY\ seeks to produce a general analysis which is
valid for an arbitrary non-Abelian group $U(N)$,
for our purpose it is more convenient to
treat  the Weinberg-Salam-Glashow model as if it were merely
the quantum electrodynamics of three types of charged particles
(scalar, spinor, and vector), plus additional interactions which will be
handled as they are needed.
The couplings are determined
by the $SU(2)\times U(1)$ symmetry present in the standard
electroweak model.  In addition,  there are the  neutral
particles (Higgs, neutrino, and $Z$) in the  broken symmetry
group which must be accounted for.
There are several reasons for examining  multiphoton processes in the
Weinberg-Salam-Glashow model apart
from the $U(N)$ formalism.  The more direct
approach   eliminates the group
factors present in the expressions derived using the $U(N)$ formalism,
simplifying some aspects of the calculation.  In addition, the
$SU(2)$--$U(1)$ mixing is explicitly accounted for.  If we were to use
$U(N)$ formalism, an appropriate linear combination of amplitudes would
be required to discuss the scattering of $W$'s, $Z$'s and photons,
instead of {\bf A}'s and $B$'s.
Furthermore,  for processes that involve
only a few $W$'s and $Z$'s, it is
more economical to take care of these particles directly rather than
by the full machinery of the $U(N)$ formalism.  For example, consider
processes such as
$$
e^{+} e^{-}\longrightarrow W^{+} W^{-} \gamma\ldots\gamma.
\eqlabel\sampleprocess
$$
The exchanged particle, which must be neutral, contributes only
a propagator to the amplitude.  This more direct approach
allows us to obtain results for somewhat more
complicated helicity configurations.

This paper lays the groundwork for considering processes
such as (\sampleprocess).  In Section 2 we
introduce currents consisting of a charged line
plus $n$ photons.  The charged line may have spin 0,
representing the would-be Goldstone bosons $\phi^{\pm}$, spin $1\over2$,
representing charged leptons, or spin 1, representing the transverse
degrees of freedom of the $W$ bosons.  All of the photons are on shell,
but only one end of the charged
line is on shell.  We exploit the Bose symmetry enjoyed by the
photons to write the recursion relations for these currents
in the form of a sum over permutations of the photons.
In Section 3, we present closed-form solutions
to the recursion relations
for two special  helicity configurations:  those with all
like-helicity bosons, and those that contain a single opposite-helicity
boson.   All of these currents may be written in terms of a single
function.  This product may be simplified by taking advantage
of the explicit sum on permutations appearing in the expression for
the current.  As in the $U(N)$ case, we find that
currents with differing spins are connected by a simple proportionality
\cite\DY.
In Section~4, we present the various on-shell scattering amplitudes
which may be obtained from the currents taken either singly or in
pairs.  When two currents are combined, we can insert a photon of
either helicity at the joint.  Thus, we are able to produce amplitudes
with as many as two unlike helicity photons.
We find that for the simplest helicity configurations leading
to non-vanishing amplitudes, the
supersymmetric-like relations remain simple proportionalities.
When more complicated amplitudes are constructed,
however,
we find that they share the same over-all structure, but  without simple
proportionalities connecting them.  We finish this section
with a discussion of processes in which some of the photons
are replaced by transversely polarized $Z$-bosons.
Section 5 contains a few concluding remarks.

\chapter{THE RECURSION RELATIONS}

In this section we will review the recursion relation of Berends and
Giele~\cite\BG\ for
a charged fermion line with an
arbitrary number of photons attached.
We will then generalize their work to include the cases of a  charged
vector ($W_T$) or scalar ($W_L$) line.


\section{The fermion currents}

We start by defining the tree-level fermion current $\positronon
\arglt{p}{1}{2}{n}$.
Taking the convention that all momenta  flow into the diagram, we define
this current to have an on-shell positron of momentum $p$,
$n$ on-shell photons of momenta $k_1, k_2,\ldots,k_n$, and an off-shell
electron with momentum
$q=-(p+k_1+k_2+\cdots+k_n)\equiv  -[p+\kappa(1,n)]$.
The $n$ photons are attached in all possible ways.
Berends and Giele \cite\BG\ obtain the following
recursion relation for this quantity:
$$
\positronon\arglt{p}{1}{2}{n}=-e \sum_{j=1}^n
\positronon\argltbreak{p}{1}{j{-}1}{j{+}1}{n}
\slash{\eps}(j){1\over{\slash p+\slash \kappa(1,n)}}.
\eqlabel\recursionposonbreak
$$
In the massless limit being considered here, the helicity of each
fermion line is conserved.  Thus, when we translate to multispinor
formalism using the replacement rules (\msreplacefermion),
we obtain two distinct quantities with very similar recursion relations:
$$
\eqalign{
{{\positronon}_{\dot\alpha}}&\argltpos{p}{1}{2}{n}=
\cr & =
-e\sqrt2
\sum_{j=1}^n \positronon_{\dot\beta}\argltposbreak{p}{1}{j{-}1}{j{+}1}{n}
{\bar{\eps}}^{\dot\beta\beta}(j)
{{[p+\kappa(1,n)]_{\beta\dot\alpha}}\over{[p+\kappa(1,n)]^2}}
}
\eqlabel\recursionposRHonbreak
$$
$$
\eqalign{
{{\positronon}^{\alpha}}&\argltneg{p}{1}{2}{n}=
\cr & =
-e\sqrt2
\sum_{j=1}^n \positronon^{\beta}\argltnegbreak{p}{1}{j{-}1}{j{+}1}{n}
{\eps}_{\beta\dot\beta}(j)
{{[\bar p+\bar \kappa(1,n)]^{\dot\beta\alpha}}
\over{[p+\kappa(1,n)]^2}},
}
\eqlabel\recursionposLHonbreak
$$
where the superscript signs denote the helicity of the positron.
As we will see later, the main difference between the solutions of
(\recursionposRHonbreak) and (\recursionposLHonbreak) is their
spinor structure.

We also define $\electronon\argrt{1}{2}{n}{q}$, which
has an on-shell electron of momentum $q$, $n$ on-shell
photons of momenta $k_1, k_2,\ldots,k_n$, and an off-shell
positron with momentum  $p=-[\kappa(1,n)+q]$.
This quantity is simply related to $\positronon\arglt{p}{1}{2}{n}$, as
can be seen immediately from the appropriate recursion relations, which
may be written down in analogous fashion to (\recursionposonbreak).
The connection is
$$
{{\electronon}^{\dot\alpha}}\argrtpos{1}{2}{n}{q}=
(-1)^n{{\positronon}^{\dot\alpha}}\argltpos{q}{1}{2}{n}
\eqlabel\recursioneleRHonbreak
$$
$$
{{\electronon}_{\alpha}}\argrtneg{1}{2}{n}{q}=
(-1)^n{{\positronon}_{\alpha}}\argltneg{q}{1}{2}{n},
\eqlabel\recursioneleLHonbreak
$$
which is precisely what one would anticipate from the requirements of
charge-conjugation symmetry.


\section{The vector currents}

Next, we define currents to represent the transverse polarizations of
the $W$ bosons.  The quantity
${\Wnorm}^{\mu} \arglt{P}{1}{2}{n}$ describes
an on-shell $W^{+}$ of momentum $P$, $n$ on-shell photons of momenta
$k_1,k_2,\ldots,k_n$, and an off-shell $W^{-}$ with momentum
$Q=-[P+\kappa(1,n)]$.  This definition gives us
$$
\eqalign{
{\Wnorm}^{\mu} &\arglt{P}{1}{2}{n} =
\cr & =
\sum_{j=1}^n {\Wnorm}_{\lambda} \argltbreak{P}{1}{j{-}1}{j{+}1}{n}\cr
&\qquad\times
(-ie)V^{\lambda\mu\nu}[P+\kappa(1,j{-}1)+\kappa(j{+}1,n),
-P-\kappa(1,n), k_j]\cr
&\qquad\times
{ {-i {\eps}_{\nu}(j)} \over
{[P+\kappa(1,n)]^2} } \cr
& + \sum_{j=1}^{n-1} \sum_{\ell=j+1}^{n}
{\Wnorm}_{\lambda}
\argltdblbreak{P}{1}{j{-}1}{j{+}1}{\ell{-}1}{\ell{+}1}{n} \cr
&\qquad\times
(-ie^2S^{\lambda\mu\nu\kappa})
{{ -i {\eps}_{\nu}(j){\eps}_{\kappa}(\ell) } \over
{[P+\kappa(1,n)]^2} } ,
}
\eqlabel\WTstart
$$
where we have used the notations
$$
V^{\lambda\mu\nu}(k_1,k_2,k_3)=g^{\lambda\mu}(k_1-k_2)^{\nu}
                              +g^{\mu\nu}(k_2-k_3)^{\lambda}
                              +g^{\nu\lambda}(k_3-k_1)^{\mu}
\eqlabel\Threepoint
$$
to designate the three-point vertex function and
$$
S^{\lambda\mu\nu\kappa}=2g^{\lambda\mu}g^{\nu\kappa}
                        -g^{\lambda\kappa}g^{\mu\nu}
                        -g^{\lambda\nu}g^{\mu\kappa}
\eqlabel\Fourpoint
$$
for the seagull vertex function.   The first term  $\Wnorm^{\mu}_1$
of (\WTstart) corresponds to all possible ways of adding
a single photon to a $(n-1)$-photon current,
while the second term $\Wnorm^{\mu}_2$ adds two photons
at seagull vertices to a
$(n-2)$-photon current in all possible ways.

If we insert the definition (\Threepoint) into $\Wnorm^{\mu}_1$ we obtain
$$
\eqalign{
\Wnorm^{\mu}_1 &=
{-e\over{[P+\kappa(1,n)]^2}}
\cr & \qquad\times
\sum_{j=1}^n
\Bigl\{
2{\Wnorm}^{\mu} \argltbreak{P}{1}{j{-}1}{j{+}1}{n}
\thinspace\eps(j)\cdot[P+\kappa(1,n)]
\cr& \qquad\quad
- {\eps}^{\mu}(j)
\thinspace\Wnorm \argltbreak{P}{1}{j{-}1}{j{+}1}{n}
\cdot[P+\kappa(1,n)+k_j]
\cr& \qquad\quad
+ \eps(j)\cdot \Wnorm \argltbreak{P}{1}{j{-}1}{j{+}1}{n}
\thinspace [2k_j-P-\kappa(1,n)]^{\mu}\Bigr\},
}
\eqlabel\threepointa
$$
where we have made use of the gauge condition $k_m\cdot\eps(m)=0$.
The current $\Wnorm$ is conserved, in that it satisfies the property
$$
[P+\kappa(1,n)]\cdot\Wnorm\arglt{P}{1}{2}{n}=0
\eqlabel\Wcc
$$
(see Appendix \cc).  Thus, (\threepointa) may be rewritten as
$$
\eqalign{
\Wnorm_1^{\mu}& =
{-e\over{[P+\kappa(1,n)]^2}}
\cr & \quad\times
\sum_{j=1}^n
\Bigl\{
2\eps(j)\cdot[P+\kappa(1,n)-k_j]\thinspace
{\Wnorm}^{\mu} \argltbreak{P}{1}{j{-}1}{j{+}1}{n}
\cr&\qquad
-2k_j\cdot\Wnorm \argltbreak{P}{1}{j{-}1}{j{+}1}{n}\thinspace
{\eps}^{\mu}(j)
\cr&\qquad
+\eps(j)\cdot \Wnorm \argltbreak{P}{1}{j{-}1}{j{+}1}{n}\thinspace
{ \bigl[ k_j-[P{+}\kappa(1,n){-}k_j]\bigr] }^{\mu}\Bigr\},
}
\eq
$$
which is easily recognized as being equivalent to
$$
\Wnorm^{\mu}_1=
{-e\over{[P+\kappa(1,n)]^2}}
\sum_{j=1}^n
\bigl[\eps(j),\Wnorm \argltbreak{P}{1}{j{-}1}{j{+}1}{n}
{\bigr]}^{\mu}.
\eqlabel\threepointb
$$
In (\threepointb) the square brackets, defined by
$$
\bigl[J(1),J(2)\bigr]^{\mu}= 2k_2 \cdot J(1)\thinspace J^{\mu}(2)
- 2k_1 \cdot J(2) \thinspace J^{\mu}(1)
+ J(1)\cdot J(2)\thinspace(k_1-k_2)^{\mu},
\eqlabel\sqbrak
$$
represent the same function occurring in the
discussion of $U(N)$ recursion relations \cite{\DY,\BG}.

If we perform the contractions
indicated in  the second term of (\WTstart), we obtain
$$
\eqalign{
\Wnorm_2^{\mu}&=
{{-e^2}\over{[P+\kappa(1,n)]^2}}
\cr & \quad\times
\sum_{j=1}^{n-1} \sum_{\ell=j+1}^n
\Bigl\{
2\eps(j)\cdot\eps(\ell)\thinspace
{\Wnorm}^{\mu}
\argltdblbreak{P}{1}{j{-}1}{j{+}1}{\ell{-}1}{\ell{+}1}{n}
\cr&\qquad
-\eps(\ell)\cdot\Wnorm
\argltdblbreak{P}{1}{j{-}1}{j{+}1}{\ell{-}1}{\ell{+}1}{n}
\thinspace{\eps}^{\mu}(j)
\cr&\qquad
-\eps(j)\cdot\Wnorm
\argltdblbreak{P}{1}{j{-}1}{j{+}1}{\ell{-}1}{\ell{+}1}{n}
\thinspace{\eps}^{\mu}(\ell)\Bigr\},
}
\eqlabel\seagulla
$$
which may be written
$$
\eqalign{
\Wnorm_2^{\mu}&=
{{-e^2}\over{[P+\kappa(1,n)]^2}}
\cr & \quad\times
\sum_{j=1}^{n-1} \sum_{\ell=j+1}^n
\bigl\{ \eps(j),
\Wnorm\argltdblbreak{P}{1}{j{-}1}{j{+}1}{\ell{-}1}{\ell{+}1}{n},
\eps(\ell)
{\bigr\}}^{\mu}.
}
\eqlabel\seagullb
$$
The curly brackets in (\seagullb) are again the same four-point function
appearing in references \cite\DY\ and \cite\BG , namely
$$
\eqalign{
\bigl\{ J(1),J(2),J(3) {\bigr\}}^{\mu}& =
J(1)\cdot[J(3)J^{\mu}(2)-J(2) J^{\mu}(3)]
\cr & \quad
-J(3)\cdot[J(2)J^{\mu}(1)-J(1) J^{\mu}(2)].
}
\eqlabel\curlybrak
$$
Thus, the recursion relation for $\Wnorm \arglt{P}{1}{2}{n}$ reads
$$
\eqalign{
\Wnorm&\arglt{P}{1}{2}{n} =
\cr & =
{-e\over{[P+\kappa(1,n)]^2}}
\Biggl[
\sum_{j=1}^n
\bigl[\eps(j),\Wnorm \argltbreak{P}{1}{j{-}1}{j{+}1}{n}\bigr]
\cr&\quad
+e
\sum_{j=1}^{n-1} \sum_{\ell=j+1}^{n}
\bigl\{ \eps(j),
\Wnorm\argltdblbreak{P}{1}{j{-}1}{j{+}1}{\ell{-}1}{\ell{+}1}{n},
\eps(\ell)
\bigr\}
\Biggr].
}
\eqlabel\WTrecursionnorm
$$

We define a current with an on-shell $W^{-}$
in analogous fashion, and denote it by
${\Wnorm}_{\mu} \argrt{1}{2}{n}{Q}$.
Following the same procedure as before, we obtain the recursion relation
$$
\eqalign{
\Wnorm&\argrt{1}{2}{n}{Q} =
\cr & =
{e\over{[\kappa(1,n)+Q]^2}}
\Biggl[
\sum_{j=1}^n
\bigl[\eps(j),\Wnorm \argrtbreak{1}{j{-}1}{j{+}1}{n}{Q}\bigr]
\cr&\quad
-e
\sum_{j=1}^{n-1} \sum_{\ell=j+1}^{n}
\bigl\{ \eps(j),
\Wnorm\argrtdblbreak{1}{j{-}1}{j{+}1}{\ell{-}1}{\ell{+}1}{n}{Q},
\eps(\ell)\bigr\}
\Biggr].
}
\eqlabel\WTBARrecursionnorm
$$
Intuitively, we expect to find that
$$
\Wnorm \argrt{1}{2}{n}{Q}=(-1)^n\Wnorm\arglt{Q}{1}{2}{n},
\eqlabel\Wcrossing
$$
since the only difference between these two currents is the charge of
the line to which we have attached the $n$ photons.
This is obviously true for $n=0$ and $n=1$.  For $n>1$,
we proceed inductively.  Assuming (\Wcrossing)
to be true, we have, using (\WTBARrecursionnorm),
$$
\eqalign{
\Wnorm &\argrt{1}{2}{n}{Q} =
\cr & \negthinspace\negthinspace\negthinspace\negthinspace  =
{{e}\over{[\kappa(1,n)+Q]^2}}
\Biggl[
(-1)^{n-1}
\sum_{j=1}^n
\bigl[\eps(j),\Wnorm \argltbreak{Q}{1}{j{-}1}{j{+}1}{n}\bigr]
\cr& \negthinspace\negthinspace
-(-1)^{n-2}e
\sum_{j=1}^{n-1} \sum_{\ell=j+1}^{n}
\bigl\{ \eps(j),
\Wnorm\argltdblbreak{Q}{1}{j{-}1}{j{+}1}{\ell{-}1}{\ell{+}1}{n},
\eps(\ell)\bigr\}
\Biggr]
\cr& \negthinspace\negthinspace\negthinspace\negthinspace
=(-1)^n\Wnorm\arglt{Q}{1}{2}{n},
}
\eqlabel\established
$$
which proves (\Wcrossing).


\section{The scalar currents}

Finally, we define currents to represent the longitudinal polarizations
of the $W$ bosons. By the equivalence theorem \cite\ET ,
these are associated with
the would-be Goldstone bosons ${\phi}^{+}$ and ${\phi}^{-}$.
The quantity $\PHI \arglt{P}{1}{2}{n}$ represents
an on-shell ${\phi}^{+}$ of momentum $P$, $n$ on-shell
photons of momenta
$k_1,k_2,\ldots,k_n$, and an off-shell ${\phi}^{-}$ with momentum
$Q=-[P+\kappa(1,n)]$.  The vertices in this case are so simple that we
can immediately write down the recursion relation
$$
\eqalign{
\PHI&\arglt{P}{1}{2}{n}=
\cr & =
{{-e}\over{[P+\kappa(1,n)]^2}}
\Biggl[
\sum_{j=1}^n
2[P+\kappa(1,n)]\cdot\eps(j)
\thinspace\PHI\argltbreak{P}{1}{j{-}1}{j{+}1}{n}
\cr&\thinspace\thinspace\thinspace\thinspace
+2e
\sum_{j=1}^{n-1} \sum_{\ell=j+1}^{n}
\eps(j)\cdot\eps(\ell)
\thinspace\PHI\argltdblbreak{P}{1}{j{-}1}{j{+}1}{\ell{-}1}{\ell{+}1}{n}
\Biggr],
}
\eqlabel\WLrecursionnorm
$$
where we define
$$
\PHI(P) \equiv 1.
\eqlabel\PHIzero
$$
Likewise, when the ${\phi}^{+}$ is off shell we obtain
$$
\eqalign{
\PHI&\argrt{1}{2}{n}{Q}=
\cr & =
{{-e}\over{[\kappa(1,n)+Q]^2}}
\Biggl[
\sum_{j=1}^n
2[\kappa(1,n)+Q]\cdot\eps(j)
\thinspace\PHI\argrtbreak{1}{j{-}1}{j{+}1}{n}{Q}
\cr&\thinspace\thinspace\thinspace\thinspace
+2e
\sum_{j=1}^{n-1} \sum_{\ell=j+1}^{n}
\eps(j)\cdot\eps(\ell)
\thinspace\PHI\argrtdblbreak{1}{j{-}1}{j{+}1}{\ell{-}1}{\ell{+}1}{n}{Q}
\Biggr].
}
\eqlabel\WLBARrecursion
$$
These currents are related to each other in the expected manner:
$$
\PHI\argrt{1}{2}{n}{Q}=(-1)^n\PHI\arglt{Q}{1}{2}{n}.
\eqlabel\WLcrossing
$$
The proof of (\WLcrossing) is virtually identical to the one for
(\Wcrossing).


\section{Permutation symmetric form of the recursion relations}
All of the currents which we we have defined consist of a charged
line plus $n$ photons attached to that line in all possible ways.
This means that by definition these currents are symmetric with
respect to the permutation of their photon arguments
(momentum {\it and} helicity labels together).
So, the exact order of the photon arguments doesn't matter,
only their momentum and helicity
content.  This allows us to write all of the recursion relations in a
manifestly permutation symmetric form.  If we define the notation
${{\cal{P}}(1\ldots n)}$  to tell us
to sum over all permutations of the labels $(1,\ldots ,n)$,
(\recursionposRHonbreak) becomes
$$
\eqalign{
{{\positronon}_{\dot\alpha}}&\argltpos{p}{1}{2}{n}=
\cr & =
-e\sqrt2
\sum_{\ell=1}^n
{\sum_{{\cal{P}}(1\ldots \ell{-}1 \thinspace \ell{+}1 \ldots n)}}
{
{1 }
\over
{(n-1)!}
}
\cr&\qquad\qquad\times
\positronon_{\dot\beta}(p^{+};1,\ldots,\ell{-}1,\ell{+}1,\ldots,n)
{\bar{\eps}}^{\dot\beta\beta}(\ell)
{{[p+\kappa(1,n)]_{\beta\dot\alpha}}\over{[p+\kappa(1,n)]^2}}
\cr&=
-e\sqrt2
\permsum{1}{n} {{1}\over{(n-1)!}}
\positronon_{\dot\beta}\argltpos{p}{1}{2}{n{-}1}
{\bar{\eps}}^{\dot\beta\beta}(n)
{{[p+\kappa(1,n)]_{\beta\dot\alpha}}\over{[p+\kappa(1,n)]^2}},
}
\eqlabel\recursionposRHonperm
$$
with a similar result corresponding to (\recursionposLHonbreak).
Likewise, (\WTrecursionnorm) and (\WLrecursionnorm) may be rewritten as
$$
\eqalign{
\Wnorm&\arglt{P}{1}{2}{n} =
\cr & =
{-e\over{[P+\kappa(1,n)]^2}}
\Biggl[
\permsum{1}{n}
{1\over{(n-1)!}} \thinspace
[\eps(n),\Wnorm \arglt{P}{1}{2}{n{-}1}]
\cr&\quad
+e\permsum{1}{n}
{1\over{2!\thinspace (n-2)!}} \thinspace
\{ \eps(n{-}1),\Wnorm\arglt{P}{1}{2}{n{-}2},\eps(n)
\}
\Biggr]
}
\eqlabel\WTrecursionperm
$$
and
$$
\eqalign{
\PHI&\arglt{P}{1}{2}{n}=
\cr &\negthinspace\negthinspace =
{{-e}\over{[P+\kappa(1,n)]^2}}
\Biggl[
\permsum{1}{n} \negthinspace
{1\over{(n-1)!}}
 2[P{+}\kappa(1,n)]\cdot\eps(n)
\thinspace
\PHI\arglt{P}{1}{2}{n{-}1}
\cr&\quad\negthinspace\negthinspace
+2e
\permsum{1}{n}
{1\over{2!\thinspace (n-2)!}} \thinspace
 \eps(n{-}1)\cdot\eps(n)
\thinspace\PHI\arglt{P}{1}{2}{n{-}2}
\Biggr].
}
\eqlabel\WLrecursionperm
$$
We will find that the permutation symmetric forms of the recursion
relations are often more useful than their ``direct'' counterparts.

\chapter{SOLUTIONS TO THE RECURSION RELATIONS}

In this section we present closed form solutions
to the recursion relations
for two special helicity configurations.  The first is the case where
all of the boson helicities are the same, while the second allows one
of the boson helicities to be different.  There is no restriction on the
helicity of the fermions in the fermionic currents.

As  in the $U(N)$ case \cite\DY, we will find  that there are
supersymmetric-like relations among
the currents.  That is, currents involving charged lines of
differing spins will be connected by a simple proportionality.

\section{Like helicity currents}

We consider a configuration with $n$ like helicity photons.
For concreteness, let the photons all have positive helicity.
A current with all negative
helicity photons may be obtained by complex conjugation.
Our choice of gauge spinor for this configuration is the  same one
used in the discussion of the $U(N)$ formalism \cite{\DY,\BG}.
That is,
$$
{\eps}_{\alpha\dot\alpha}(j^{+})=
{{ u_{\alpha}(g) {\bar u}_{\dot\alpha}(k_j) }
\over {\braket{j}{g}}},
\eqlabel\polallplus
$$
where $g^2=0$, but is otherwise arbitrary.
As a consequence of this choice,
$$
{\eps}_{\alpha\dot\alpha}(i^{+}) {\bar\eps}^{\dot\alpha\alpha}(j^{+})=0.
\eqlabel\seagullkill
$$

\subsection{The scalar current}

Consider first the scalar current
with $n$ like-helicity photons attached.
Because of (\seagullkill), the seagull contributions to the recursion
relation (\WLrecursionperm) all vanish, leaving just
$$
\eqalign{
\PHI&(P;1^{+},\ldots,n^{+}) =
\cr & \negthinspace\negthinspace =
-e\sqrt2 \negthinspace\negthinspace\permsum{1}{n} \negthinspace
{ 1\over{(n-1)!} }
{ { {\bar u}_{\dot\beta}(k_n)[\bar P + \bar \kappa(1,n)]^
{\dot\beta\beta} u_{\beta}(g) } \over
{ \braket{n}{g} \varsp [P+\kappa(1,n)]^2 } }
\PHI\bigl(P;1^{+},\ldots,(n{-}1)^{+}\bigr).
}
\eqlabel\WLallplusrecurs
$$
If we iterate (\WLallplusrecurs), we can obtain the following product
form for the solution
$$
\PHI(P;1^{+},\ldots,n^{+}) = (-e\sqrt2)^n \permsum{1}{n}  \PHI(P)
\prod_{\ell=1}^n
{ { {\bar u}_{\dot\beta}(k_{\ell})[\bar P + \bar \kappa(1,\ell)]^
{\dot\beta\beta} u_{\beta}(g) } \over
{ \braket{\ell}{g} \varsp [P+\kappa(1,\ell)]^2 } }.
\eqlabel\WLallplusprod
$$
The product appearing inside the permutation sum
is universal in the sense that we will encounter it several times in
our solutions for various helicity configurations.
In Appendix \prtosm, we show that
$$
\eqalign{
\Xi(j,n)&\equiv   \permsum{j}{n} \prod_{\ell=j}^{n}
{
{ {\bar u}_{\dot\beta}(k_{\ell})
[\bar P + \bar \kappa(1,\ell)]^{\dot\beta\beta} u_{\beta}(g) }
\over
{ \braket{\ell}{g} \varsp [P+\kappa(1,\ell)]^2 }
}
\cr &
=
\permsum{j}{n}
{
{[P+\kappa(1,j{-}1)]^2}
\over
{\bra{g} j,\ldots,n \ket{g}}
}
\sum_{\ell=j}^n
u^{\alpha}(g)
{{\pole}_{\alpha}}^{\beta}(P,1,2,\ldots,\ell)
u_{\beta}(g).
}
\eqlabel\XIid
$$
In (\XIid) we have defined the pole factor
$$
{{\pole}_{\alpha}}^{\beta}(P,1,2,\ldots,\ell) \equiv
{
{(k_{\ell})_{\alpha\dot\alpha}
[\bar P + \bar \kappa(1,\ell)]^{\dot\alpha\beta}}
\over
{[P+\kappa(1,\ell{-}1)]^2
[P+\kappa(1,\ell)]^2}
}
\eqlabel\poledef
$$
as well as the notation
$$
\bra{g} 1,2,\ldots,n \ket{g} \equiv
\braket{g}{1} \braket{1}{2} \cdots \braket{n}{g}.
\eqlabel\stringdef
$$
We note the following basic properties of
$\bra{g} 1,2,\ldots,n \ket{g'}$:
$$
\bra{g}\thinspace\ket{g'} \equiv \braket{g}{g'}
\newlett
$$
$$
\bra{g} 1,2,\ldots,j{-}1 \ket{j} \bra{j} j{+}1,j{+}2,\ldots,n \ket{g'}
= \bra{g} 1,2,\ldots,n \ket{g'}
\lett
$$
$$
\bra{g'} n,n{-}1,\ldots,1 \ket{g} =
(-1)^{n-1} \bra{g} 1,2,\ldots,n \ket{g'}.
\lett
$$
Two useful identities involving the pole factor are developed in
Appendix \polepropappendix.

In the present case, we notice that (\WLallplusprod) is simply
$n$ powers of  the coupling constant times
$\Xi(1,n)$.  Thus, we may write
$$
\eqalign{
\PHI&(P;1^{+},\ldots,n^{+}) =
(-e\sqrt2)^n  \Xi(1,n) \cr& =
(-e\sqrt2)^n
\permsum{1}{n}
{
{P^2}
\over
{\bra{g} 1,\ldots,n \ket{g}}
}
\sum_{\ell=1}^n
u^{\alpha}(g)
{{\pole}_{\alpha}}^{\beta}(P,1,2,\ldots,\ell)
u_{\beta}(g).
}
\eqlabel\XIspecial
$$
Because $P^2=0$, the only surviving piece of (\XIspecial) is
the $\ell=1$ term, as it is the only term with a  compensating
pole $1/P^2$.  So, after a few minor cancellations, we end up
with the much simpler form
$$
\PHI(P;1^{+},\ldots,n^{+}) =
(-e\sqrt2)^n
\permsum{1}{n}
{
{\braket{P}{g}}
\over
{\bra{P} 1,\ldots,n \ket{g}}
}.
\eqlabel\PHIallplussoln
$$

\subsection{The fermion currents}

Next, let us consider the fermion current ${\positronon}_{\dot\alpha}
(p^{+};1^{+},\ldots,n^{+})$.  Using the choice (\polallplus) for the
polarization spinors, (\recursionposRHonperm) reads
$$
\eqalign{
{\positronon}_{\dot\alpha}&(p^{+};1^{+},\ldots,n^{+}) =
\cr & =
-e \sqrt2 \permsum{1}{n}  { 1\over{(n-1)!} }
{\bar u}^{\dot\beta}(k_n)
{\positronon}_{\dot\beta}\bigl(p^{+};1^{+},\ldots,(n{-}1)^{+}\bigr)
{
{ u^{\beta}(g)[p+\kappa(1,n)]_{\beta\dot\alpha}}
\over
{\braket{n}{g}\varsp [p+\kappa(1,n)]^2}
}.
}
\eqlabel\RHallplusrecurs
$$
We may write the zero-photon current as
$$
{\positronon}_{\dot\alpha}(p^{+}) = {\bar u}_{\dot\alpha}(p)
=
{
{ u^{\beta}(g) p_{\beta\dot\alpha} }
\over
{ \braket{g}{p} }
}.
\eqlabel\RHzero
$$
The presence of an explicit factor of
$u^{\beta}(g)[p+\kappa(1,n)]_{\beta\dot\alpha}$ on the
right hand side of (\RHallplusrecurs)
suggests the following ansatz for the spinor structure
of ${\positronon}_{\dot\alpha}(p^{+};1^{+},\ldots,n^{+})$:
$$
{\positronon}_{\dot\alpha}(p^{+};1^{+},\ldots,n^{+}) =
u^{\beta}(g)[p+\kappa(1,n)]_{\beta\dot\alpha}
Y(p^{+};1^{+},\ldots,n^{+}).
\eqlabel\RHspinor
$$
This is obviously true for $n=0$ with
$$
Y(p^{+})={ {-1}\over{\braket{p}{g}} }.
\eqlabel\Yzero
$$
The inductive proof for this ansatz is trivial.  Assuming (\RHspinor) to
be correct for $n-1$ photons, (\RHallplusrecurs) becomes
$$
\eqalign{
{\positronon}_{\dot\alpha}&(p^{+};1^{+},\ldots,n^{+}) =
\cr & =
-e \sqrt2 \permsum{1}{n}  { 1\over{(n-1)!} }
{
{ u^{\beta}(g)[p+\kappa(1,n{-}1)]_{\beta\dot\beta}{\bar u}^{\dot\beta}(k_n)}
\over
{ \braket{n}{g} \varsp [p+\kappa(1,n)]^2 }
}
\cr&\qquad\times
Y(p^{+};1^{+},\ldots,(n{-}1)^{+})
u^{\alpha}(g)[p+\kappa(1,n)]_{\alpha\dot\alpha}.
}
\eq
$$
So, we see that the ansatz is true provided that
$$
\eqalign{
Y&(p^{+};1^{+},\ldots,n^{+}) =
\cr & =
-e\sqrt2 \permsum{1}{n}
{ 1\over{(n-1)!} }
{ { {\bar u}_{\dot\beta}(k_n)[\bar p + \bar \kappa(1,n)]^
{\dot\beta\beta} u_{\beta}(g) } \over
{ \braket{n}{g} \varsp [p+\kappa(1,n)]^2 } }
Y\bigl(p^{+};1^{+},\ldots,(n{-}1)^{+}\bigr).
}
\eqlabel\Yallplusrecurs
$$
We immediately recognize (\Yallplusrecurs) as the same recursion
relation (\WLallplusrecurs) obtained for
$\PHI(P;1^{+},\ldots,n^{+})$.  Thus we can immediately write
$$
Y(p^{+};1^{+},\ldots,n^{+}) =
{ {Y(p^{+})} \over {\PHI(p)} }\PHI(p;1^{+},\ldots,n^{+}).
\eqlabel\Ysoln
$$
This is the first of the supersymmetry relations among the currents.
So, the final result for
${\positronon}_{\dot\alpha}(p^{+};1^{+},\ldots,n^{+})$
reads
$$
{\positronon}_{\dot\alpha}(p^{+};1^{+},\ldots,n^{+}) =
(-e\sqrt2)^n \permsum{1}{n}
{
{-u^{\beta}(g)[p+\kappa(1,n)]_{\beta\dot\alpha}}
\over
{ \bra{p} 1,2,\ldots,n \ket{g} }
}.
\eqlabel\RHallplussoln
$$

For fermions of negative helicity, we have the following recursion
relation:
$$
\eqalign{
{\positronon}^{\alpha}&(p^{-};1^{+},\ldots,n^{+}) =
\cr & =
-e \sqrt2 \permsum{1}{n}  { 1\over{(n-1)!} }
{\positronon}^{\beta}\bigl(p^{-};1^{+},\ldots,(n{-}1)^{+}\bigr)
u_{\beta}(g)
{
{ {\bar u}_{\dot\beta}(k_n)[\bar p+\bar\kappa(1,n)]^{\dot\beta\alpha}}
\over
{\braket{n}{g}\varsp [p+\kappa(1,n)]^2}
}.
}
\eqlabel\LHallplusrecurs
$$
In this case, we define
$$
Z(p^{-};1^{+},\ldots,n^{+}) \equiv
u_{\alpha}(g){\positronon}^{\alpha}(p^{-};1^{+},\ldots,n^{+}),
\eqlabel\Zdef
$$
because (\LHallplusrecurs) implies that $Z(p^{-};1^{+},\ldots,n^{+})$
satisfies a simple recursion relation, namely
$$
\eqalign{
Z&(p^{-};1^{+},\ldots,n^{+}) =
\cr & =
-e \sqrt2 \permsum{1}{n}  { 1\over{(n-1)!} }
Z(p^{-};1^{+},\ldots,(n{-}1)^{+})
{
{ {\bar u}_{\dot\beta}(k_n)
[\bar p+\bar\kappa(1,n)]^{\dot\beta\alpha}u_{\alpha}(g)}
\over
{\braket{n}{g}\varsp [p+\kappa(1,n)]^2}
}.
}
\eqlabel\Zrecurs
$$
We again recognize the recursion (\WLallplusrecurs), this time satisfied
by $Z(p^{-};1^{+},\ldots,n^{+})$.  Hence
$$
Z(p^{-};1^{+},\ldots,n^{+}) =
{ {Z(p^{-})} \over {\PHI(p)} }\PHI(p;1^{+},\ldots,n^{+}),
\eq
$$
where $Z(p^{-})\equiv \braket{p}{g}$, according to (\Zdef).
Note that we still must insert the result for
$Z(p^{-};1^{+},\ldots,n^{+})$ back into (\LHallplusrecurs) to obtain
${\positronon}^{\alpha}(p^{-};1^{+},\ldots,n^{+})$.  However,
this is a straightforward algebraic manipulation which closely
follows the procedure used in Appendix \prtosm { } ({\it cf.}
equations (\Dstart)--(\Dend)), although with not all of the
complications encountered there.  The result is
$$
{\positronon}^{\alpha}(p^{-};1^{+},\ldots,n^{+}) =
(-e\sqrt2)^n \permsum{1}{n}
{
{u^{\alpha}(p) \varsp \braket{p}{g}}
\over
{ \bra{p} 1,2,\ldots,n \ket{g} }
}.
\eqlabel\LHallplussoln
$$

Berends and Giele \cite\BG{ }
have presented a solution for the fermion current with
all like photon helicities which
is already summed over permutations, with the gauge choice $g=p$.  It is
not hard to generalize their proof to the gauge used here.
The results of doing this yield
$$
{\positronon}_{\dot\alpha}(p^{+};1^{+},\ldots,n^{+}) =
-(-e\sqrt2)^n u^{\beta}(g)[p+\kappa(1,n)]_{\beta\dot\alpha}
{
{{\braket{p}{g}}^{n-1}}
\over
{\prod\limits_{j=1}^n \bra{p} j \ket{g} }
}.
\eqlabel\RHallplussummed
$$
Comparison of (\RHallplussummed) with (\RHallplussoln) recovers the
identity [\ref{M. Mangano, Nucl. Phys. {\bf B309}, 461 (1988).}
\refname\IDENTITYref{\kern-.7em}]
$$
\permsum{1}{n}
{
{1}
\over
{\bra{p} 1,2,\ldots,n \ket{g} }
}
=
{
{{\braket{p}{g}}^{n-1}}
\over
{\prod\limits_{j=1}^n \bra{p} j \ket{g} }
}.
\eqlabel\permsumid
$$


\subsection{The transverse $W$ current}

Finally, we consider the transverse $W$ current
$\Wms (P^{+};1^{+},\ldots,n^{+})$.  We must choose a gauge spinor
for the on-shell $W$ boson.  The natural choice is to set
$$
{\Wms}_{\alpha\dot\alpha}(P^{+})=
{{ u_{\alpha}(g) {\bar u}_{\dot\alpha}(P) }
\over {\braket{P}{g}}},
\eqlabel\polallplusW
$$
so that $\Wms(P^{+})$ contracted into any of the other polarizations
vanishes, as in (\seagullkill).  This choice of polarization spinor
has two nice consequences:
$$
\eps(m^{+})\cdot\Wnorm(P^{+};1^{+},\ldots,n^{+})=0
\eq
$$
and
$$
\bigl\{ \eps(\ell^{+}),
\Wnorm(P^{+};1^{+},\ldots,n^{+}), \eps(m^{+}) \bigr\}
 = 0.
\eq
$$
These relations are true for any choice of $\ell$, $m$ and $n$.
These simplifications allow us to write the recursion relation
(\WTrecursionperm) in a shorter form:
$$
\eqalign{
{\Wms}_{\alpha\dot\alpha}(P^{+};1^{+},\ldots,n^{+})& =
-e\sqrt2
\permsum{1}{n}   { 1\over{(n-1)!} }
{
{[P+\kappa(1,n)]_{\beta\dot\beta}}
\over
{[P+\kappa(1,n)]^2}
}
\cr & \quad\times
\Biggl[
{\bar\eps}^{\dot\beta\beta}(n^{+})
{\Wms}_{\alpha\dot\alpha}\bigl(P^{+};1^{+},\ldots,(n{-}1)^{+}\bigr)
\cr &  \quad\quad
-
{\Wbar}^{\dot\beta\beta}\bigl(P^{+};1^{+},\ldots,(n{-}1)^{+}\bigr)
{\eps}_{\alpha\dot\alpha}(n^{+})
\Biggr].
}
\eqlabel\WTallplusstart
$$
Noting that we may rewrite (\polallplusW) as
$$
{\Wms}_{\alpha\dot\alpha}(P^{+})=
{
{u_{\alpha}(g)u^{\beta}(g)P_{\beta\dot\alpha}}
\over
{\braket{P}{g} \varsp \braket{g}{P} }
},
\eqlabel\WTzero
$$
and from (\WTallplusstart) we have
$$
{\Wms}_{\alpha\dot\alpha}(P^{+};1^{+})=-e\sqrt2
{
{u_{\alpha}(g)u^{\beta}(g)(P+k_1)_{\beta\dot\alpha}}
\over
{\braket{g}{P} \braket{P}{1} \braket{1}{g} }
},
\eqlabel\WTone
$$
suggests the following ansatz:
$$
{\Wms}_{\alpha\dot\alpha}(P^{+};1^{+},\ldots,n^{+})=
u_{\alpha}(g)u^{\beta}(g)[P+\kappa(1,n)]_{\beta\dot\alpha}
X(P^{+};1^{+},\ldots,n^{+}).
\eqlabel\Xansatz
$$
We prove this ansatz and obtain a recursion relation for
$X(P^{+};1^{+},\ldots,n^{+})$ by induction.
For $n=0$ and $n=1$, (\WTzero) and (\WTone) tells us that
$$
X(P^{+})= { {-1}\over{{\braket{P}{g}}^2} }
\newlett
$$
$$
X(P^{+};1^{+})=-e\sqrt2
{  {-1}\over{\braket{P}{g}\braket{P}{1}\braket{1}{g}} }.
\lett
$$
For $n\ge2$,
$$
\eqalign{
{\Wms}_{\alpha\dot\alpha}&(P^{+};1^{+},\ldots,n^{+})=
-e\sqrt2 \permsum{1}{n}  { 1\over{(n-1)!} }
{
{X(P^{+};1^{+},\ldots,(n{-}1)^{+})}
\over
{\braket{n}{g}\varsp[P+\kappa(1,n)]^2}
} \cr&
\times
\Biggl\{ u^{\beta}(g)[P+\kappa(1,n)]_{\beta\dot\beta}
{\bar u}^{\dot\beta}(k_n) u_{\alpha}(g)u^{\gamma}(g)
[P+\kappa(1,n{-}1)]_{\gamma\dot\alpha}
\cr &\qquad
+u^{\beta}(g)[P+\kappa(1,n)]_{\beta\dot\beta}
[\bar P+\bar \kappa(1,n{-}1)]^{\dot\beta\gamma}u_{\gamma}(g)
u_{\alpha}(g){\bar u}_{\dot\alpha}(k_n) \Biggr\}.
}
\eqlabel\midstep
$$
The quantity in curly brackets may be simplified
using (\slashsqr) to give
$$
\eqalign{&
u_{\alpha}(g)
\bigl\{
u^{\beta}(g)
[ P+\kappa(1,n)]_{\beta\dot\beta} {\bar u}^{\dot\beta}(k_n)
u^{\gamma}(g)[P+\kappa(1,n{-}1)]_{\gamma\dot\alpha}
\cr & \qquad
+ u^{\beta}(g)k_{n\beta\dot\beta}
[\bar P+\bar \kappa(1,n)]^{\dot\beta\gamma}
u_{\gamma}(g) {\bar u}_{\dot\alpha}(k_n) \bigr\}
\cr &
= u_{\alpha}(g)
{\bar u}_{\dot\beta}(k_n)
[\bar P + \bar \kappa(1,n)]^{\dot\beta\gamma}
u_{\gamma}(g)
\cr & \qquad\times
\bigl\{
u^{\beta}(g)[P+\kappa(1,n{-}1)]_{\beta\dot\alpha}
+ u^{\beta}(g)k_{n\beta\dot\alpha} \bigr\}
\cr &
= u_{\alpha}(g)u^{\beta}(g)
[P+\kappa(1,n)]_{\beta\dot\alpha}\thinspace
{\bar u}_{\dot\beta}(k_n)
[\bar P + \bar \kappa(1,n)]^{\dot\beta\gamma}
u_{\gamma}(g).
}
\eqlabel\SQbrakets
$$
If we insert (\SQbrakets) into  (\midstep), we see that the following
recursion relation for $X(P^{+};1^{+},\ldots,n^{+})$ is implied:
$$
\eqalign{
X&(P^{+};1^{+},\ldots,n^{+}) =
\cr & =
-e \sqrt2 \permsum{1}{n}  { 1\over{(n-1)!} }
{
{ {\bar u}_{\dot\beta}(k_n)[P+\kappa(1,n)]^{\dot\beta\beta}u_{\beta}(g)}
\over
{\braket{n}{g}\varsp [P+\kappa(1,n)]^2}
}
X\bigl(P^{+};1^{+},\ldots,(n{-}1)^{+}\bigr).
}
\eqlabel\Xrecurs
$$
Since this is the same recursion as (\WLallplusrecurs), we have yet
another supersymmetry relation:
$$
X(P^{+};1^{+},\ldots,n^{+}) =
{ {X(P^{+})} \over {\PHI(P)} }\PHI(P;1^{+},\ldots,n^{+}).
\eq
$$
Putting everything together, we have
$$
{\Wms}_{\alpha\dot\alpha}(P^{+};1^{+},\ldots,n^{+}) =
(-e\sqrt2)^n  \permsum{1}{n}
{
{- u_{\alpha}(g)u^{\beta}(g)[P+\kappa(1,n)]_{\beta\dot\alpha} }
\over
{ \braket{P}{g} \varsp \bra{P} 1,2,\ldots,n \ket{g} }
}.
\eqlabel\WTallplussoln
$$

\section{Currents with one unlike helicity}

We now allow one
of the gauge bosons to have opposite helicity from the rest.  This leads
us to investigate $\PHI(P;I^{-},2^{+},\ldots,n^{+})$,
${\positronon}(p^{+};I^{-},2^{+},\ldots,n^{+})$,
${\positronon}(p^{-};I^{-},2^{+},\ldots,n^{+})$,
${\Wms}(P^{+};I^{-},2^{+},\ldots,n^{+})$,
and ${\Wms}(P^{-};1^{+},2^{+},\ldots,n^{+})$.
The first four of these currents turn out to be closely related, while
the fifth is somewhat different.  In each of the five cases it will
be convenient to choose the gauge momentum $g$ of the positive helicity
bosons to be equal to the momentum of the negative helicity boson
\cite{\DY,\BG}.
The gauge momentum of the negative helicity boson is left as an
arbitrary null vector $h$.


\subsection{The scalar current}
We begin our discussion with the scalar current
$\PHI(P;I^{-},2^{+},\ldots,n^{+})$.  Note that the negative helicity
photon is labelled with momentum $k_I$, not $k_1$.  This distinction
will be useful later.

The gauge choice described in the introduction to this section reads
$$
{\eps}_{\alpha\dot\alpha}(I^{-})=
{{ u_{\alpha}(k_I) {\bar u}_{\dot\alpha}(h) }
\over {\braket{I}{h}}^{*}}
\newlettlabel\poloneminusA
$$
$$
{\eps}_{\alpha\dot\alpha}(j^{+})=
{{ u_{\alpha}(k_I) {\bar u}_{\dot\alpha}(k_j) }
\over {\braket{j}{I}}}.
\lett
$$
Note that, as when we were considering currents with all
like helicities,
all dot products between pairs of polarization vectors vanish:
$$
{\eps}_{\alpha\dot\alpha}(I^{-}) {\bar\eps}^{\dot\alpha\alpha}(j^{+})=
{\eps}_{\alpha\dot\alpha}(j^{+}) {\bar\eps}^{\dot\alpha\alpha}(m^{+})=0.
\eqlabel\seagullkillB
$$
As a result, the recursion relation for this current will not contain
any seagull terms.  Because the helicity of the first photon is different
from the others, however, it is convenient to recast the surviving
piece of (\WLrecursionperm) in a form which separates out this
photon explicitly:
$$
\eqalign{
\PHI&(P;I^{-},2^{+},\ldots,n^{+})=
\cr &  =
{-e\sqrt2}\negthinspace\negthinspace\permsum{2}{n}\negthinspace
{
{\bar u_{\dot\alpha}(h)
[\bar P {+} \bar k_I {+} \bar \kappa(2,n)]^{\dot\alpha\alpha}
u_{\alpha}(k_I)}
\over
{(n{-}1)!\thinspace {\braket{I}{h}}^{*} \varsp [P{+}k_I{+}\kappa(2,n)]^2}
}
\PHI (P;2^{+},\ldots,n^{+})
\cr& \quad
{-e\sqrt2}\negthinspace\negthinspace\permsum{2}{n}\negthinspace
{
{\bar u_{\dot\alpha}(k_n)
[\bar P {+} \bar k_I {+} \bar \kappa(2,n)]^{\dot\alpha\alpha}
u_{\alpha}(k_I)}
\over
{(n{-}2)! \thinspace \braket{n}{I} \varsp [P{+}k_I{+}\kappa(2,n)]^2}
}
\PHI \bigl(P;I^{-},2^{+},\ldots,(n{-}1)^{+}\bigr).
}
\eqlabel\WLoneminusrecurs
$$
Note that we have made use of the permutation symmetry among the positive
helicity photons to relabel the $n-1$ terms involving $\eps(2^{+})$,{ }
$\eps(3^{+})$,$\ldots$,{ }$\eps(n^{+})$ onto a single term.   The
one-photon current implied by this expression is
$$
\PHI(P;I^{-})=-(-e\sqrt2)
{
{ {\braket{h}{P}}^{*} }
\over
{ { \bra{h} I \ket{P} }^{*} }
}.
\eqlabel\PHIone
$$

If we insert the known
solution for $\PHI (P;2^{+},\ldots,n^{+})$ (equation (\PHIallplussoln)
with $g=k_I$) into (\WLoneminusrecurs), we get
$$
\eqalign{
\PHI&(P;I^{-},2^{+},\ldots,n^{+})=
\cr & =
(-e\sqrt2)^n \permsum{2}{n}
{
{\bar u_{\dot\alpha}(P)
[\bar P + \bar k_I + \bar \kappa(2,n)]^{\dot\alpha\alpha}
u_{\alpha}(k_I)}
\over
{{\braket{I}{P}}^{*} \varsp
[P+k_I+\kappa(2,n)]^2 }
}
{
{\braket{P}{I}}
\over
{\bra{P} 2,\ldots,n \ket{I} }
} \cr & \quad
{-e\sqrt2}\negthinspace \permsum{2}{n}
{
{\bar u_{\dot\alpha}(k_n)
[\bar P + \bar k_I + \bar \kappa(2,n)]^{\dot\alpha\alpha}
u_{\alpha}(k_I)}
\over
{(n{-}2)! \thinspace \braket{n}{I} \varsp [P+k_I+\kappa(2,n)]^2}
}
\PHI (P;I^{-},2^{+},\ldots,(n{-}1)^{+}),
}
\eqlabel\oneminunstart
$$
where we have set $h=P$.  This restriction on the gauge of the negative
helicity photon will be removed later.

We iterate equation (\oneminunstart)  until the right hand
side contains $\PHI(P;I^{-})$, which, according to
(\PHIone), vanishes when $h=P$.  This gives us
\def \fA{(-e\sqrt2)^n \permsum{2}{n}}
\def \fB{
{
{\bar u_{\dot\alpha}(P)
[\bar P + \bar k_I + \bar \kappa(2,n)]^{\dot\alpha\alpha}
u_{\alpha}(k_I)}
\over
{{\braket{I}{P}}^{*} \varsp [P+k_I+\kappa(2,n)]^2 }
}
}
\def \fC{
{
{\braket{P}{I}}
\over
{\bra{P} 2,\ldots,n \ket{I} }
}
}
\def \firstterm{ \fA \fB \fC }
\def \termfirst{ \fA \fC \fB }
$$
\eqalign{
\PHI&(P;I^{-},2^{+},\ldots,n^{+})=
\cr & =
\firstterm\cr & +
(-e\sqrt2)^n \permsum{2}{n} \sum_{j=3}^n
{
{\bar u_{\dot\alpha}(P)
[\bar P + \bar k_I + \bar \kappa(2,j{-}1)]^{\dot\alpha\alpha}
u_{\alpha}(k_I)}
\over
{{\braket{I}{P}}^{*} \varsp
[P+k_I+\kappa(2,j{-}1)]^2 }
}
{
{\braket{P}{I}}
\over
{\bra{P} 2,\ldots,j{-}1 \ket{I} }
} \cr &
\qquad\qquad\qquad\qquad\times
\permsum{j}{n}
\prod_{\ell=j}^{n}
{
{\bar u_{\dot\alpha}(k_{\ell})
[\bar P + \bar k_I + \bar \kappa(2,\ell)]^{\dot\alpha\alpha}
u_{\alpha}(k_I)}
\over
{(n-j+1)! \thinspace \braket{\ell}{I} \varsp [P+k_I+\kappa(2,\ell)]^2}
}.
}
\eqlabel\BIGGmess
$$
We have inserted
$$
\permsum{j}{n} {1\over{(n-j+1)!}}  = 1
\eq
$$
into (\BIGGmess) in order to obtain an explicit factor of $\Xi(j,n)$,
as defined in (\XIid).  Thus, we may  convert the product appearing in
 (\BIGGmess) into a sum, producing
$$
\eqalign{
\PHI&(P;I^{-},2^{+},\ldots,n^{+})=
\cr & =
\termfirst\cr & +
(-e\sqrt2)^n \permsum{2}{n} \sum_{j=3}^n \sum_{\ell=j}^n
{
{\braket{P}{I}
\bar u_{\dot\gamma}(P)
[\bar P + \bar k_I + \bar \kappa(2,j{-}1)]^{\dot\gamma\gamma}
u_{\gamma}(k_I)}
\over
{\bra{P} 2,\ldots,j{-}1 \ket{I} \varsp
\bra{I} j,\ldots,n\ket{I}\varsp
{\braket{I}{P}}^{*}  }
} \cr &
\qquad\qquad\qquad\qquad\times
u^{\alpha}(k_I)
{{\pole}_{\alpha}}^{\beta}(P,I,2,\ldots,\ell)
u_{\beta}(k_I).
}
\eq
$$
Interchanging the order of summations over $j$ and $\ell$
in the second term and forcing
it to have the same denominator structure as the first term
yields
$$
\eqalign{
\PHI&(P;I^{-},2^{+},\ldots,n^{+})=
\cr & =
\termfirst\cr &\quad +
(-e\sqrt2)^n \permsum{2}{n} \sum_{\ell=3}^n \sum_{j=3}^{\ell}
{
{\braket{P}{I}}
\over
{\bra{P} 2,\ldots,n \ket{I}}
}
{ 1\over { {\braket{I}{P}}^{*} } }
{
{\braket{j{-}1}{j}}
\over
{\bra{j{-}1} I \ket{j} }
}
\cr & \qquad\quad\times
\bar u_{\dot\gamma}(P)
{\bar \kappa}^{\dot\gamma\gamma}(2,j{-}1)
u_{\gamma}(k_I)
u^{\alpha}(k_I)
{{\pole}_{\alpha}}^{\beta}(P,I,2,\ldots,\ell)
u_{\beta}(k_I).
}
\eqlabel\readytosumj
$$
We have used the Weyl equation to
reduce $[\bar P + \bar k_I + \bar \kappa(2,j{-}1)]^{\dot\gamma\gamma}$
to ${\bar \kappa}^{\dot\gamma\gamma}(2,j{-}1)$ in the second term.
We now perform the sum on $j$, using (\linkidsummed):
$$
\eqalign{
\sum_{j=3}^{\ell} \sum_{m=2}^{j-1}
{
{\braket{j{-}1}{j}}
\over
{\bra{j{-}1} I \ket{j} }
}
\bar k_m^{\dot\gamma\gamma}u_{\gamma}(k_I) &=
\sum_{m=2}^{\ell-1} \sum_{j=m+1}^{\ell}
{
{\braket{j{-}1}{j}}
\over
{\bra{j{-}1} I \ket{j} }
}
\bar u^{\dot\gamma}(k_m) \braket{m}{I}
\cr &
= \sum_{m=2}^{\ell-1}
{
{\braket{m}{\ell}}
\over
{\bra{m} I \ket{\ell} }
}
\bar u^{\dot\gamma}(k_m) \braket{m}{I}
\cr &
= { 1 \over {\braket{I}{\ell}} }
{\bar \kappa }^{\dot\gamma\gamma}(2,\ell{-}1)
u_{\gamma}(k_{\ell}).
}
\eqlabel\jsum
$$
Inserting (\jsum) into (\readytosumj) produces
$$
\eqalign{
\PHI&(P;I^{-},2^{+},\ldots,n^{+})=
\cr & =
\termfirst\cr & \quad+
(-e\sqrt2)^n \permsum{2}{n} \sum_{\ell=3}^n
{
{\braket{P}{I}}
\over
{\bra{P} 2,\ldots,n \ket{I}}
}
{ 1\over { {\braket{I}{P}}^{*} } }\cr &
\qquad\qquad\qquad\qquad\times
\bar u_{\dot\gamma}(P)
{\bar \kappa }^{\dot\gamma\gamma}(2,\ell{-}1)
{{\pole}_{\gamma}}^{\beta}(P,I,2,\ldots,\ell)
u_{\beta}(k_I).
}
\eqlabel\yetanothermess
$$
To obtain (\yetanothermess) we have exploited the fact that
$u_{\alpha}(k_I){{\pole}_{\alpha}}^{\beta}(P,I,2,\ldots,\ell)
u_{\beta}(k_I)$ contains a factor of $\braket{I}{\ell}$.

At this stage, we would like to apply (\splitid)
to simplify the second term in (\yetanothermess).
However,
(\yetanothermess) doesn't quite contain the expression
required by (\splitid).
Thus, we use the Weyl equation to introduce
$\bar P^{\dot\gamma\gamma}$
and $\bar k_{\ell}^{\dot\gamma\gamma}$, and  subtract the appropriate
piece to include $\bar k_I^{\dot\gamma\gamma}$.  This procedure yields
$$
\eqalign{
\PHI&(P;I^{-},2^{+},\ldots,n^{+})=
\cr &\negthinspace\negthinspace\negthinspace =
\termfirst\cr & +
(-e\sqrt2)^n \permsum{2}{n} \sum_{\ell=3}^n
{
{\braket{P}{I}}
\over
{\bra{P} 2,\ldots,n \ket{I}}
}
{ 1\over { {\braket{I}{P}}^{*} } }\cr &
\qquad\qquad\qquad\times
\bar u_{\dot\gamma}(P)
[\bar P + \bar k_I +\bar \kappa (2,\ell)]^{\dot\gamma\gamma}
{{\pole}_{\gamma}}^{\beta}(P,I,2,\ldots,\ell)
u_{\beta}(k_I)
\cr &
-(-e\sqrt2)^n \permsum{2}{n} \sum_{\ell=3}^n
{
{\braket{P}{I}}
\over
{\bra{P} 2,\ldots,n \ket{I}}
}
u^{\alpha}(k_I){{\pole}_{\alpha}}^{\beta}(P,I,2,\ldots,\ell)
u_{\beta}(k_I).
}
\eqlabel\doesiteverend
$$
The second term of (\doesiteverend) now contains
the appropriate combination of factors.
When we insert (\splitid) into (\doesiteverend), and
do the (trivial) sum on $\ell$, we discover that
the first term in (\doesiteverend) cancels one of the two
terms that results.  The remainder is
$$
\eqalign{
\PHI&(P;I^{-},2^{+},\ldots,n^{+})=
\cr & =
(-e\sqrt2)^n \permsum{2}{n}
{
{\braket{P}{I}}
\over
{\bra{P} 2,\ldots,n \ket{I}}
}
{
{\braket{I}{P} \varsp
\bar u_{\dot\alpha}(P)
[\bar P +\bar k_I +\bar k_2]^{\dot\alpha\alpha}
u_{\alpha}(k_I)}
\over
{ (P+k_I)^2(P+k_I+k_2)^2 }
} \cr &
-(-e\sqrt2)^n \permsum{2}{n}
{
{\braket{P}{I}}
\over
{\bra{P} 2,\ldots,n \ket{I}}
}
\sum_{\ell=3}^n
u^{\alpha}(k_I){{\pole}_{\alpha}}^{\beta}(P,I,2,\ldots,\ell)
u_{\beta}(k_I).
}
\eqlabel\whew
$$
If we examine the numerator in the first term we find that we can write
$$
\eqalign{
\braket{I}{P} \varsp
\bar u_{\dot\alpha}(P)
[\bar P +\bar k_I +\bar k_2&]^{\dot\alpha\alpha}
u_{\alpha}(k_I)  =
\cr & =
-u^{\alpha}(k_I)[P+k_I+k_2]_{\alpha\dot\alpha}
\bar P^{\dot\alpha\beta} u_{\beta}(k_I)  \cr&
=-u^{\alpha}(k_I)k_{2\alpha\dot\alpha}
\bar P^{\dot\alpha\beta} u_{\beta}(k_I)  \cr&
=-u^{\alpha}(k_I)k_{2\alpha\dot\alpha}
[\bar P + \bar k_I + \bar k_2] ^{\dot\alpha\beta} u_{\beta}(k_I),
}
\eqlabel\createPItwo
$$
where we have exploited the antisymmetry of the contractions and
used the Weyl equation.  It is clear from (\createPItwo) that the
first term of (\whew) simply extends the sum appearing in the
second term to $\ell=2$.  Thus,
$$
\eqalign{
\PHI&(P;I^{-},2^{+},\ldots,n^{+})=
\cr &\negthinspace =
-(-e\sqrt2)^n
\negthinspace\negthinspace\negthinspace\permsum{2}{n}\negthinspace
{
{\braket{P}{I}}
\over
{\bra{P} 2,\ldots,n \ket{I}}
}
\sum_{\ell=2}^n
u^{\alpha}(k_I){{\pole}_{\alpha}}^{\beta}(P,I,2,\ldots,\ell)
u_{\beta}(k_I).
}
\eqlabel\PHIoneminussolnhfixed
$$
Note that this expression is valid only for $n \geq 2$.  For
$n=1$ we must use (\PHIone).

We now remove the restriction $h=P$, as promised earlier.  In the
following, let all symbols with a hat denote quantities written
in the gauge $h=P$, while symbols with no hat denote the gauge
where $h$ is simply an arbitrary null vector.  Note that since
$h$ is the gauge spinor for the negative helicity photon,
quantities not containing this photon are invariant under the
gauge change being considered here.   In particular,
$$
\widehat{\PHI}(P;2^{+},\ldots,n^{+})={\PHI}(P;2^{+},\ldots,n^{+}),
\newlettlabel\invars
$$
and
$$
{\widehat{\eps}}_{\alpha\dot\alpha}(j^{+}) =
{\eps}_{\alpha\dot\alpha}(j^{+}).
\lett
$$
Also of note is the relation
$$
{\eps}_{\alpha\dot\alpha}(I^{-}) =
{
{ u_{\alpha}(k_I) \bar u_{\dot\alpha}(h) }
\over
{ {\braket{I}{h}}^{*} }
}
=
{\widehat{\eps}}_{\alpha\dot\alpha}(I^{-}) -
u_{\alpha}(k_I) \bar u_{\dot\alpha}(k_I)
{
{ {\braket{h}{P}}^{*} }
\over
{ { \bra{h}I\ket{P} }^{*} }
},
\eqlabel\shiftneg
$$
where we have used the Schouten identity (\fierz) to extract
the term containing
${\widehat{\eps}}_{\alpha\dot\alpha}(I^{-})$.

We expect to find that the new solution is the same as the old
solution, plus a term which vanishes when $h=P$.  So, we write
$$
{\PHI}(P;I^{-},2^{+},\ldots,n^{+})\equiv
\widehat{\PHI}(P;I^{-},2^{+},\ldots,n^{+})
+\gauge (P;I^{-},2^{+},\ldots,n^{+}).
\eqlabel\gform
$$
If we insert the gauge changing relations (\shiftneg) and (\gform)
into the recursion relation (\WLoneminusrecurs) we obtain
$$
\eqalign{
\PHI&(P;I^{-},2^{+},\ldots,n^{+})=
\cr & \negthinspace\negthinspace\negthinspace=
e\sqrt2 \negthinspace\permsum{2}{n}
{
{2k_I \cdot [P+k_I+\kappa(2,n)]}
\over
{(n-1)!\thinspace [P+k_I+\kappa(2,n)]^2 }
}
{
{ {\braket{h}{P}}^{*} }
\over
{ { \bra{h}I\ket{P} }^{*} }
}
{\PHI}(P;2^{+},\ldots,n^{+})
\cr&\thinspace
-e\sqrt2 \negthinspace\permsum{2}{n}
{
{{\widehat{\eps}}_{\alpha\dot\alpha}(I^{-})
[\bar P + \bar k_I + \bar\kappa(2,n)]^{\dot\alpha\alpha}}
\over
{(n-1)!\thinspace [P+k_I+\kappa(2,n)]^2 }
}
\widehat{\PHI}(P;2^{+},\ldots,n^{+})
\cr&\thinspace
-e\sqrt2 \negthinspace\permsum{2}{n}
{
{{\widehat{\eps}}_{\alpha\dot\alpha}(n^{+})
[\bar P + \bar k_I + \bar\kappa(2,n)]^{\dot\alpha\alpha}}
\over
{ (n-2)!\thinspace[P+k_I+\kappa(2,n)]^2 }
}
\widehat{\PHI}\bigl(P;I^{-},2^{+},\ldots,(n{-}1)^{+}\bigr)
\cr&\thinspace
-e\sqrt2 \negthinspace\permsum{2}{n}
{
{{\eps}_{\alpha\dot\alpha}(n^{+})
[\bar P + \bar k_I + \bar\kappa(2,n)]^{\dot\alpha\alpha}}
\over
{(n-2)!\thinspace [P+k_I+\kappa(2,n)]^2 }
}
\gauge\bigl(P;I^{-},2^{+},\ldots,(n{-}1)^{+}\bigr) .
}
\eqlabel\noname
$$
Using (\invars), we have written (\noname) in a way which makes
it obvious that
its second and third terms are  simply the recursion relation
for $\widehat{\PHI}(P;I^{-},2^{+},\ldots,n^{+})$.  Thus, we have
the following recursion
relation for $\gauge\bigl(P;I^{-},2^{+},\ldots,(n{-}1)^{+}\bigr)$:
$$
\eqalign{
\gauge&(P;I^{-},2^{+},\ldots,n^{+})=
\cr & =
e\sqrt2
{
{ {\braket{h}{P}}^{*} }
\over
{ { \bra{h}I\ket{P} }^{*} }
}
{
{2k_I \cdot [P+k_I+\kappa(2,n)]}
\over
{ [P+k_I+\kappa(2,n)]^2 }
}
{\PHI}(P;2^{+},\ldots,n^{+})
\cr& \quad
-e\sqrt2 \permsum{2}{n}
{
{\bar u_{\dot\alpha}(k_n)
[\bar P + \bar k_I + \bar\kappa(2,n)]^{\dot\alpha\alpha}
u_{\alpha}(k_I)}
\over
{(n-2)!\thinspace\braket{n}{I} \varsp [P+k_I+\kappa(2,n)]^2 }
}
\gauge\bigl(P;I^{-},2^{+},\ldots,(n{-}1)^{+}\bigr),
}
\eqlabel\Grecurs
$$
where we have performed the  trivial permutation sum appearing
in the first term of (\noname).

Since $\PHI(P)=\widehat{\PHI}(P)$, we have for our starting point
the relation
$$
\gauge(P)=0.
\eq
$$
Furthermore, the single-photon current (\PHIone) tells us that
$$
\gauge(P;I^{-})=e\sqrt2
{
{ {\braket{h}{P}}^{*} }
\over
{ { \bra{h} I \ket{P} }^{*} }
}.
\eq
$$
Using these starting points in the recursion relation (\Grecurs) gives
the following result for $n=2$ without much difficulty:
$$
\gauge(P;I^{-},2^{+}) = e\sqrt2
{
{ {\braket{h}{P}}^{*} }
\over
{ { \bra{h}I\ket{P} }^{*} }
}
{\PHI}(P;2^{+}).
\eq
$$
Thus, we are led to the ansatz
$$
\gauge(P;I^{-},2^{+},\ldots,n^{+}) = e\sqrt2
{
{ {\braket{h}{P}}^{*} }
\over
{ { \bra{h}I\ket{P} }^{*} }
}
{\PHI}(P;2^{+},\ldots,n^{+}).
\eqlabel\Gsoln
$$
This ansatz is easily proven.  Proceeding inductively,  from
(\Grecurs) we have
$$
\eqalign{
\gauge&(P;I^{-},2^{+},\ldots,n^{+})=
\cr & =
e\sqrt2
{
{ {\braket{h}{P}}^{*} }
\over
{ { \bra{h}I\ket{P} }^{*} }
}
{
{2k_I \cdot [P+k_I+\kappa(2,n)]}
\over
{ [P+k_I+\kappa(2,n)]^2 }
}
{\PHI}(P;2^{+},\ldots,n^{+})
\cr& \quad
-e\sqrt2 \permsum{2}{n}
{
{\bar u_{\dot\alpha}(k_n)
[\bar P +   \bar\kappa(2,n)]^{\dot\alpha\alpha}
u_{\alpha}(k_I)}
\over
{(n-2)! \thinspace \braket{n}{I} \varsp [P+k_I+\kappa(2,n)]^2 }
}
\cr & \qquad\qquad\qquad\quad\times
(e\sqrt2)
{
{ {\braket{h}{P}}^{*} }
\over
{ { \bra{h}I\ket{P} }^{*} }
}
{\PHI}\bigl(P;2^{+},\ldots,(n{-}1)^{+}\bigr),
}
\eqlabel\Gstart
$$
where we have used the Weyl equation to eliminate
$ {\bar k_I}^{\dot\alpha\alpha} $ from the numerator of the second
term.  Now, according to the recursion relation for
${\PHI}(P;2^{+},\ldots,n^{+})$, we have
$$
\eqalign{
\PHI&(P;2^{+},\ldots,n^{+}) =
\cr & =
-e\sqrt2 \permsum{2}{n}
{ { {\bar u}_{\dot\beta}(k_n)[\bar P + \bar \kappa(2,n)]^
{\dot\beta\beta} u_{\beta}(I) } \over
{ (n-2)!\thinspace\braket{n}{I} \varsp [P+\kappa(2,n)]^2 } }
\PHI\bigl(P;2^{+},\ldots,(n{-}1)^{+}\bigr).
}
\eq
$$
We recognize that, up to a factor symmetric under ${{\cal{P}}(2\ldots n)}$,
this is precisely the quantity appearing in the second term of
(\Gstart).  Therefore, we may write
$$
\eqalign{
\gauge&(P;I^{-},2^{+},\ldots,n^{+})=
\cr & =
e\sqrt2
{
{ {\braket{h}{P}}^{*} }
\over
{ { \bra{h}I\ket{P} }^{*} }
}
{
{2k_I \cdot [P+k_I+\kappa(2,n)]}
\over
{ [P+k_I+\kappa(2,n)]^2 }
}
{\PHI}(P;2^{+},\ldots,n^{+})  \cr&\quad
+e\sqrt2
{
{ {\braket{h}{P}}^{*} }
\over
{ { \bra{h}I\ket{P} }^{*} }
}
{
{[P+\kappa(2,n)]^2}
\over
{[P+k_I+\kappa(2,n)]^2 }
}
{\PHI}(P;2^{+},\ldots,n^{+}).
}
\eqlabel\Gdone
$$
The combination of these two terms involves
$$
2k_I \cdot [P+\kappa(2,n)] + [P+\kappa(2,n)]^2 =
[P+k_I+\kappa(2,n)]^2
\eq
$$
because of the condition $k_I^2=0$.  Hence, the denominator cancels
and we are left with the expression (\Gsoln).

We may combine (\Gsoln) with (\PHIoneminussolnhfixed) to produce an
expression which is the correct current for all $n$:
$$
\eqalign{
\PHI&(P;I^{-},2^{+},\ldots,n^{+})=
\cr & =
-(-e\sqrt2)^n \permsum{2}{n}
{
{\braket{P}{I}}
\over
{\bra{P} 2,\ldots,n \ket{I}}
}
\Biggl\{
{
{ {\braket{h}{P}}^{*} }
\over
{ { \bra{h}I\ket{P} }^{*} }
} \cr&\qquad\qquad\qquad\qquad
+ (1-{\delta}_{n1})
\sum_{j=2}^n
u^{\alpha}(k_I){{\pole}_{\alpha}}^{\beta}(P,I,2,\ldots,j)
u_{\beta}(k_I)
\Biggr\}.
}
\eqlabel\PHIoneminussoln
$$


\subsection{The fermion currents}

We will now examine the two fermion currents in the case where one
of the photons has the opposite helicity from the rest.  Again,
we will label this photon by ``$I$'' instead of ``1''.  The gauge choice
will be the same as for the scalar current, namely (\poloneminusA).
If we explicitly separate out the first photon from
the permutation form of the recursion relation (\recursionposRHonperm),
we have
$$
\eqalign{
{{\positronon}_{\dot\alpha}}&(p^{+};I^{-},2^{+},\ldots,n^{+})=
\cr & =
-e\sqrt2
\permsum{2}{n}
{{1}\over{(n-1)!}}
\bar u^{\dot\beta}(h)
\positronon_{\dot\beta}(p^{+};2^{+},\ldots,n^{+})
\cr & \qquad\qquad\qquad\qquad \times
{
{u^{\beta}(k_I)[p+k_I+\kappa(2,n)]_{\beta\dot\alpha}}
\over
{{\braket{I}{h}}^{*} \varsp [p+k_I+\kappa(2,n)]^2}
}
\cr& \quad
-e\sqrt2
\permsum{2}{n}
{{1}\over{(n-2)!}}
\bar u^{\dot\beta}(k_n)
\positronon_{\dot\beta}\bigl(p^{+};I^{-},2^{+},\ldots,(n{-}1)^{+}\bigr)
\cr & \qquad\qquad\qquad\qquad \times
{
{u^{\beta}(k_I)[p+k_I+\kappa(2,n)]_{\beta\dot\alpha}}
\over
{\braket{n}{I}\varsp [p+k_I+\kappa(2,n)]^2}
}.
}
\eqlabel\RHnegstart
$$
Since there is an explicit factor
$u^{\beta}(k_I)[p+k_I+\kappa(2,n)]_{\beta\dot\alpha}$ ready to
be factored out of the right hand side of (\RHnegstart), we make the
ansatz
$$
{\positronon}_{\dot\alpha}(p^{+};I^{-},2^{+},\ldots,n^{+}) =
u^{\beta}(k_I)
[p+k_I+\kappa(2,n)]_{\beta\dot\alpha}
Y(p^{+};I^{-},2^{+},\ldots,n^{+}).
\eqlabel\RHspinorneg
$$
Equation (\Yzero) for $Y(p^{+})$ still applies here as it is independent
of the photons.  The ansatz is obviously correct, provided
we take
$$
\eqalign{
Y&(p^{+};I^{-},2^{+},\ldots,n^{+})=
\cr &\negthinspace\negthinspace\negthinspace =
-e\sqrt2  \negthinspace\negthinspace
\permsum{2}{n}
{
{\bar u_{\dot\beta}(h)
[\bar p+\bar k_I+\bar\kappa(2,n)]^{\dot\beta\beta}
u_{\beta}(k_I)}
\over
{(n-1)!\thinspace{\braket{I}{h}}^{*} \varsp [p+k_I+\kappa(2,n)]^2}
}
Y(p^{+};2^{+},\ldots,n^{+})
\cr&
-e\sqrt2  \negthinspace\negthinspace
\permsum{2}{n}
{
{\bar u_{\dot\beta}(k_n)
[\bar p+\bar k_I+\bar\kappa(2,n)]^{\dot\beta\beta}
u_{\beta}(k_I)}
\over
{(n-2)!\thinspace\braket{n}{I}\varsp [p+k_I+\kappa(2,n)]^2}
}
Y\bigl(p^{+};I^{-},2^{+},\ldots,(n{-}1)^{+}\bigr),
}
\eqlabel\RHneg
$$
as implied by  inserting (\RHspinorneg) into (\RHnegstart).
This recursion relation  is precisely the expression  (\WLoneminusrecurs)
we obtained for
$\PHI(P;I^{-},2^{+},\ldots,n^{+})$, implying the proportionality
$$
Y(p^{+};I^{-},2^{+},\ldots,n^{+}) =
{{Y(p^{+})}\over{\PHI(p)}}
\PHI(p;I^{-},2^{+},\ldots,n^{+}).
\eq
$$
Hence,
$$
\eqalign{
{\positronon}_{\dot\alpha}&(p^{+};I^{-},2^{+},\ldots,n^{+}) =
\cr & =
(-e\sqrt2)^n \permsum{2}{n}
{
{u^{\beta}(k_I)
[p+k_I+\kappa(2,n)]_{\beta\dot\alpha}}
\over
{\bra{p} 2,\ldots,n \ket{I}}
}
\Biggl\{
{
{ {\braket{h}{p}}^{*} }
\over
{ { \bra{h}I\ket{p} }^{*} }
}
\cr&\qquad\qquad\qquad\qquad
+ (1-{\delta}_{n1})
\sum_{j=2}^n
u^{\alpha}(k_I){{\pole}_{\alpha}}^{\beta}(p,I,2,\ldots,j)
u_{\beta}(k_I)
\Biggr\}.
}
\eqlabel\RHoneminussoln
$$

For the current with the other helicity, we have
$$
\eqalign{
{\positronon}^{\alpha}&(p^{-};I^{-},2^{+},\ldots,n^{+}) =
\cr & =
-e \sqrt2 \permsum{2}{n}  { 1\over{(n-1)!} }
{\positronon}^{\beta}(p^{-};2^{+},\ldots,n^{+})u_{\beta}(k_I)
\cr & \qquad\qquad\qquad\qquad\times
{
{ {\bar u}_{\dot\beta}(h)
[\bar p+\bar k_I+\bar \kappa(2,n)]^{\dot\beta\alpha}}
\over
{{\braket{I}{h}}^{*}\varsp [p+k_I+\kappa(2,n)]^2}
}
\cr&\quad
-e \sqrt2 \permsum{2}{n}  { 1\over{(n-2)!} }
{\positronon}^{\beta}
\bigl(p^{-};I^{-},2^{+},\ldots,(n{-}1)^{+}\bigr)
u_{\beta}(k_I)
\cr & \qquad\qquad\qquad\qquad\times
{
{ {\bar u}_{\dot\beta}(k_n)
[\bar p+\bar k_I+\bar\kappa(2,n)]^{\dot\beta\alpha}}
\over
{\braket{n}{I}\varsp [p+k_I+\kappa(2,n)]^2}
}.
}
\eqlabel\LHoneminusrecurs
$$
As was the case when we considered all like photon helicities, the
natural quantity to solve for first is
$$
Z(p^{-};I^{-},2^{+},\ldots,n^{+}) \equiv
u_{\alpha}(g){\positronon}^{\alpha}(p^{-};I^{-},2^{+},\ldots,n^{+}),
\eqlabel\Zdefneg
$$
since it satisfies the  recursion relation
$$
\eqalign{
Z&(p^{-};I^{-},2^{+},\ldots,n^{+}) =
\cr & =
-e \sqrt2 \permsum{2}{n}
Z(p^{-};2^{+},\ldots,n^{+})
{
{ {\bar u}_{\dot\beta}(h)
[\bar p+\bar k_I+\bar\kappa(2,n)]^{\dot\beta\alpha}u_{\alpha}(k_I)}
\over
{(n-1)!\thinspace{\braket{I}{h}}^{*}\varsp [p+k_I+\kappa(2,n)]^2}
} \cr&
-e \sqrt2 \permsum{2}{n}
Z\bigl(p^{-};I^{-},2^{+},\ldots,(n{-}1)^{+}\bigr)
{
{ {\bar u}_{\dot\beta}(k_n)
[\bar p+\bar k_I+\bar\kappa(2,n)]^{\dot\beta\alpha}u_{\alpha}(k_I)}
\over
{(n-2)!\thinspace\braket{n}{I}\varsp [p+k_I+\kappa(2,n)]^2}
}
}
\eqlabel\Zminusrecurs
$$
with $Z(p^{-})=\braket{p}{g}$.  But this recursion relation is the same
as (\WLoneminusrecurs), implying that
$$
Z(p^{-};I^{-},2^{+},\ldots,n^{+}) =
{ {Z(p^{-})} \over {\PHI(p)} }\PHI(p;I^{-},2^{+},\ldots,n^{+}).
\eq
$$
When we insert our solution for $Z(p^{-};I^{-},2^{+},\ldots,n^{+})$
back into (\LHoneminusrecurs) to obtain
${\positronon}^{\alpha}(p^{-};I^{-},2^{+},\ldots,n^{+})$,
we find
$$
\eqalign{
{\positronon}^{\alpha}&(p^{-};I^{-},2^{+},\ldots,n^{+}) =
\cr & =
(-e\sqrt2)^n \permsum{2}{n}
{
{-\braket{p}{I}}
\over
{\bra{p} 2,\ldots,n \ket{I}}
}
\Biggl\{
{
{ {\braket{h}{p}}^{*} }
\over
{ { \bra{h}I\ket{p} }^{*} }
}u^{\alpha}(p)
-{
{\braket{p}{I} u^{\alpha}(k_I)}
\over
{[p+k_I+\kappa(2,n)]^2}
}
\cr&\qquad\qquad
- (1-{\delta}_{n1}) \braket{p}{I}
\sum_{j=2}^n
{
{ u^{\beta}(k_I)
[p+k_I+\kappa(2,j)]_{\beta\dot\alpha}
\bar k_j^{\dot\alpha\alpha} }
\over
{ [p+k_I+\kappa(2,j{-}1)]^2 [p+k_I+\kappa(2,j)]^2 }
}
\Biggr\}.
}
\eqlabel\LHoneminussoln
$$
To obtain (\LHoneminussoln), a series of steps very much like those
performed in {Appendix~\prtosm}{ }({\it cf.}
equations (\Dstart)--(\Dend)) is required.


\subsection{The transverse $W$ currents}

Since the $W$ boson is a vector particle, we must consider its helicity
along with the helicity of the photons.  Thus, the one unlike helicity
can appear either as the helicity of one of the photons, or as the
helicity of the $W$ itself.  Consider first the case where it is one
of the photons which has the opposing helicity.  If we choose
$$
{\eps}_{\alpha\dot\alpha}(I^{-})=
{{ u_{\alpha}(k_I) {\bar u}_{\dot\alpha}(h) }
\over {\braket{I}{h}}^{*}}
\newlett
$$
$$
{\eps}_{\alpha\dot\alpha}(j^{+})=
{{ u_{\alpha}(k_I) {\bar u}_{\dot\alpha}(k_j) }
\over {\braket{j}{I}}}
\lett
$$
$$
{\Wms}_{\alpha\dot\alpha}(P^{+})=
{{ u_{\alpha}(k_I) {\bar u}_{\dot\alpha}(P) }
\over {\braket{P}{I}}},
\lett
$$
then we find that the same conditions hold as held when we considered
a current with purely positive helicities.  Thus, we may begin with
the simplified version of the transverse $W$ recursion relation
(\WTallplusstart):
$$
\eqalign{
{\Wms}_{\alpha\dot\alpha}&(P^{+};I^{-},2^{+},\ldots,n^{+})=
\cr & =
-e\sqrt2
\permsum{2}{n}
{ 1\over{(n-1)!} }
{
{[P+k_I+\kappa(2,n)]_{\beta\dot\beta}
}
\over
{[P+k_I+\kappa(2,n)]^2}
}
\cr & \qquad\qquad\qquad\times
\Biggl[
{\bar\eps}^{\dot\beta\beta}(I^{-})
{\Wms}_{\alpha\dot\alpha}(P^{+};2^{+},\ldots,n^{+})
\cr & \qquad\qquad\qquad\qquad
-
{\Wbar}^{\dot\beta\beta}(P^{+};2^{+},\ldots,n^{+})
{\eps}_{\alpha\dot\alpha}(I^{-})
\Biggr] \cr & \quad
-e\sqrt2
\permsum{2}{n}
{ 1\over{(n-2)!} }
{
{[P+k_I+\kappa(2,n)]_{\beta\dot\beta}
}
\over
{[P+k_I+\kappa(2,n)]^2}
}
\cr & \qquad\qquad\qquad\times
\Biggl[
{\bar\eps}^{\dot\beta\beta}(n^{+})
{\Wms}_{\alpha\dot\alpha}\bigl(P^{+};I^{-},2^{+},\ldots,(n{-}1)^{+}\bigr)
\cr & \qquad\qquad\qquad\qquad
-
{\Wbar}^{\dot\beta\beta}\bigl(P^{+};I^{-},2^{+},\ldots,(n{-}1)^{+}\bigr)
{\eps}_{\alpha\dot\alpha}(n^{+})
\Biggr].
}
\eqlabel\WToneminusstart
$$
Again, we have explicitly written out the part of the permutation
sum involving the first photon.  The one-photon current is, from
(\WToneminusstart)
$$
{\Wms}_{\alpha\dot\alpha}(P^{+};I^{-})=
-e\sqrt2  \thinspace
u_{\alpha}(k_I)u^{\beta}(k_I)[P+k_I]_{\beta\dot\alpha}
{ 1\over{ {\braket{P}{I}}^2 } }
{
{ {\braket{h}{P}}^{*} }
\over
{ { \bra{h}I\ket{P} }^{*} }
}.
\eq
$$
This suggests that we try the same ansatz for the spinor structure used
in the like helicity case:
$$
\eqalign{
{\Wms}_{\alpha\dot\alpha}&(P^{+};I^{-},2^{+},\ldots,n^{+})=
\cr & =
u_{\alpha}(k_I)u^{\beta}(k_I)[P+k_I+\kappa(2,n)]_{\beta\dot\alpha}
X(P^{+};I^{-},2^{+},\ldots,n^{+}).
}
\eqlabel\Xnegansatz
$$
In fact, a small amount of reflection on the proof of (\Xansatz)
reveals that the conditions which caused it to hold are still
met in this case.  As a result, the proof of (\Xnegansatz) follows
equations  (\Xansatz)--(\Xrecurs) without major modification.
The resulting recursion relation for
$X(P^{+};I^{-},2^{+},\ldots,n^{+})$ is
given by
$$
\eqalign{
X&(P^{+};I^{-},2^{+},\ldots,n^{+})=
\cr & =
{-e\sqrt2}\permsum{2}{n}
{
{\bar u_{\dot\alpha}(h)
[\bar P + \bar k_I + \bar \kappa(2,n)]^{\dot\alpha\alpha}
u_{\alpha}(k_I)}
\over
{(n-1)! \thinspace {\braket{I}{h}}^{*} \varsp [P+k_I+\kappa(2,n)]^2}
}
X (P^{+};2^{+},\ldots,n^{+})\cr&
{-e\sqrt2}\permsum{2}{n}
{
{\bar u_{\dot\alpha}(k_n)
[\bar P + \bar k_I + \bar \kappa(2,n)]^{\dot\alpha\alpha}
u_{\alpha}(k_I)}
\over
{(n-2)!\thinspace\braket{n}{I} \varsp [P+k_I+\kappa(2,n)]^2}
}
X \bigl(P^{+};I^{-},2^{+},\ldots,(n{-}1)^{+}\bigr).
}
\eqlabel\Xoneminusrecurs
$$
Once again, we discover a proportionality between this vector
current and the scalar current:
$$
X(P^{+};I^{-},2^{+},\ldots,n^{+}) =
{{X(p^{+})}\over{\PHI(P)}}
\PHI(P;I^{-},2^{+},\ldots,n^{+}).
\eq
$$
allowing us to immediately write down the final result:
$$
\eqalign{
{\Wms}_{\alpha\dot\alpha}&(P^{+};I^{-},2^{+},\ldots,n^{+})=
\cr & =
{ {(-e\sqrt2)^n}\over{\braket{P}{I}} } \permsum{2}{n}
{
{u_{\alpha}(k_I)u^{\beta}(k_I)[P+k_I+\kappa(2,n)]_{\beta\dot\alpha}}
\over
{\bra{P} 2,\ldots,n \ket{I}}
}
\Biggl\{
{
{ {\braket{h}{P}}^{*} }
\over
{ { \bra{h}I\ket{P} }^{*} }
}
\cr&\qquad\qquad\qquad\quad
+ (1-{\delta}_{n1})
\sum_{j=2}^n
u^{\gamma}(k_I){{\pole}_{\gamma}}^{\delta}(P,I,2,\ldots,j)
u_{\delta}(k_I)
\Biggr\}.
}
\eqlabel\WToneminussoln
$$

The final helicity combination we consider is the one where the $W$
boson has the differing helicity, while all of the photons have the
same helicity.  In this case, the appropriate gauge choice reads
$$
{\Wms}_{\alpha\dot\alpha}(P^{-})=
{
{ u_{\alpha}(P) {\bar u}_{\dot\alpha}(h) }
\over
{{\braket{P}{h}}^{*}}
},
\newlettlabel\negW
$$
$$
{\eps}_{\alpha\dot\alpha}(j^{+})=
{
{ u_{\alpha}(P) {\bar u}_{\dot\alpha}(k_j) }
\over
{\braket{j}{P}}
}.
\lett
$$
Since all dot products between the polarization vectors vanish with
this choice, we can once again use the simplified form
(\WTallplusstart) of the
recursion relation, which becomes
$$
\eqalign{
{\Wms}_{\alpha\dot\alpha}&(P^{-};1^{+},\ldots,n^{+}) =
\cr & =
-e\sqrt2
\permsum{1}{n}
{ 1\over{(n-1)!} }
{
{[P+\kappa(1,n)]_{\beta\dot\beta}}
\over
{[P+\kappa(1,n)]^2}
}
\cr & \qquad\qquad\qquad\times
\Biggl[
{\bar\eps}^{\dot\beta\beta}(n^{+})
{\Wms}_{\alpha\dot\alpha}\bigl(P^{-};1^{+},\ldots,(n{-}1)^{+}\bigr)
\cr & \qquad\qquad\qquad\qquad
-
{\Wbar}^{\dot\beta\beta}\bigl(P^{-};1^{+},\ldots,(n{-}1)^{+}\bigr)
{\eps}_{\alpha\dot\alpha}(n^{+})
\Biggr].
}
\eqlabel\negWTallplusstart
$$
Notice, however, that our usual ansatz,
(\Xansatz) with $g=P$, does not work for $n=0$.  A direct calculation
of the one-photon current from (\negWTallplusstart) gives
$$
\eqalign{
{\Wms}_{\alpha\dot\alpha}(P^{-};1^{+})& =
-e\sqrt2 \thinspace \thinspace
{
{u_{\alpha}(P)\bar u_{\dot\alpha}(k_1)}
\over
{[P+k_1]^2}
}
{
{ {\braket{h}{1}}^{*} }
\over
{ {\braket{P}{h}}^{*} }
} \cr&
= -e\sqrt2  \thinspace \thinspace
{
{ u_{\alpha}(P)u^{\beta}(P)[P+k_1]_{\beta\dot\alpha}}
\over
{\braket{P}{1} \varsp [P+k_1]^2}
}
{
{ {\braket{h}{1}}^{*} }
\over
{ {\braket{P}{h}}^{*} }
},
}
\eqlabel\negWTallplusone
$$
which  {\it does} satisfy the ansatz with
$$
X(P^{-};1^{+})=
{
{ {\braket{h}{1}}^{*} }
\over
{ {\braket{P}{h}}^{*} }
}
{
{-e\sqrt2}
\over
{\braket{P}{1} \varsp [P+k_1]^2}
}.
\eqlabel\Xseed
$$
A somewhat more lengthy calculation for the two-photon current produces
$$
\eqalign{
{\Wms}_{\alpha\dot\alpha}&(P^{-};1^{+},2^{+})=
\cr & =
(-e\sqrt2)^2 {\sum_{{\cal{P}}(12)}}
{
{ {\braket{h}{1}}^{*} }
\over
{ {\braket{P}{h}}^{*} }
}
{
{u_{\alpha}(P)u^{\beta}(P)[P{+}k_1{+}k_2]_{\beta\dot\alpha} }
\over
{ \braket{P}{1} \bra{P}2\ket{P} }
}
u^{\gamma}(P) {{\pole}_{\gamma}}^{\delta}(P,1,2) u_{\delta}(P).
}
\eqlabel\alsoran
$$
This also satisfies the ansatz, so we propose
$$
{\Wms}_{\alpha\dot\alpha}(P^{-};1^{+},\ldots,n^{+})=
u_{\alpha}(P)u^{\beta}(P)[P+\kappa(1,n)]_{\beta\dot\alpha}
X(P^{-};1^{+},\ldots,n^{+})
\eqlabel\negXansatz
$$
for $n\geq1$.  A proof very similar to the one in the all like helicity
case (equations (\Xansatz)--(\Xrecurs)) shows this ansatz to be true
with
$$
\eqalign{
X&(P^{-};1^{+},\ldots,n^{+}) =
\cr & =
-e \sqrt2 \permsum{1}{n}
{
{ {\bar u}_{\dot\beta}(k_n)[P+\kappa(1,n)]^{\dot\beta\alpha}u_{\beta}(P)}
\over
{(n-1)!\thinspace\braket{n}{P}\varsp [P+\kappa(1,n)]^2}
}
X\bigl(P^{-};1^{+},\ldots,(n{-}1)^{+}\bigr).
}
\eqlabel\negXrecurs
$$
We may iterate (\negXrecurs) to obtain
$$
\eqalign{
X&(P^{-};1^{+},\ldots,n^{+}) =
\cr & =
(-e\sqrt2)^{n-1}
\permsum{1}{n} X(P^{-};1^{+})
\prod_{\ell=2}^n
{ { {\bar u}_{\dot\beta}(k_{\ell})[\bar P + \bar \kappa(1,\ell)]^
{\dot\beta\beta} u_{\beta}(P) } \over
{ \braket{\ell}{P} \varsp [P+\kappa(1,\ell)]^2 } }.
}
\eqlabel\WLallplusprod
$$
We may convert this to an expression containing $\Xi(2,n)$ by
inserting
$$
\permsum{2}{n} {1\over{(n-1)!}} =1.
\eq
$$
So, using (\XIid) with $j=2$  for $\Xi(2,n)$ and (\Xseed) for
$X(P^{-};1^{+})$, we see  that (\negXansatz) yields
$$
\eqalign{
{\Wms}_{\alpha\dot\alpha}&(P^{-};1^{+},\ldots,n^{+}) =
\cr & =
(-e\sqrt2)^n  \permsum{1}{n}
{
{ {\braket{h}{1}}^{*} }
\over
{ {\braket{P}{h}}^{*} }
}
{
{u_{\alpha}(P)u^{\beta}(P)[P+\kappa(1,n)]_{\beta\dot\alpha} }
\over
{ \braket{P}{1} \bra{P}2,\ldots,n\ket{P} }
}
\cr & \qquad\qquad\qquad\times
\sum_{\ell=2}^n
u^{\gamma}(P) {{\pole}_{\gamma}}^{\delta}(P,1,\ldots,\ell)
u_{\delta}(P).
}
\eqlabel\negWTrestallplussoln
$$
This expression is valid for $n\geq2$.  For $n=0$ and $n=1$ we must use
(\negW a)  and (\negWTallplusone) respectively.

We note in passing that it
is possible to eliminate the gauge momentum $h$ from (\alsoran) by
explicitly working out the permutation sum.  The iteration process
producing (\WLallplusprod) can then be terminated with $X(P^{-},
1^{+},2^{+})$, yielding a form of the $n$-photon current that
is explicitly independent of the gauge momentum for $n\geq3$.
The cost, however, is a current which does not take its
``universal'' form until $n=3$, complicating computations which sum
over various currents.  Since the zero- and one-photon currents
still contain $h$, it is best to stick with (\negW a),{ }
(\negWTallplusone)
and (\negWTrestallplussoln).

\chapter{AMPLITUDES}

A significant number of amplitudes for specific helicity configurations
may be obtained from the currents derived in the last section.
In this paper we will consider those amplitudes that may be derived
from a single current, or the combination of two currents.
Amplitudes involving three or more currents will be discussed
elsewhere [\ref{
G. Mahlon, Cornell preprint CLNS 92/1154, 1992.}
\refname\thirdpaper {\kern-.7em}].

The total number of unlike helicities must
be at least two \cite\REVIEWx, or else the amplitude vanishes.
Fermion lines or
scalar lines each count as one of the
required unlike helicities.  In the massless
limit, fermion and
antifermion must have opposite helicities to couple.
The scalars, although carrying no helicity themselves, may
be thought of as the superposition of a pair of opposing
helicities~\cite\DY.


\section{Amplitudes Involving One Current}

The simplest imaginable scheme for computing an amplitude from
the currents is to simply remove the propagator for the off-shell
particle, and put that particle on shell, supplying a polarization
spinor, if required.   Thus, for a scalar line we have
$$
\eqalign{
\amp&(P^{0},Q^{0};1,\ldots,n)=
\lim_{Q^2\to0} iQ^2
\PHI(P;1,\ldots,n)
\Big\vert_{Q\equiv -P-\kappa(1,n)},
}
\eqlabel\scalarsample
$$
while fermion lines produce the forms
$$
\eqalign{
\amp&(p^{+},q^{-};1,\ldots,n)=
\lim_{q^2\to0} -i
{\positronon}_{\dot\alpha}(p^{+};1,\ldots,n)
\bar q^{\dot\alpha\alpha} u_{\alpha}(q)
\Big\vert_{q\equiv -p-\kappa(1,n)},
}
\newlettlabel\fermionLsample
$$
$$
\eqalign{
\amp&(p^{-},q^{+};1,\ldots,n)=
\lim_{q^2\to0} -i
{\positronon}^{\alpha}(p^{-};1,\ldots,n)
q_{\alpha\dot\alpha} \bar u^{\dot\alpha}(q)
\Big\vert_{q\equiv -p-\kappa(1,n)},
}
\lett
$$
and vector lines give
$$
\eqalign{
\amp&(P,Q;1,\ldots,n)=
\lim_{Q^2\to0} iQ^2
\bar\eps^{\dot\alpha\alpha}(Q)
\Wms_{\alpha\dot\alpha}(P;1,\ldots,n)
\Big\vert_{Q\equiv -P-\kappa(1,n)}.
}
\eqlabel\vectorsample
$$
Similar expressions apply when the corresponding anti-particle
currents replace the currents that appear above.

To illustrate the application of this method, consider the process
producing $n$ like-helicity photons from a pair of transverse $W$'s of
the other helicity.  This amplitude is given by
$$
\eqalign{
\amp&(P^{-},Q^{-};1^{+},\ldots,n^{+})=
\cr & =
\lim_{Q^2\to0} iQ^2
\bar\eps^{\dot\alpha\alpha}(Q^{-})
\Wms_{\alpha\dot\alpha}(P^{-};1^{+},\ldots,n^{+})
\Big\vert_{Q\equiv -P-\kappa(1,n)}.
}
\eqlabel\amptwonegWTstart
$$

Elementary kinematics requires that there be at least
two photons present, allowing
us to use (\negWTrestallplussoln) for
$\Wms(P^{-};1^{+},\ldots,n^{+})$.
Further, note that the on-shell condition $Q^2=0$ causes
every term in the sum on $\ell$ appearing in (\negWTrestallplussoln)
to vanish, except for $\ell=n$, which has a matching pole.  Thus,
(\amptwonegWTstart) becomes
$$
\eqalign{
\amp&(P^{-},Q^{-};1^{+},\ldots,n^{+})=
\cr &\negthinspace\negthinspace\negthinspace =\negthinspace
{
{-i(-e\sqrt2)^n}
\over
{ {\braket{Q}{h^\prime}}^{*} }
}
\negthinspace\negthinspace\negthinspace\negthinspace
\permsum{1}{n}\negthinspace\negthinspace
{
{ {\braket{h}{1}}^{*} }
\over
{ {\braket{h}{P}}^{*} }
}
{
{ \braket{Q}{P} u^{\beta}(P)Q_{\beta\dot\alpha}
\bar u^{\dot\alpha}(h^\prime)}
\over
{ \braket{P}{1} \bra{P}2,\ldots,n\ket{P} }
}
{
{ u^{\gamma}(P) k_{n\gamma\dot\gamma}
\bar Q^{\dot\gamma\delta} u_{\delta}(P) }
\over
{ (k_n + Q)^2 }
},
}
\eqlabel\yucky
$$
where $h^\prime$ is the gauge momentum appearing in $\eps(Q^{-})$,
and we have used momentum conservation to eliminate $P+\kappa(1,n)$
in favor of $Q$.  Use of (\spinnorm) and (\antisym) allows us to
cancel many of the factors appearing in (\yucky), producing
$$
\eqalign{
\amp&(P^{-},Q^{-};1^{+},\ldots,n^{+})=
\cr &\qquad\qquad =
i(-e\sqrt2)^n
\permsum{1}{n}
{
{ {\braket{h}{1}}^{*} }
\over
{ {\braket{h}{P}}^{*} }
}
{
{ {\braket{P}{Q}}^3 }
\over
{ \braket{P}{1} }
}
{
1
\over
{\bra{P}2,\ldots,n\ket{Q}}
}.
}
\eqlabel\sumME
$$

The next step is to use (\permsumid) to convert the denominator
appearing in  (\sumME) to a form which is explicitly invariant
under ${{\cal{P}}(1\ldots n)}$.  Noting that
$$
1=\permsum{2}{n} { 1\over{(n-1)!} },
\eq
$$
we have
$$
\eqalign{
\amp&(P^{-},Q^{-};1^{+},\ldots,n^{+})=
\cr & =
i(-e\sqrt2)^n
\permsum{1}{n}
{ 1\over{(n-1)!} }
{
{ {\braket{h}{1}}^{*} }
\over
{ {\braket{h}{P}}^{*} }
}
{
{ {\braket{P}{Q}}^3 }
\over
{ \braket{P}{1} }
}
\permsum{2}{n}
{
1
\over
{\bra{P}2,\ldots,n\ket{Q}}
}
\cr&=
i(-e\sqrt2)^n
\permsum{1}{n}
{ 1\over{(n-1)!} }
{
{ {\braket{h}{1}}^{*} \braket{1}{Q}}
\over
{ {\braket{h}{P}}^{*} }
}
{
{ {\braket{P}{Q}}^3 }
\over
{ \bra{P}1\ket{Q} }
}
{
{ {\braket{P}{Q}}^{n-2} }
\over
{ \prod\limits_{j=2}^n \bra{P}j\ket{Q} }
}
\cr&=
-i(-e\sqrt2)^n
{
{ {\braket{P}{Q}}^{n+1} }
\over
{ \prod\limits_{j=1}^n \bra{P}j\ket{Q} }
}
{ 1\over {  {\braket{h}{P}}^{*}  } }
\permsum{1}{n}
{
{\bar u_{\dot\alpha}(h) \bar k_1^{\dot\alpha\alpha} u_{\alpha}(Q)}
\over
{ (n-1)! }
}.
}
\eq
$$
The remaining permutation sum is trivial.  It simply produces
$(n-1)!$ copies of $\bar u_{\dot\alpha}(h)
\bar \kappa^{\dot\alpha\alpha}(1,n) u_{\alpha}(Q) = -
\bar u_{\dot\alpha}(h) (\bar P + \bar Q)^{\dot\alpha\alpha}
u_{\alpha}(Q)$.  Thus
$$
\amp(P^{-},Q^{-};1^{+},\ldots,n^{+}) =
-i(-e\sqrt2)^n
{
{ {\braket{P}{Q}}^{n+2} }
\over
{ \prod\limits_{j=1}^n \bra{P}j\ket{Q} }
}.
\eq
$$

The remaining non-vanishing amplitudes which may be computed
from a single current are done in much the same manner.   The
results all take the form
$$
\amp(P,Q;1,2^{+},\ldots,n^{+}) =
i(-e\sqrt2)^n
{
{ {\braket{P}{Q}}^{n-2} }
\over
{ \prod\limits_{j=1}^n \bra{P}j\ket{Q} }
}
f(P,Q,1)
\eqlabel\onecurrentamps
$$
where the function $f(P,Q,1)$ depends on the spin of the charged line
and the helicity combination involved.  Values of $f(P,Q,1)$ are
given in Table \Tlabel\onecurrenttable .  In Table~1, the helicities
and identities of the charged particles are based on the assumption
that their momenta are flowing {\it into}\ the diagram.
\noadvancetable{
{  
\begintable
$P$        \vb$Q$        \vb$k_1$             \| $f(P,Q,1)$ \crthick
$ W_L^{+} \thinspace$\vb$ W_L^{-}\thinspace$\vb$ \gamma_{\down} $\|
$ {\braket{P}{1}}^2 {\braket{1}{Q}}^2$ \cr
$\bar e_{\down}$\vb$ e_{\up} $\vb$ \gamma_{\down} $\|
$-{\braket{P}{1}}^3 \braket{1}{Q} \thinspace $\cr
$ \bar e_{\up} $\vb$e_{\down}$\vb$ \gamma_{\down} $\|
$ {\braket{P}{1}}   {\braket{1}{Q}}^3 $\cr
$W^{+}_{\down} $\vb$W^{-}_{\down}$\vb$\gamma_{\up}  $\|
$ -{\braket{P}{Q}}^4 $ \cr
$W^{+}_{\down} $\vb$W^{-}_{\up}$\vb$\gamma_{\down}  $\|
$ -{\braket{P}{1}}^4 $ \cr
$W^{+}_{\up} $\vb$W^{-}_{\down}$\vb$\gamma_{\down}  $\|
$ -{\braket{1}{Q}}^4 $
\endtable
}}{Helicity functions for $W^{+}W^{-}$
and $e^{+}e^{-}$ annihilation into photons} 

The supersymmetric-like relations among the various currents are
still in evidence in the results for these amplitudes:  they are
connected by simple proportionalities.


\section{Amplitudes Involving Two Currents}

We now consider the computation of amplitudes by joining
a pair of currents at a vertex.  Since there is always one particle
left off shell at the vertex in such a construction, this
allows us to compute certain amplitudes with the equivalent
of three unlike helicities.  In choosing which pairs of currents
may be joined in this manner, we must be sure to be consistent
in our choices in gauge momenta.  For example, we cannot
join $\Wms(P^{-};I^{-},2^{+},\ldots,m^{+})$ and
$\Wms\bigl((m+1)^{+},\ldots,n^{+};Q^{-}\bigr)$ in an attempt to obtain
$\amp(P^{-},Q^{-};I^{-},2{+},\ldots,n^{+})$, because the former
current requires the gauge momentum of the positive helicity photons
to be~$P$, while the latter current requires that momentum to
be~$Q$.  We must choose combinations with compatible gauge conditions.
In spite of these restrictions, it is possible to form a
number of amplitudes in this manner which cannot be obtained by
the methods of the previous section.  Of particular interest is
the possibility of joining currents of differing spins, allowing
for the neutral particles in a process to be neutrinos, Higgs
bosons, or longitudinally polarized~$Z$'s.

\subsection{Annihilation of two longitudinal $W$'s into  photons}

As an example of how to link two currents together to form an
amplitude, we compute the matrix element for two longitudinally
polarized
$W$'s annihilating into a pair of negative helicity photons
plus $n-2$ positive helicity photons.  This process has been
selected because it contains essentially all of the complications
which must be overcome in a computation of an amplitude from
two currents.

We begin by defining the object $G_{\mu}(P^0,Q^0;1,2,\ldots,n) $,
a charged scalar line with $n$  on-shell photons
plus one off-shell photon attached.   Taking
all momenta to be flowing inward, the positively charged scalar
has momentum $P$, the negatively charged scalar has momentum $Q$,
the on-shell photons have momenta $k_1$,$\thinspace k_2$,
$\ldots\thinspace$,$\thinspace k_n$,
and the off-shell photon has momentum $-[P+\kappa(1,n)+Q]$.  We
do not include a propagator for the off-shell photon since it
is not required to obtain an amplitude.
Realizing that the off-shell photon may be attached to either
a three- or four-point vertex, we have, by definition,
$$
\eqalign{
G_{\mu}&(P^0,Q^0;1,2,\ldots,n) =
\cr & =
ie \permsum{1}{n}  \sum_{m=0}^n
{ 1\over{m!(n-m)!} }
\PHI(P;1,2,\ldots,m)
\Bigl\{ \bigl[P+\kappa(1,m)\bigr]
\cr & \qquad\qquad\qquad\qquad\quad
- \bigl[\kappa(m{+}1,n)+Q\bigr] \Bigr\}_{\mu}
\PHI(m{+}1,\ldots,n;Q)
\cr & \quad
+ 2ie^2 \permsum{1}{n}  \sum_{m=1}^n
{ 1\over{(m-1)!(n-m)!} }
\PHI(P;1,2,\ldots,m{-}1)
\cr & \qquad\qquad\qquad\qquad\qquad\qquad\qquad\times
{\eps}_{\mu}(m)
\PHI(m{+}1,\ldots,n;Q).
}
\eqlabel\gluescalar
$$
If we let
$$
\dblscalar_{\alpha\dot\alpha} \equiv
{1\over{\sqrt2}}(\sigma^{\mu})_{\alpha\dot\alpha}G_{\mu}
\eq
$$
and rewrite the permutation sums with the first photon explicitly
separated out, we obtain
$$
\eqalign{
{\dblscalar}_{\alpha\dot\alpha}&(P^0,Q^0;I^{-},2^{+},\ldots,n^{+}) =
\cr & =
-{i\over2}(-e\sqrt2) \permsum{2}{n} \sum_{m=1}^n
{ 1\over{(m{-}1)!(n{-}m)!} }
\PHI\bigl(P;I^{-},2^{+},\ldots,m^{+}\bigr)
\cr & \qquad\qquad\qquad\qquad\times
\bigl\{ [P+k_I+\kappa(2,m)]
- [\kappa(m{+}1,n)+Q] \bigr\}_{\alpha\dot\alpha}
\cr & \qquad\qquad\qquad\qquad\times
\PHI\bigl((m{+}1)^{+},\ldots,n^{+};Q \bigr)
\cr &
-{i\over2}(-e\sqrt2) \permsum{2}{n} \sum_{m=1}^n
{ 1\over{(m{-}1)!(n{-}m)!} }
\PHI\bigl(P;(m{+}1)^{+},\ldots,n^{+}\bigr)
\cr & \qquad\qquad\qquad\qquad\times
\bigl\{ [P+\kappa(m{+}1,n)]
- [k_I+\kappa(2,m)+Q] \bigr\}_{\alpha\dot\alpha}
\cr & \qquad\qquad\qquad\qquad\times
\PHI\bigl(I^{-},2^{+},\ldots,m^{+};Q \bigr)
\cr &
+ i(-e\sqrt2)^2 \permsum{2}{n}  \sum_{m=1}^{n-1}
{ 1\over{(m{-}1)!(n{-}m{-}1)!} }
\PHI\bigl(P;I^{-},2^{+},\ldots,m^{+}\bigr)
\cr & \qquad\qquad\qquad\qquad\times
{\eps}_{\alpha\dot\alpha}\bigl((m{+}1)^{+}\bigr)
\PHI\bigl((m{+}2)^{+},\ldots,n^{+};Q \bigr)
\cr &
+ i(-e\sqrt2)^2 \permsum{2}{n}  \sum_{m=1}^{n-1}
{ 1\over{(m{-}1)!(n{-}m{-}1)!} }
\PHI\bigl(P;(m{+}2)^{+},\ldots,n^{+}\bigr)
\cr & \qquad\qquad\qquad\qquad\times
{\eps}_{\alpha\dot\alpha}\bigl((m{+}1)^{+}\bigr)
\PHI\bigl(I^{-},2^{+},\ldots,m^{+};Q \bigr)
\cr &
+ i(-e\sqrt2)^2 \permsum{2}{n}  \sum_{m=1}^{n}
{ 1\over{(m{-}1)!(n{-}m)!} }
\PHI\bigl(P;2^{+},\ldots,m^{+}\bigr)
\cr & \qquad\qquad\qquad\qquad\times
{\eps}_{\alpha\dot\alpha}\bigl(I^{-}\bigr)
\PHI\bigl((m{+}1)^{+},\ldots,n^{+};Q \bigr).
}
\eqlabel\youmustbejoking
$$
Note that we have shifted the summation over $m$ in two of the seagull
contributions.  Although (\youmustbejoking) looks extremely
complicated, it is clear where each term comes from.  First, the
negative helicity photon could be on the same side of the vertex
as the $\phi^{+}$  (momentum $P$).  This is represented by the
first and third terms.
Or, it could be on the same side as the $\phi^{-}$ (momentum $Q$), as
in the second and fourth
terms.  Finally, it could be located precisely at a seagull
vertex, the source of the fifth term.

The amplitude we wish to compute is given by
$$
\eqalign{
\amp&(P^0,Q^0;I^{-},II^{-},3^{+},\ldots,n^{+}) \equiv
\cr & \qquad\qquad\equiv
\bar\eps^{\dot\alpha\alpha}(II^{-})
{\dblscalar}_{\alpha\dot\alpha}(P^0,Q^0;I^{-},3^{+},\ldots,n^{+})
\cr &\qquad\qquad =
\bar\eps^{\dot\alpha\alpha}(I^{-})
{\dblscalar}_{\alpha\dot\alpha}(P^0,Q^0;II^{-},3^{+},\ldots,n^{+}).
}
\eqlabel\sewstart
$$
Because of the way we have defined $\dblscalar$, we do not need to
symmetrize between the two negative helicity photons; either
form in (\sewstart) will ultimately give us the same expression
which is explicitly symmetric under the interchange of $I$ and $II$.

We now select a gauge in which to do the calculation.  Our choice is
$$
{\eps}_{\alpha\dot\alpha}(I^{-})=
{{ u_{\alpha}(k_I) {\bar u}_{\dot\alpha}(h) }
\over {\braket{I}{h}}^{*}}
\newlettlabel\sewscalargauge
$$
$$
{\eps}_{\alpha\dot\alpha}(II^{-})=
{{ u_{\alpha}(k_{II}) {\bar u}_{\dot\alpha}(h) }
\over {\braket{II}{h}}^{*}}
\lett
$$
$$
{\eps}_{\alpha\dot\alpha}(j^{+})=
{{ u_{\alpha}(k_I) {\bar u}_{\dot\alpha}(k_j) }
\over {\braket{j}{I}}}.
\lett
$$
The selection of the same gauge momentum $h$ for both negative
helicity photons causes the last term in (\youmustbejoking) to
give no contribution to the amplitude.  The other two seagull
terms do contribute, however, and must be dealt with.
\eject

The difference of momenta appearing in the
first two terms of (\youmustbejoking) may be rewritten using momentum
conservation.  For example,
$$
\eqalign{
\bar\eps^{\dot\alpha\alpha}(II^{-})
\bigl\{ [P+k_I+\kappa&(2,m)]-[\kappa(m{+}1,n)+Q] \bigr\}_{\alpha\dot\alpha}
=
\cr & =
\bar\eps^{\dot\alpha\alpha}(II^{-})
\bigl\{ -k_{II} - 2[\kappa(m{+}1,n)+Q] \bigr\}_{\alpha\dot\alpha}
\cr &
= -2
\bar\eps^{\dot\alpha\alpha}(II^{-})
[\kappa(m{+}1,n)+Q]_{\alpha\dot\alpha}.
}
\eq
$$
Thus, (\sewstart) becomes
$$
\eqalign{
\amp&(P^0,Q^0;I^{-},II^{-},3^{+},\ldots,n^{+}) =
\cr & =
i(-e\sqrt2) \permsum{3}{n} \sum_{m=2}^n
{ 1\over{(m-2)!(n-m)!} }
\bar\eps^{\dot\alpha\alpha}(II^{-})
[\kappa(m{+}1,n)+Q]_{\alpha\dot\alpha}
\cr & \qquad\qquad\qquad\qquad\times
\PHI\bigl(P;I^{-},3^{+},\ldots,m^{+}\bigr)
\PHI\bigl((m{+}1)^{+},\ldots,n^{+};Q \bigr)
\cr &
-i(-e\sqrt2) \permsum{3}{n} \sum_{m=2}^n
{ 1\over{(m-2)!(n-m)!} }
\bar\eps^{\dot\alpha\alpha}(II^{-})
[P+\kappa(m{+}1,n)]_{\alpha\dot\alpha}
\cr & \qquad\qquad\qquad\qquad\times
\PHI\bigl(P;(m{+}1)^{+},\ldots,n^{+}\bigr)
\PHI\bigl(I^{-},3^{+},\ldots,m^{+};Q \bigr)
\cr &
+ i(-e\sqrt2)^2 \permsum{3}{n}  \sum_{m=2}^{n-1}
{ 1\over{(m-2)!(n-m-1)!} }
\bar\eps^{\dot\alpha\alpha}(II^{-})
{\eps}_{\alpha\dot\alpha}\bigl((m{+}1)^{+}\bigr)
\cr & \qquad\qquad\qquad\qquad\times
\PHI\bigl(P;I^{-},3^{+},\ldots,m^{+}\bigr)
\PHI\bigl((m{+}2)^{+},\ldots,n^{+};Q \bigr)
\cr &
+ i(-e\sqrt2)^2 \permsum{3}{n}  \sum_{m=2}^{n-1}
{ 1\over{(m-2)!(n-m-1)!} }
\bar\eps^{\dot\alpha\alpha}(II^{-})
{\eps}_{\alpha\dot\alpha}\bigl((m{+}1)^{+}\bigr)
\cr & \qquad\qquad\qquad\qquad\times
\PHI\bigl(P;(m{+}2)^{+},\ldots,n^{+}\bigr)
\PHI\bigl(I^{-},3^{+},\ldots,m^{+};Q \bigr).
}
\eqlabel\getready
$$

The first step is to eliminate the seagull terms.  We do this by
inserting the appropriate expressions for the like helicity currents
(equations (\PHIallplussoln) and (\WLcrossing)) into
(\getready).  After some minor
rearrangement we have
$$
\eqalign{
\amp&(P^0,Q^0;I^{-},II^{-},3^{+},\ldots,n^{+}) =
\cr& =
{i\over{  {\braket{II}{h}}^{*}  } }
\permsum{3}{n} \sum_{m=2}^n
{
{ (-e\sqrt2)^{n-m+1} }
\over
{(m-2)!}
}
\PHI\bigl(P;I^{-},3^{+},\ldots,m^{+}\bigr)
\cr & \qquad\qquad\qquad\qquad\times
{
{ \braket{I}{Q}
\bar u_{\dot\alpha}(h)
[\bar\kappa(m{+}1,n)+\bar Q]^{\dot\alpha\alpha}
u_{\alpha}(k_{II}) }
\over
{ \bra{I} m{+}1,\ldots,n \ket{Q} }
}
\cr & \quad
+{i\over{  {\braket{II}{h}}^{*}  } }
\permsum{3}{n} \sum_{m=2}^{n-1}
{
{ (-e\sqrt2)^{n-m+1} }
\over
{(m-2)!}
}
\PHI\bigl(P;I^{-},3^{+},\ldots,m^{+}\bigr)
\cr & \qquad\qquad\qquad\qquad\times
{
{ \braket{I}{Q} \braket{I}{II} {\braket{h}{m{+}1}}^{*} }
\over
{ \braket{I}{m{+}1} \bra{I} m{+}2,\ldots,n \ket{Q} }
}
\cr & \quad
-{i\over{  {\braket{II}{h}}^{*}  } }
\permsum{3}{n} \sum_{m=2}^n
{
{ (-e\sqrt2)^{n-m+1} }
\over
{(m-2)!}
}
\PHI\bigl(I^{-},3^{+},\ldots,m^{+};Q\bigr)
\cr & \qquad\qquad\qquad\qquad\times
{
{ \braket{P}{I}
\bar u_{\dot\alpha}(h)
[\bar P +\bar\kappa(m{+}1,n)]^{\dot\alpha\alpha}
u_{\alpha}(k_{II}) }
\over
{ \bra{P} n,n{-}1,\ldots,m{+}1 \ket{I} }
}
\cr & \quad
-{i\over{  {\braket{II}{h}}^{*}  } }
\permsum{3}{n} \sum_{m=2}^{n-1}
{
{ (-e\sqrt2)^{n-m+1} }
\over
{(m-2)!}
}
\PHI\bigl(I^{-},3^{+},\ldots,m^{+};Q\bigr)
\cr & \qquad\qquad\qquad\qquad\times
{
{ \braket{P}{I} \braket{I}{II} {\braket{h}{m{+}1}}^{*} }
\over
{ \bra{P} n,n{-}1,\ldots,m{+}2 \ket{I} \braket{m{+}1}{I} }
}.
}
\eqlabel\GO
$$
Let us call the first term in (\GO) $\amp_1$.
We begin by using the
Schouten identity (\fierz) to decompose $\amp_1$ into two pieces,
obtaining
$$
\eqalign{
\amp_1&=
{i\over{  {\braket{II}{h}}^{*}  } }
\permsum{3}{n} \sum_{m=2}^n
{
{ (-e\sqrt2)^{n-m+1} }
\over
{(m-2)!}
}
\PHI\bigl(P;I^{-},3^{+},\ldots,m^{+}\bigr)
\cr & \qquad\qquad\qquad\qquad\times
{
{ \braket{II}{Q}
\bar u_{\dot\alpha}(h)
[\bar\kappa(m{+}1,n)+\bar Q]^{\dot\alpha\alpha}
u_{\alpha}(k_{I}) }
\over
{ \bra{I} m{+}1,\ldots,n \ket{Q} }
}
\cr & \quad
+{i\over{  {\braket{II}{h}}^{*}  } }
\permsum{3}{n} \sum_{m=2}^{n-1}
{
{ (-e\sqrt2)^{n-m+1} }
\over
{(m-2)!}
}
\PHI\bigl(P;I^{-},3^{+},\ldots,m^{+}\bigr)
\cr & \qquad\qquad\qquad\qquad\times
{
{ \braket{I}{II}
\bar u_{\dot\alpha}(h)
\bar\kappa^{\dot\alpha\alpha} (m{+}1,n)
u_{\alpha}(Q) }
\over
{ \bra{I} m{+}1,\ldots,n \ket{Q} }
}
\cr& \equiv \amp_{1A} + \amp_{1B}.
}
\eqlabel\killthegullone
$$
Notice that we have eliminated the vanishing $m=n$ part from the sum
in $\amp_{1B}$ and used the Weyl equation to dispose of
$\bar Q^{\dot\alpha\alpha}$.

We now exploit the freedom allowed by having expressions
in the form of a permutation sum to demonstrate that
$\amp_{1B}$ in (\killthegullone) cancels the
second term of (\GO).  In particular,
we relabel each term  in the sum
$\kappa(m{+}1,n)=k_{m+1}+\cdots+k_n$ as $k_{m+1}$, applying
the following set of cyclic relabellings to the photon momenta
to the coefficients of each term:
$$
\eqalign{ &
\{ m{+}1 \} \rightarrow
\{ m{+}1 \} \cr &
\{ m{+}1,m{+}2\} \rightarrow
\{ m{+}2,m{+}1\} \cr &
\{ m{+}1,m{+}2,m{+}3\} \rightarrow
\{ m{+}2,m{+}3,m{+}1\} \cr&
\qquad\vdots\cr&
\{ m{+}1,m{+}2,\ldots,n\} \rightarrow
\{m{+}2,m{+}3,\ldots, n,m{+}1\}.
}
\eq
$$
This procedure, which is similar to the one employed in
Appendix~\prtosm\ ({\it cf.}\ equations
(\RELABELstart)--(\RELABELend)), yields
$$
\eqalign{
\amp_{1B}&=
{i\over{  {\braket{II}{h}}^{*}  } }
\permsum{3}{n} \sum_{m=2}^{n-1}
{
{ (-e\sqrt2)^{n-m+1} }
\over
{(m-2)!}
}
\PHI\bigl(P;I^{-},3^{+},\ldots,m^{+}\bigr)
\cr & \times
\braket{I}{II}
{\braket{m{+}1}{h}}^{*}
\braket{m{+}1}{Q}
\Biggl\{
{1\over { \bra{I} m{+}1,m{+}2,\ldots,n \ket{Q} }}
\cr & \quad
+{1\over {\bra{I} m{+}2,m{+}1,m{+}3,\ldots,n \ket{Q} }}  + \cdots
+{1\over {\bra{I} m{+}2,\ldots,n,m{+}1 \ket{Q} }} \Biggr\}.
}
\eqlabel\secondmess
$$
The expression in curly brackets may be summed using (\linkidsummed):
$$
\eqalign{ &
\Biggl[
{ {\braket{I}{m{+}2}}\over{\bra{I}m{+}1\ket{m{+}2}} } +
{ {\braket{m{+}2}{m{+}3}}\over{\bra{m{+}2}m{+}1\ket{m{+}3}} } +
\cdots +
{ {\braket{n}{Q}}\over{\bra{n}m{+}1\ket{Q}} }
\Biggr]
\cr & \qquad\qquad\qquad\times
{1\over { \bra{I} m{+}2,\ldots,n \ket{Q} }}
=
{ {\braket{I}{Q}}\over{\bra{I}m{+}1\ket{Q}} }
{1\over { \bra{I} m{+}2,\ldots,n \ket{Q} }}.
}
\eqlabel\curlysum
$$
Insertion of (\curlysum) into (\secondmess) gives
$$
\eqalign{
\amp_{1B}=
{i\over{  {\braket{II}{h}}^{*}  } }
\permsum{3}{n} &\sum_{m=2}^{n-1}
{
{ (-e\sqrt2)^{n-m+1} }
\over
{(m-2)!}
}
\PHI\bigl(P;I^{-},3^{+},\ldots,m^{+}\bigr)
\cr & \qquad\qquad\times
{
{ \braket{I}{Q} \braket{I}{II} {\braket{m{+}1}{h}}^{*} }
\over
{ \braket{I}{m{+}1} \bra{I} m{+}2,\ldots,n \ket{Q} }
}.
}
\eq
$$
which indeed cancels the second term of (\GO), thanks to the
antisymmetry of the spinor inner product.

After applying an entirely analogous procedure to the third and
fourth terms of (\GO), we are left with
$$
\eqalign{
\amp&(P^0,Q^0;I^{-},II^{-},3^{+},\ldots,n^{+}) =
\cr& =
{i\over{  {\braket{II}{h}}^{*}  } }
\permsum{3}{n} \sum_{m=2}^n
{
{ (-e\sqrt2)^{n-m+1} }
\over
{(m-2)!}
}
\PHI\bigl(P;I^{-},3^{+},\ldots,m^{+}\bigr)
\cr & \qquad\qquad\qquad\qquad\times
{
{ \braket{II}{Q}
\bar u_{\dot\alpha}(h)
[\bar\kappa(m{+}1,n)+\bar Q]^{\dot\alpha\alpha}
u_{\alpha}(k_{I}) }
\over
{ \bra{I} m{+}1,\ldots,n \ket{Q} }
}
\cr & \quad -
{i\over{  {\braket{II}{h}}^{*}  } }
\permsum{3}{n} \sum_{m=2}^n
{
{ (-e\sqrt2)^{n-m+1} }
\over
{(m-2)!}
}
\PHI\bigl(I^{-},3^{+},\ldots,m^{+};Q\bigr)
\cr & \qquad\qquad\qquad\qquad\times
{
{ \braket{P}{II}
\bar u_{\dot\alpha}(h)
[\bar P+\bar\kappa(m{+}1,n)]^{\dot\alpha\alpha}
u_{\alpha}(k_{I}) }
\over
{ \bra{P}n,n{-}1,\ldots,m{+}1 \ket{I} }
}.
}
\eqlabel\nextplug
$$
Equations (\PHIoneminussoln) and (\WLcrossing) for the currents
containing a negative helicity photon may now be inserted
into the amplitude to obtain
$$
\eqalign{
\amp&(P^0,Q^0;I^{-},II^{-},3^{+},\ldots,n^{+}) =
\cr& =
-
{
{ i(-e\sqrt2)^{n} }
\over
{  {\braket{II}{h}}^{*}  }
}
\permsum{3}{n} \sum_{m=2}^n
{
{ \bar u_{\dot\alpha}(h)
[\bar\kappa(m{+}1,n)+\bar Q]^{\dot\alpha\alpha}
u_{\alpha}(k_I) }
\over
{ \bra{P} 3,\ldots,m \ket{I}
  \bra{I} m{+}1,\ldots,n \ket{Q} }
}
\cr & \qquad\qquad\qquad\qquad\qquad\qquad\times
\braket{P}{I} \braket{II}{Q}
{
{  {\braket{h}{P}}^{*}  }
\over
{  {\bra{h}I\ket{P}}^{*}  }
}
\cr & \quad
-
{
{ i(-e\sqrt2)^{n} }
\over
{  {\braket{II}{h}}^{*}  }
}
\permsum{3}{n} \sum_{m=2}^n
{
{ \bar u_{\dot\alpha}(h)
[\bar P+\bar\kappa(m{+}1,n)]^{\dot\alpha\alpha}
u_{\alpha}(k_I) }
\over
{ \bra{P} n,n{-}1,\ldots,m{+}1 \ket{I}
  \bra{I} m,m{-}1,\ldots,3 \ket{Q} }
}
\cr & \qquad\qquad\qquad\qquad\qquad\qquad\times
\braket{P}{II} \braket{I}{Q}
{
{  {\braket{h}{Q}}^{*}  }
\over
{  {\bra{h}I\ket{Q}}^{*}  }
}
\cr & \quad
-
{
{ i(-e\sqrt2)^{n} }
\over
{  {\braket{II}{h}}^{*}  }
}
\permsum{3}{n} \sum_{m=3}^n \sum_{j=3}^m
{
{ \bar u_{\dot\alpha}(h)
[\bar\kappa(m{+}1,n)+\bar Q]^{\dot\alpha\alpha}
u_{\alpha}(k_I) }
\over
{ \bra{P} 3,\ldots,m \ket{I}
  \bra{I} m{+}1,\ldots,n \ket{Q} }
}
\cr & \qquad\qquad\qquad\qquad\qquad\qquad\times
\braket{P}{I} \braket{II}{Q}
u^{\gamma}(k_I)
{{\pole}_{\gamma}}^{\delta}(P,I,3,\ldots,j)
u_{\delta}(k_I)
\cr & \quad
-
{
{ i(-e\sqrt2)^{n} }
\over
{  {\braket{II}{h}}^{*}  }
}
\permsum{3}{n} \sum_{m=3}^n \sum_{j=3}^m
{
{ \bar u_{\dot\alpha}(h)
[\bar P+\bar\kappa(m{+}1,n)]^{\dot\alpha\alpha}
u_{\alpha}(k_I) }
\over
{ \bra{P} n,n{-}1,\ldots,m{+}1 \ket{I}
  \bra{I} m,m{-}1,\ldots,3 \ket{Q} }
}
\cr & \qquad\qquad\qquad\qquad\qquad\qquad\times
\braket{P}{II} \braket{I}{Q}
u^{\gamma}(k_I)
{{\pole}_{\gamma}}^{\delta}(Q,I,3,\ldots,j)
u_{\delta}(k_I).
}
\eqlabel\stagetwo
$$
Call  the four terms in (\stagetwo) $\amp_1, \amp_2, \amp_3,$ and
$\amp_4$, respectively.

In order to do the sum on $m$ in $\amp_1$, we make the definitions
$k_2\equiv P$ and $k_{n+1}\equiv Q$.
Then, the sums may be written as
$$
\eqalign{
\sum_{m=2}^n \sum_{\ell=m+1}^{n+1}
{
{\braket{m}{m{+}1}}
\over
{\bra{m}I\ket{m{+}1}}
}
\bar u_{\dot\alpha}(h)&
\bar k_{\ell}^{\dot\alpha\alpha}
u_{\alpha}(k_I) =
\cr & =
\sum_{\ell=3}^{n+1} \sum_{m=2}^{\ell-1}
{
{\braket{m}{m{+}1}}
\over
{\bra{m}I\ket{m{+}1}}
}
{\braket{\ell}{h}}^{*}
\braket{\ell}{I}
\cr&
= {1\over{\braket{P}{I}}}
\bar u_{\dot\alpha}(h)
[\bar \kappa(3,n)+\bar Q]^{\dot\alpha\alpha}
u_{\alpha}(P)
\cr&
= {{-1}\over{\braket{P}{I}}}
\bar u_{\dot\alpha}(h)
(\bar k_I + \bar k_{II} )^{\dot\alpha\alpha}
u_{\alpha}(P),
}
\eqlabel\sumMone
$$
where we have used (\linkidsummed) to do the $m$ summation, and momentum
conservation plus the Weyl equation to obtain the last line.  Thus,
we have
$$
\eqalign{
\amp_1 &=
{
{ i(-e\sqrt2)^{n} }
\over
{  {\braket{II}{h}}^{*}  }
}
\permsum{3}{n}
{
{  \braket{II}{Q}
\bar u_{\dot\alpha}(h)
(\bar k_I + \bar k_{II} )^{\dot\alpha\alpha}
u_{\alpha}(P) }
\over
{ \bra{P} 3,\ldots,n \ket{Q} }
}
{
{  {\braket{h}{P}}^{*}  }
\over
{  {\bra{h}I\ket{P}}^{*}  }
}
\cr&
=-
{
{ i(-e\sqrt2)^{n} }
\over
{  {\braket{II}{h}}^{*}  }
}
\permsum{3}{n}
{
{  \braket{P}{I} \braket{II}{Q}  }
\over
{ \bra{P} 3,\ldots,n \ket{Q} }
}
{
{  {\braket{h}{P}}^{*}  }
\over
{  {\braket{P}{I}}^{*}  }
}
\cr&\quad
+{ i(-e\sqrt2)^{n} }
\permsum{3}{n}
{
{ \bra{P}II\ket{Q}   }
\over
{ \bra{P} 3,\ldots,n \ket{Q} }
}
{
{  {\braket{P}{h}}^{*}  }
\over
{  {\bra{P}I\ket{h}}^{*}  }
}.
}
\eqlabel\Mone
$$
Performing the corresponding steps on $\amp_2$ yields
$$
\eqalign{
\amp_2 =&
-
{
{ i(-e\sqrt2)^{n} }
\over
{  {\braket{II}{h}}^{*}  }
}
\permsum{3}{n}
{
{  \braket{P}{II} \braket{I}{Q}  }
\over
{ \bra{P} 3,\ldots,n \ket{Q} }
}
{
{  {\braket{h}{Q}}^{*}  }
\over
{  {\braket{I}{Q}}^{*}  }
}
\cr&{ }
+{ i(-e\sqrt2)^{n} }
\permsum{3}{n}
{
{ \bra{P}II\ket{Q}   }
\over
{ \bra{P} 3,\ldots,n \ket{Q} }
}
{
{  {\braket{h}{Q}}^{*}  }
\over
{  {\bra{h}I\ket{Q}}^{*}  }
},
}
\eqlabel\Mtwo
$$
where the final step has been to relabel $\{n,n{-}1,\ldots,3\}
\longrightarrow \{3,4,\ldots,n\}$.

In $\amp_3$, we must interchange the sum on $m$ with the sum on $j$
before proceeding.  This has the effect of modifying the result
(\sumMone) used to simplify $\amp_1$ as follows:
$$
\eqalign{
\sum_{m=j}^n \sum_{\ell=m+1}^{n+1}&
{
{\braket{m}{m{+}1}}
\over
{\bra{m}I\ket{m{+}1}}
}
\bar u_{\dot\alpha}(h)
\bar k_{\ell}^{\dot\alpha\alpha}
u_{\alpha}(k_I) =
\cr & =
{1\over{\braket{I}{j}}}
\bar u_{\dot\alpha}(h)
[\bar P + \bar k_I + \bar\kappa(3,j) + \bar k_{II}]^{\dot\alpha\alpha}
u_{\alpha}(k_j)
\cr&
={\braket{II}{h}}^{*}
{{\braket{II}{j}}\over{\braket{I}{j}}}
+{1\over{\braket{I}{j}}}
\bar u_{\dot\alpha}(h)
[\bar P + \bar k_I + \bar\kappa(3,j) ]^{\dot\alpha\alpha}
u_{\alpha}(k_j).
}
\eqlabel\sumMthree
$$
Thus, using the fact that $u^{\gamma}(k_{I})
{{\pole}_{\gamma}}^{\delta}(P,I,3,\ldots,j)
u_{\delta}(k_I)$ contains a factor of $\braket{I}{j}$, we have
$$
\eqalign{
\amp_3=&
-
{ i(-e\sqrt2)^{n} }
\permsum{3}{n}  \sum_{j=3}^n
{
{ \braket{P}{I} \braket{II}{Q} }
\over
{ \bra{P} 3,\ldots,n \ket{Q} }
}
\cr & \qquad\qquad\qquad\qquad\qquad\qquad\times
u^{\gamma}(k_{II})
{{\pole}_{\gamma}}^{\delta}(P,I,3,\ldots,j)
u_{\delta}(k_I)
\cr&
-
{
{ i(-e\sqrt2)^{n} }
\over
{  {\braket{II}{h}}^{*}  }
}
\permsum{3}{n}  \sum_{j=3}^n
{
{ \braket{P}{I} \braket{II}{Q}
 }
\over
{ \bra{P} 3,\ldots,n \ket{Q} }
}
\cr & \qquad\qquad\times
\bar u_{\dot\alpha}(h)
[\bar P + \bar k_I + \bar\kappa(3,j)]^{\dot\alpha\alpha}
{{\pole}_{\alpha}}^{\delta}(P,I,3,\ldots,j)
u_{\delta}(k_I).
}
\eqlabel\Mthreea
$$
The second term in (\Mthreea) may be rewritten as the difference
of two terms using (\splitid) and then summed over $j$.  This gives
$$
\eqalign{
\amp_3=&
-
{ i(-e\sqrt2)^{n} }
\permsum{3}{n}  \sum_{j=3}^n
{
{ \braket{P}{I} \braket{II}{Q} }
\over
{ \bra{P} 3,\ldots,n \ket{Q} }
}
\cr & \qquad\qquad\qquad\qquad\qquad\times
u^{\gamma}(k_{II})
{{\pole}_{\gamma}}^{\delta}(P,I,3,\ldots,j)
u_{\delta}(k_I)
\cr&
-
{
{ i(-e\sqrt2)^{n} }
\over
{  {\braket{II}{h}}^{*}  }
}
\permsum{3}{n}
{
{ \braket{P}{I} \braket{II}{Q}
 }
\over
{ \bra{P} 3,\ldots,n \ket{Q} }
}
\Biggl\{
{
{ \bar u_{\dot\alpha}(h)
  [\bar P + \bar k_I]^{\dot\alpha\alpha}
  u_{\alpha}(k_I) }
\over
{ [P+k_I]^2 }
}
\cr&\qquad\qquad\qquad\qquad\qquad
-
{
{ \bar u_{\dot\alpha}(h)
  [\bar P + \bar k_I + \bar\kappa(3,n)]^{\dot\alpha\alpha}
  u_{\alpha}(k_I) }
\over
{ [P+k_I+\kappa(3,n)]^2 }
}
\Biggr\}.
}
\eqlabel\Mthreeb
$$
Using momentum conservation to write $P+k_I+\kappa(3,n)=-(k_{II}+Q)$
produces
$$
\eqalign{
\amp_3=&
-
{ i(-e\sqrt2)^{n} }
\permsum{3}{n}  \sum_{j=3}^n
{
{ \braket{P}{I} \braket{II}{Q} }
\over
{ \bra{P} 3,\ldots,n \ket{Q} }
}
\cr & \qquad\qquad\qquad\qquad\qquad\times
u^{\gamma}(k_{II})
{{\pole}_{\gamma}}^{\delta}(P,I,3,\ldots,j)
u_{\delta}(k_I)
\cr&
+
{
{ i(-e\sqrt2)^{n} }
\over
{  {\braket{II}{h}}^{*}  }
}
\permsum{3}{n}
{
{ \braket{P}{I} \braket{II}{Q}
 }
\over
{ \bra{P} 3,\ldots,n \ket{Q} }
}
{
{  {\braket{h}{P}}^{*}  }
\over
{  {\braket{P}{I}}^{*}  }
}
\cr &
+
{ i(-e\sqrt2)^{n} }
\permsum{3}{n}
{
{ \bra{P}I\ket{Q}
 }
\over
{ \bra{P} 3,\ldots,n \ket{Q} }
}
{
{  {\braket{h}{Q}}^{*}  }
\over
{  {\bra{h}II\ket{Q}}^{*}  }
}
\cr &
+
i(-e\sqrt2)^n
\permsum{3}{n}
{
{ \braket{P}{I} \braket{I}{II}
 }
\over
{ \bra{P} 3,\ldots,n \ket{Q} }
}
{
{  1  }
\over
{  {\braket{II}{Q}}^{*}  }
}.
}
\eqlabel\Mthree
$$
When we treat $\amp_4$ in the analogous manner, we obtain
$$
\eqalign{
\amp_4=&+
{ i(-e\sqrt2)^{n} }
\permsum{3}{n}  \sum_{j=3}^n
{
{ \braket{P}{II} \braket{I}{Q} }
\over
{ \bra{P} n,n{-}1,\ldots,3\ket{Q} }
}
\cr & \qquad\qquad\qquad\qquad\qquad\times
u^{\gamma}(k_{II})
{{\pole}_{\gamma}}^{\delta}(Q,I,3,\ldots,j)
u_{\delta}(k_I)
\cr&
+
{
{ i(-e\sqrt2)^{n} }
\over
{  {\braket{II}{h}}^{*}  }
}
\permsum{3}{n}
{
{ \braket{P}{II} \braket{I}{Q}
 }
\over
{ \bra{P} 3,\ldots,n \ket{Q} }
}
{
{  {\braket{h}{Q}}^{*}  }
\over
{  {\braket{I}{Q}}^{*}  }
}
\cr &
+
{ i(-e\sqrt2)^{n} }
\permsum{3}{n}
{
{ \bra{P}I\ket{Q}
 }
\over
{ \bra{P} 3,\ldots,n \ket{Q} }
}
{
{  {\braket{P}{h}}^{*}  }
\over
{  {\bra{P}II\ket{h}}^{*}  }
}
\cr &
-
{ i(-e\sqrt2)^{n} }
\permsum{3}{n}
{
{ \braket{I}{II} \braket{I}{Q}
 }
\over
{ \bra{P} 3,\ldots,n \ket{Q} }
}
{
{  1  }
\over
{  {\braket{P}{II}}^{*}  }
},
}
\eqlabel\Mfour
$$
where we have relabelled $\{n,n{-}1,\ldots,3\}
\longrightarrow \{3,4,\ldots,n\}$ in those terms in which
such relabelling is trivial.

We now collect terms from (\Mone), (\Mtwo), (\Mthree), and (\Mfour).
There are several terms which cancel immediately.  There are also
two pairs of terms which combine using the complex-conjugated form of
(\linkidnosum), eliminating a factor of $h$ from the numerator and
denominator in the process. The result of all of this is
$$
\eqalign{
\amp&(P^0,Q^0;I^{-},II^{-},3^{+},\ldots,n^{+}) =
\cr& =
-
{ i(-e\sqrt2)^{n} }
\permsum{3}{n}  \sum_{j=3}^n
{
{ \braket{P}{I} \braket{II}{Q} }
\over
{ \bra{P} 3,\ldots,n \ket{Q} }
}
\cr & \qquad\qquad\qquad\qquad\qquad\times
u^{\gamma}(k_{II})
{{\pole}_{\gamma}}^{\delta}(P,I,3,\ldots,j)
u_{\delta}(k_I)
\cr& \quad
+
{ i(-e\sqrt2)^{n} }
\permsum{3}{n}  \sum_{j=3}^n
{
{ \braket{P}{II} \braket{I}{Q} }
\over
{ \bra{P} n,n{-}1,\ldots,3\ket{Q} }
}
\cr & \qquad\qquad\qquad\qquad\qquad\times
u^{\gamma}(k_{II})
{{\pole}_{\gamma}}^{\delta}(Q,I,3,\ldots,j)
u_{\delta}(k_I)
\cr&\quad
+
{ i(-e\sqrt2)^{n} }
\permsum{3}{n}
{
{ \bra{P}II\ket{Q} }
\over
{ \bra{P} 3,\ldots,n\ket{Q} }
}
{
{  {\braket{P}{Q}}^{*}  }
\over
{  {\bra{P}I\ket{Q}}^{*}  }
}
\cr&\quad
+
{ i(-e\sqrt2)^{n} }
\permsum{3}{n}
{
{ \bra{P}I\ket{Q} }
\over
{ \bra{P}3,\ldots,n\ket{Q} }
}
{
{  {\braket{P}{Q}}^{*}  }
\over
{  {\bra{P}II\ket{Q}}^{*}  }
}
\cr&\quad
+
{ i(-e\sqrt2)^{n} }
\permsum{3}{n}
{
{ \braket{P}{I} \braket{I}{II} }
\over
{ \bra{P} 3,\ldots,n\ket{Q} }
}
{
{  1  }
\over
{  {\braket{II}{Q}}^{*}  }
}
\cr&\quad
-
{ i(-e\sqrt2)^{n} }
\permsum{3}{n}
{
{ \braket{I}{II} \braket{I}{Q} }
\over
{ \bra{P}3,\ldots,n\ket{Q} }
}
{
{  1  }
\over
{  {\braket{P}{II}}^{*}  }
}.
}
\eqlabel\notquitedone
$$

We should like to cast (\notquitedone) into a form which is explicitly
symmetric under the interchange $I\leftrightarrow II$.
Let us call the second term of this expression $\amp_2^{\prime}$.
This is the only term in (\notquitedone) in which the ordering
of momenta in the denominator is from $n$ to 3, rather than from
3 to $n$.
If we write
$$
\{n,n{-}1,\ldots,j{+}1,j,\ldots,3\}
\longrightarrow
\{3,4,\ldots,n{-}j{+}2,n{-}j{+}3,\ldots,n\}
\eq
$$
in order to make the denominator in $\amp_2^{\prime}$ look like the other
denominators, we obtain
$$
\eqalign{
\amp_2^{\prime}=
{ i(-e\sqrt2)^{n} }
\permsum{3}{n} & \sum_{j=3}^n
{
{ \braket{P}{II} \braket{I}{Q} }
\over
{ \bra{P} 3,\ldots,n\ket{Q} }
}
\cr & \times
u^{\gamma}(k_{II})
{{\pole}_{\gamma}}^{\delta}(Q,I,n,n{-}1,\ldots,n{-}j{+}3)
u_{\delta}(k_I).
}
\eqlabel\fixit
$$
The definition  (\poledef) of $\pole$ in conjunction with
momentum conservation tells us that
$$
\eqalign{
{{\pole}_{\gamma}}^{\delta}&(Q,I,n,n{-}1,\ldots,n{-}j{+}3) =
\cr & =
{
{ (k_{n-j+3})_{\gamma\dot\gamma}
  [\bar Q + \bar k_I + \bar\kappa(n{-}j{+}3,n)]^{\dot\gamma\delta} }
\over
{ [Q+k_I+\kappa(n{-}j{+}4,n)]^2
  [Q+k_I+\kappa(n{-}j{+}3,n)]^2 }
}
\cr&
= -{
{ (k_{n-j+3})_{\gamma\dot\gamma}
  [\bar P + \bar k_{II} + \bar\kappa(3,n{-}j{+}2)]^{\dot\gamma\delta} }
\over
{ [P+k_{II}+\kappa(3,n{-}j{+}3)]^2
  [P+k_{II}+\kappa(3,n{-}j{+}2)]^2 }
}
\cr&
= -{{\pole}_{\gamma}}^{\delta}(P,II,3,\ldots,n{-}j{+}3).
}
\eq
$$
So, letting $j^\prime=n-j+3$, (\fixit) becomes
$$
\eqalign{
\amp_2^{\prime}=
{- i(-e\sqrt2)^{n} }
\permsum{3}{n} & \sum_{j'=3}^n
{
{ \braket{P}{II} \braket{I}{Q} }
\over
{ \bra{P} 3,\ldots,n\ket{Q} }
}
\cr & \times
u^{\gamma}(k_{II})
{{\pole}_{\gamma}}^{\delta}(P,II,3,\ldots,j^\prime)
u_{\delta}(k_I).
}
\eqlabel\xxx
$$
In order to make $\amp_2^{\prime}$ look like the first term
of (\notquitedone)
with $I \leftrightarrow II$, we must use (\reverseid) to exchange
$u^{\gamma}(k_{II}){{\pole}_{\gamma}}^{\delta}u_{\delta}(k_I)$
for $u^{\gamma}(k_I){{\pole}_{\gamma}}^{\delta}u_{\delta}(k_{II})$.
Note that since the difference of two terms resulting from the
transformation may be summed, we are left with
$$
\eqalign{
\amp_2^{\prime}&=
- i(-e\sqrt2)^{n}
\permsum{3}{n}  \sum_{j=3}^n
{
{ \braket{P}{II} \braket{I}{Q} }
\over
{ \bra{P} 3,\ldots,n\ket{Q} }
}
\cr & \qquad\qquad\qquad\qquad\qquad\times
u^{\gamma}(k_{II})
{{\pole}_{\gamma}}^{\delta}(P,II,3,\ldots,j)
u_{\delta}(k_I)
\cr & \quad
- i(-e\sqrt2)^{n}
\permsum{3}{n}
{
{ \braket{P}{II} \braket{I}{Q} \braket{II}{I} }
\over
{ \bra{P}3,\ldots,n\ket{Q} }
}
\cr & \qquad\qquad\qquad\qquad\qquad\times
\Biggl\{
{ 1\over{[P+k_{II}]^2} }
-
{ 1\over{[P+k_{II}+\kappa(3,n)]^2} }
\Biggr\}
\cr &
= - i(-e\sqrt2)^{n}
\permsum{3}{n}  \sum_{j=3}^n
{
{ \braket{P}{II} \braket{I}{Q} }
\over
{ \bra{P} 3,\ldots,n\ket{Q} }
}
\cr & \qquad\qquad\qquad\qquad\qquad\times
u^{\gamma}(k_{II})
{{\pole}_{\gamma}}^{\delta}(P,II,3,\ldots,j)
u_{\delta}(k_I)
\cr & \quad
+ i(-e\sqrt2)^{n}
\permsum{3}{n}
{
{  \braket{I}{Q} \braket{I}{II} }
\over
{ \bra{P}3,\ldots,n\ket{Q} }
}
{ 1\over{{\braket{P}{II}}^{*}} }
\cr & \quad
- i(-e\sqrt2)^{n}
\permsum{3}{n}
{
{  \braket{P}{II} \braket{I}{II} }
\over
{ \bra{P}3,\ldots,n\ket{Q} }
}
{ 1\over{{\braket{I}{Q}}^{*}} }.
}
\eqlabel\finally
$$
The first term of (\finally) is the $I \leftrightarrow II$ partner
for the first term of (\notquitedone).  The second term of
(\finally) cancels the last term of (\notquitedone).  The final
term of (\finally) is the $I \leftrightarrow II$ partner for the
fifth term of (\notquitedone), the relative minus sign reflecting
the antisymmetry of the spinor inner product, $\braket{I}{II}=
-\braket{II}{I}$.
Note that the third and fourth terms in (\notquitedone) already
form a $I \leftrightarrow II$ symmetric pair.  Thus, we have
$$
\eqalign{
\amp&(P^0,Q^0;I^{-},II^{-},3^{+},\ldots,n^{+}) =
\cr& =
-
{ i(-e\sqrt2)^{n} }
\sum_{{\cal{P}}(I\thinspace\thinspace II)} \permsum{3}{n}
{
{ \braket{P}{I} \braket{II}{Q} }
\over
{ \bra{P} 3,\ldots,n \ket{Q} }
}
\cr & \qquad\qquad\qquad\qquad\qquad\times
\sum_{j=3}^n
u^{\gamma}(k_{II})
{{\pole}_{\gamma}}^{\delta}(P,I,3,\ldots,j)
u_{\delta}(k_I)
\cr& \quad
+
{ i(-e\sqrt2)^{n} }
\sum_{{\cal{P}}(I\thinspace\thinspace II)} \permsum{3}{n}
{
{ \bra{P}II\ket{Q} }
\over
{ \bra{P} 3,\ldots,n\ket{Q} }
}
{
{  {\braket{P}{Q}}^{*}  }
\over
{  {\bra{P}I\ket{Q}}^{*}  }
}
\cr& \quad
-
{ i(-e\sqrt2)^{n} }
\sum_{{\cal{P}}(I\thinspace\thinspace II)} \permsum{3}{n}
{
{ \braket{P}{II} \braket{I}{II} }
\over
{ \bra{P} 3,\ldots,n\ket{Q} }
}
{
{  1  }
\over
{  {\braket{I}{Q}}^{*}  }
},
}
\eqlabel\nowitsdoneNOT
$$
where we have collected the terms differing by $I \leftrightarrow II$
into a sum over ${{\cal{P}}(I\thinspace\thinspace II)}$.  Finally,
a bit of algebra to combine the last two terms of (\nowitsdoneNOT),
plus judicious use of momentum conservation produces
$$
\eqalign{
\amp&(P^0,Q^0;I^{-},II^{-},3^{+},\ldots,n^{+}) =
\cr& =
-
{ i(-e\sqrt2)^{n} }
\sum_{{\cal{P}}(I\thinspace\thinspace II)} \permsum{3}{n}
{
{ \braket{P}{I}  }
\over
{ \bra{P} 3,\ldots,n \ket{Q} }
}
\Biggl[
{
{ \bar u_{\dot\alpha}(P)
\bar \kappa^{\dot\alpha\alpha}(3,n)
u_{\alpha}(k_{I})}
\over
{ {\bra{P}II\ket{Q}}^{*} }
}
\cr & \qquad\qquad\qquad\qquad +
\braket{II}{Q}
\sum_{j=3}^n
u^{\alpha}(k_{II})
{{\pole}_{\alpha}}^{\beta}(P,I,3,\ldots,j)
u_{\beta}(k_I)
\Biggr].
}
\eqlabel\scalartwominusamp
$$


\subsection{Other amplitudes involving two currents of the same spin}

There are three other  amplitudes which may be constructed by
joining two charged lines with the same spin that could not
be obtained from a single current.  First,
we may consider combining a pair of transverse $W$ currents to form
$\amp(P^{+},Q^{+};1^{-},2^{-},3^{+},\ldots,n^{+})$.
The gauge choice to be made is the
same as in the longitudinal $W$ case, namely (\sewscalargauge).
With that choice, recall that all dot products between polarization
vectors vanish, except for
$$
{\eps}_{\alpha\dot\alpha}(II^{-})
{\bar\eps}^{\dot\alpha\alpha}(j^{+}) =
{
{ \braket{I}{II} {\braket{j}{h}}^{*} }
\over
{ \braket{j}{I} {\braket{II}{h}}^{*} }
}.
\eqlabel\novanish
$$
It was this non-zero product which caused seagull terms to appear
in that computation.  In the vector case, however, the seagull
vertex takes the form of the curly bracket function (\curlybrak).
As a result, the potential seagull contributions contain
quantities like
$\bigl\{\Wnorm(P^{+};I^{-},2^{+},\ldots,(j{-}1)^{+}),
\eps(j^{+}),\Wnorm((j{+}1)^{+},\ldots,n^{+})\bigr\}
\cdot \eps(II^{-})$,
which vanish because of the spinor structure of the currents
involved ({\it cf.} (\Xansatz) and (\Xnegansatz)).
Thus, this computation is considerably easier than the one
described in the previous section.
The result is
$$
\amp(P^{+},Q^{+};1^{-},2^{-},3^{+},\ldots,n^{+}) =
-i(-e\sqrt2)^n
{
{ {\braket{P}{Q}}^{n-2} }
\over
{ \prod\limits_{j=1}^n \bra{P}j\ket{Q} }
}
{\braket{1}{2}}^4
,
\eqlabel\WTtwonegphotons
$$
which fits the form of (\onecurrentamps).
In fact, the amplitudes for the scattering of two
transverse $W$'s and $n$ photons where any two particles have negative
helicity and $n-2$ particles have positive helicity may be summarized
by
$$
\amp(P,Q;1,2,\ldots,n)=
-i(-e\sqrt2)^n
{
{ {\braket{P}{Q}}^{n-2} }
\over
{ \prod\limits_{j=1}^n \bra{P}j\ket{Q} }
}
{\braket{N_1}{N_2}}^4
,
\eq
$$
where $N_1$ and $N_2$ are the momenta of the two negative helicity
bosons.
This is not surprising, since
it is possible to use the $U(N)$ formalism \cite\DY\ to obtain
this amplitude.

One may also join two fermion currents.  Note that  because
of helicity conservation, a
positive helicity positron current must be combined with a
negative helicity electron current and conversely.
The two new amplitudes we may obtain this way are
$\amp(p^{-},q^{+};I^{-},II^{-},3^{+},\ldots,n^{+})$ and
$\amp(p^{+},q^{-};I^{-},II^{-},3^{+},\ldots,n^{+})$.
Although there are no seagulls in these amplitudes, there is
one minor complication.  Consider what the expression
(\LHoneminussoln) implies for
${\positronon}(p^{-};I^{-},3^{+},\ldots,m^{+})$:
$$
\eqalign{
{\positronon}^{\alpha}&(p^{-};I^{-},3^{+},\ldots,m^{+}) =
\cr &\negthinspace\negthinspace\negthinspace\negthinspace\negthinspace =
(-e\sqrt2)^{m-1}\negthinspace\negthinspace\negthinspace\negthinspace
\permsum{3}{m}
\negthinspace
{
{-\braket{p}{I}}
\over
{\bra{p} 3,\ldots,m \ket{I}}
}
\Biggl\{
{
{ {\braket{h}{p}}^{*} }
\over
{ { \bra{h}I\ket{p} }^{*} }
}u^{\alpha}(p)
-{
{\braket{p}{I} u^{\alpha}(k_I)}
\over
{[p{+}k_I{+}\kappa(3,m)]^2}
}
\cr& \qquad
- (1-{\delta}_{m2}) \braket{p}{I}
\sum_{j=3}^m
{
{ u^{\beta}(k_I)
[p+k_I+\kappa(3,j)]_{\beta\dot\alpha}
\bar k_j^{\dot\alpha\alpha} }
\over
{ [p+k_I+\kappa(3,j{-}1)]^2 [p+k_I+\kappa(3,j)]^2 }
}
\Biggr\}.
}
\eqlabel\snagdemo
$$
When we join a pair of fermion currents to form an amplitude,
we obtain  quantities like
${\positronon}^{\alpha}(p^{-};I^{-},3^{+},\ldots,m^{+})
\eps_{\alpha\dot\alpha}(II^{-})
{\electronon}^{\dot\alpha}\bigl((m{+}1)^{+},\ldots,n^{+}\bigr)$.
Consequently,
the following  two features of (\snagdemo) are relevant.
First, there is an ``extra'' term proportional to
$u^{\alpha}(k_I)$.   It does not vanish when it is contracted into
$u_{\alpha}(k_{II})$.
The problem with this term is the
factor $[P+k_I+\kappa(3,m)]^{-2}$. It  will prevent us from
summing over $m$, as in (\sumMone).  Second, note that term
involving
the sum over poles will contain
$u^{\alpha}(k_{II}){{\pole}_{\alpha}}^{\beta}(P,I,3,\ldots,j)
u_{\beta}(k_I)$.  This is bad because it has an implicit
factor $\braket{II}{j}$ rather than $\braket{I}{j}$, blocking
a necessary cancellation which occurred in (\Mthreea).
These two problems solve each other.  Using (\reverseid)
to exchange
$u^{\alpha}(k_{II})
{{\pole}_{\alpha}}^{\beta}(P,I,3,\ldots,j)
u_{\beta}(k_I)$
for
$u^{\alpha}(k_{I})
{{\pole}_{\alpha}}^{\beta}(P,I,3,\ldots,j)
u_{\beta}(k_{II})$
not only gives us a sum over poles which contains $\braket{I}{j}$,
but it produces a term which exactly cancels the contribution
involving $[P+k_I+\kappa(3,m)]^{-2}$.  There is one other term
produced, but it may be handled by the techniques similar to those
used in (\sumMone).
Other than the differences just outlined, the computations go
through exactly as for the scalar case, with the results
$$
\eqalign{
\amp&(p^{-},q^{+};I^{-},II^{-},3^{+},\ldots,n^{+}) =
\cr& =
{ i(-e\sqrt2)^{n} }
\sum_{{\cal{P}}(I\thinspace\thinspace II)} \permsum{3}{n}
{
{ {\braket{p}{II}}  }
\over
{ \bra{p} 3,\ldots,n \ket{q} }
}
\Biggl[
{
{ \bar u_{\dot\alpha}(q)
\bar \kappa^{\dot\alpha\alpha}(3,n)
u_{\alpha}(k_{II})}
\over
{ {\bra{p}I\ket{q}}^{*} }
}
\cr & \qquad\qquad\qquad\qquad
+
{\braket{p}{II}}
\sum_{j=3}^n
u^{\alpha}(k_I)
{{\pole}_{\alpha}}^{\beta}(p,II,3,\ldots,j)
u_{\beta}(k_I)
\Biggr]
}
\eqlabel\LRspinortwonegphotons
$$
and
$$
\eqalign{
\amp&(p^{+},q^{-};I^{-},II^{-},3^{+},\ldots,n^{+}) =
\cr& =
{ i(-e\sqrt2)^{n} }
\negthinspace\negthinspace
\sum_{{\cal{P}}(I\thinspace\thinspace II)}
\permsum{3}{n}
\negthinspace
{
{ - {\braket{II}{q}} }
\over
{ \bra{p} n,n{-}1,\ldots,3 \ket{q} }
}
\Biggl[
{
{ \bar u_{\dot\alpha}(p)
\bar \kappa^{\dot\alpha\alpha}(3,n)
u_{\alpha}(k_{II})}
\over
{ {\bra{p}I\ket{q}}^{*} }
}
\cr& \qquad\qquad\qquad\qquad
-{\braket{II}{q}}
\sum_{j=3}^n
u^{\alpha}(k_I)
{{\pole}_{\alpha}}^{\beta}(q,II,3,\ldots,j)
u_{\beta}(k_I)
\Biggr].
}
\eqlabel\RLspinortwonegphotons
$$
We see the effects of the supersymmetric-like relations in
that (\LRspinortwonegphotons) and
(\RLspinortwonegphotons) have the same over-all structure
as (\scalartwominusamp), the corresponding amplitude involving scalars.
However, these quantities are no longer related by a simple
proportionality.


\subsection{Amplitudes involving two currents of differing spins}

\def\mixspin{4.2.3}
Of the three possible ways to pair two currents of differing spins,
only two are interesting in the massless limit.  Conservation of
angular momentum demands that all fermion lines appearing in a graph
must be unbroken.  The restoration of the unbroken $SU(2) \times U(1)$
symmetry at high energies manifests itself as conservation of weak
hypercharge, requiring  that all scalar lines remain unbroken,
preserving the total number of scalar particles.
Thus, the combinations we will consider here are the joining of
scalar with vector, and spinor with vector.
The former  describes processes like
$W_L^{+}  W_T^{-} \longrightarrow H \thinspace \gamma\cdots\gamma$,
$W_L^{+}  W_T^{-} \longrightarrow Z_L \thinspace \gamma\cdots\gamma$,
plus crossed reactions, while the latter encompasses
$e \bar\nu \longrightarrow W_T^{-}\thinspace  \gamma\cdots\gamma$
and related processes.
The known currents allow us to compute these amplitudes when all of
the vector particles have the same helicity or there is only one vector
with a differing helicity.  Because the off-shell particle is a
scalar or a fermion in this type of amplitude pairing, it is not
possible to obtain amplitudes with two differing vector
boson helicities.

The computation of these amplitudes goes through in much the same
fashion as the scalar case previously discussed, so no further
details are necessary.  All of the results may be cast into a common
form if we give positrons and $W_L^{+}$'s momentum $P$, electrons
and $W_L^{-}$'s momentum $Q$, and neutral particles momenta $k_j$.
The photons will have momenta $k_1$ through $k_n$, while the momentum
of the remaining neutral (neutrino, Higgs, or $Z_L$) will be
$k_0\equiv-[P+\kappa(1,n)+Q]$.

Amplitudes in which all of the vector bosons have the same helicity vanish.
The amplitudes with one opposite helicity vector boson have the form
$$
\amp(P,Q,0;1,2^{+},\ldots,n^{+}) =
{{ig}\over{\sqrt2}}
(-e\sqrt2)^n
{
{ {\braket{P}{Q}}^{n-1} }
\over
{ \prod\limits_{j=0}^n \bra{P}j\ket{Q} }
}
f(P,Q,0,1),
\eqlabel\mixedtemplate
$$
where $g$ is the $SU(2)$ coupling constant and $f(P,Q,0,1)$ is
a function which depends on the helicity combination.
Recall that for amplitudes involving neutrinos, certain helicity
combinations do not occur in the standard model.
Table~\Tlabel\mixedamptable { }summarizes the results.  Again,
note that the charges and helicities are based on inwardly-directed
momenta.

In order to obtain amplitudes with a $Z_L$ instead of a
Higgs boson, multiply by $\mp  i$ when a $W_L^{\pm}$
is incoming.

\subsection{Processes involving $Z$'s instead of photons}

Substitution of transversely polarized $Z$'s for photons
is simple in processes for which the diagram consists of a single
charged line with uniform spin along its entire length.
In this case, the Feynman rules are such that the only change
is in the coupling constant:  the $Z$'s do not have cubic or
quartic vertices among themselves.  Everything else about the
amplitude is unchanged.  The fact that $Z$'s are not the same
as photons does not show up until one considers the integrated
cross section, at which point the factors inserted to eliminate
the overcounting of identical particles will be different.
As a result, Table\Tlabel\photontoZ\ may be used to write
down amplitudes involving any combination of photons and
$Z$'s given the corresponding amplitude containing only
photons.
\global\advance\tableno by -1
\noadvancetable{
{
\begingroup
{\begintable
$P$\vb$Q$\vb$k_0$\vb$k_1$\|$f(P,Q,0,1)$ \crthick
$   W^{+}_L  \thinspace  $\vb$   W^{-}_{\up} \thinspace $\vb
$  H\thinspace$\vb$\gamma_{\down}\thinspace$\|
${\braket{P}{1}}^2 {\braket{0}{1}}^2$ \cr
$   W^{+}_L\ts$\vb$  W^{-}_{\down}\ts$\vb$ H\thinspace $\vb
$ \gamma_{\up}\ts $\|
${\braket{P}{Q}}^2 {\braket{0}{Q}}^2$ \cr
$ W^{+}_{\up}\ts$\vb$  W^{-}_L  \ts  $\vb$ H\thinspace $\vb
$\gamma_{\down}\ts$\|
${\braket{1}{0}}^2 {\braket{1}{Q}}^2$ \cr
$W^{+}_{\down}\ts$\vb$  W^{-}_L \ts  $\vb$ H\thinspace $\vb
$ \gamma_{\up}\ts $\|
${\braket{P}{0}}^2 {\braket{P}{Q}}^2$ \cr
$\bar e_{\up} $\vb$   W^{-}_{\up}\ts$\vb$ \nu_{\down} $\vb
$\gamma_{\down}\ts$\|
$ \sqrt2 {\braket{P}{1}}   {\braket{0}{1}}^3$ \cr
$\bar e_{\up} $\vb$  W^{-}_{\down}\ts$\vb$ \nu_{\down} $\vb
$ \gamma_{\up}\ts $\|
$ \sqrt2 {\braket{P}{Q}}   {\braket{0}{Q}}^3$ \cr
$ W^{+}_{\up}\ts$\vb$   e_{\down}     $\vb$\bar\nu_{\up}$\vb
$\gamma_{\down}\ts$\|
$ \sqrt2 {\braket{1}{0}}   {\braket{1}{Q}}^3$ \cr
$W^{+}_{\down}\ts$\vb$   e_{\down}     $\vb$\bar\nu_{\up}$\vb
$ \gamma_{\up}\ts $\|
$  \sqrt2 {\braket{P}{0}}   {\braket{P}{Q}}^3$
\endtable}
\endgroup}
}{Helicity functions for processes
involving a Higgs or neutrino}
\global\advance\tableno by 1
\noadvancetable{
\begintable
$\amp(W_L^{+} W_L^{-} \rightarrow mZ_T \thinspace n\gamma ) =
(\cot 2\theta_W)^m \amp(W_L^{+} W_L^{-} \rightarrow (m+n)\gamma)$\cr
$\amp(e_L \bar e_R \rightarrow mZ_T \thinspace n\gamma ) =
(\cot 2\theta_W)^m \amp(e_L \bar e_R \rightarrow (m+n)\gamma
)$\cr
$\amp(e_R \bar e_L \rightarrow mZ_T \thinspace n\gamma ) =
(-\tan\theta_W)^m \amp(e_R \bar e_L \rightarrow (m+n)\gamma
)$\cr
$\amp(W_T^{+} W_T^{-} \rightarrow mZ_T \thinspace n\gamma ) =
(\cot \theta_W)^m \amp(W_T^{+} W_T^{-} \rightarrow (m+n)\gamma)$
\endtable
}{Relation between amplitudes
containing $Z$'s instead of photons}

In contrast, it is much more difficult to substitute $Z$'s for
photons when the charged line has mixed spin, because the shift
in coupling constant depends upon the spin of the line on which
the substitution takes place.  Furthermore, $Z$'s couple to
the Higgs scalar and the neutrinos, requiring that additional
currents be derived.  While it is easy to write down expressions
for these new currents,
it is not easy to combine them with the existing currents
in order to produce explicitly gauge-invariant amplitudes.
To illustrate this point, suppose we wish to compute the
amplitude for the process
$$
\bar e \thinspace \nu \thinspace W^{-} \thinspace
\longrightarrow Z_T Z_T \cdots Z_T.
\eq
$$
The current for $n$ $Z$'s attached to a transverse $W$ line is the
same as that with $n$ photons with $(-e\sqrt2)^n$ replaced
by $(-g\sqrt2)^n$, and that for $n$ $Z's$ attached to
the neutrino line is the same as that for $n$ photons attached
to an electron line with $({g\over{\sqrt2}} \sec \theta_W)^n$
replacing $(-e\sqrt2)^n$.  The coupling constant for the
positron current  becomes $(-g\sqrt2 + {g\over{\sqrt2}}
\sec \theta_W)^n$.
To make  gauge invariance explicit, one must
use the binomial expansion on this coupling constant.  At
this point, however, the sums linking the various currents
together become very difficult  to perform.

We have, however, been able to obtain amplitudes corresponding
to the processes in section \mixspin\ in which a single negative
helicity photon has been replaced by a negative helicity $Z$.
The results  fit in the form
$$
\amp(P,Q,0;1^{-},2^{+},\ldots,n^{+}) =
{{ig}\over{\sqrt2}}
(-e\sqrt2)^n
{
{ {\braket{P}{Q}}^{n-2} }
\over
{ \prod\limits_{j=2}^n \bra{P}j\ket{Q} }
}
f(P,Q,0,1),
\eqlabel\mixedtemplate
$$
with the helicity functions given in Table \Tlabel\mixedampZtable.
Note that $k_1$ is the momentum of the $Z$, rather than a photon
momentum.  Also, recall that the charge and helicity labels on the
particles correspond to inwardly-flowing momenta.

\noadvancetable{
{\begintable
$P$\vb$Q$\vb$k_0$\vb$k_1$\|
$f(P,Q,0,1)$ \crthick
$   W^{+}_L  \thinspace  $\vb$   W^{-}_{\up} \thinspace  $\vb
$  H\thinspace$\vb$Z_{\down}\thinspace$\|
$\cot \theta_W
{
{ {\braket{P}{1}}^2 \braket{0}{1} }
\over
{ {\braket{P}{0}} \braket{1}{Q} }
}
- \cot 2\theta_W
{
{ \braket{0}{1} \braket{P}{1} }
\over
{ \braket{0}{Q} }
}
$\cr
$ W^{+}_{\up}\ts$\vb$  W^{-}_L  \ts  $\vb$ H\thinspace $\vb
$Z_{\down}\ts$\|
$-\cot \theta_W
{
{ {\braket{1}{Q}}^2 \braket{0}{1} }
\over
{ \braket{P}{1} \braket{0}{Q} }
}
+ \cot 2\theta_W
{
{ \braket{0}{1} \braket{1}{Q} }
\over
{ \braket{P}{0} }
}
$\cr
$\bar e_{\up} $\vb$   W^{-}_{\up}\ts$\vb$ \nu_{\down} $\vb
$Z_{\down}\ts$\|
$\sqrt2 \biggl[
\cot \theta_W
{
{ \braket{P}{1} {\braket{1}{0}}^2  }
\over
{ \braket{P}{0} \braket{1}{Q} }
}
- \cot 2\theta_W
{
{ {\braket{1}{0}}^2  }
\over
{ \braket{0}{Q} }
}
\biggr]
$\cr
$ W^{+}_{\up}\ts$\vb$   e_{\down}     $\vb$\bar\nu_{\up}$\vb
$Z_{\down}\ts$\|
$\sqrt2 \biggl[
\cot \theta_W
{
{  {\braket{1}{0}}^3  }
\over
{ \braket{P}{1} \braket{0}{Q} }
}
- \cot 2\theta_W
{
{ {\braket{1}{Q}}^2  }
\over
{ \braket{P}{0} }
}
\biggr]$
\endtable
}}{Helicity functions for processes
involving a Higgs or neutrino plus a $Z$}

\chapter{CONCLUSION}

We have taken a ``direct'' approach to the application of multispinor
formalism and the equivalence theorem to the electroweak sector
of the standard model.  By considering the electrodynamics of spin $0$,
spin $1\over2$, and spin 1 particles, we have obtained amplitudes
involving the production of an arbitrary number of photons for
certain helicity configurations, namely those with up to two unlike
vector boson helicities.  An important calculational technique has been
to write the various expressions involved in the form of a permutation
sum, in order to effectively exploit the Bose symmetry of the photons.
We have seen that in some cases, the replacement of photons
by  $Z$'s  is trivial, but that  in other cases, additional $Z$'s
require additional work.  Finally, we have considered
processes including
a single Higgs, longitudinally polarized $Z$, or neutrino.

Before experimentally measurable quantities can be extracted from
these amplitudes, there are a few issues which must still  be
addressed.
Differential cross sections involve the square of the amplitude.
In the cases where we have been able to
actually perform the permutation sum, squaring the amplitude
is trivial.
Amplitudes which still contain the permutation sum could
still be squared numerically, but the feasibility of
such a procedure needs to be investigated.
A second question involves the finite masses of the particles.
There are special regions of phase space where
some of the invariants formed from pairs of momenta are
of  the same order as the neglected masses.  The corrections
to the amplitudes presented here are potentially large
in such regions of phase space.
A related question focuses on the infrared divergences
present in the amplitudes.  In principle, the satisfactory
treatment of these involves a knowledge of loop diagrams.
Finally,  current experimental capabilities preclude the measurement
of helicity-projected amplitudes.  Thus, it would be desirable
to obtain a complete set of helicity amplitudes, in order to
be able to sum over helicities.
Hence, even though we have obtained helicity  amplitudes for
a wide variety of processes, there are clearly
many other important questions to be addressed
by further research into this subject.

\bigskip\bigskip
\noindent
{\chapterfont Acknowledgements:}

\medskip
\noindent
This work was supported in part by the National
Science Foundation.

\appendix{MULTISPINOR CONVENTIONS}

Below we list the important results of application of
Weyl-van der Waerden spinor calculus to gauge theories.
Readers interested
in the details should refer to references \cite\DY\ and
[\ref{For a brief introduction to properties of two-component
Weyl-van der Waerden spinors, see, for example, M. F. Sohnius,
Phys. Reports {\bf 128}, 39 (1985).}].

We use the Weyl basis
$$
\gamma^\mu =
\pmatrix{  0      & \sigma^\mu \cr
         \bar\sigma^\mu &      0   \cr},
\eqlabel\weylbasis
$$
for the Dirac matrices.  In (\weylbasis), $\sigma^\mu$
and $\bar\sigma^\mu$ refer to the convenient Lorentz-covariant
grouping of the $2\times2$ Pauli matrices plus the unit matrix:
$$
\sigma^{\mu} \equiv (1, \vec\sigma),
\newlettlabel\sigmamus
$$
$$
\bar\sigma^{\mu} \equiv (1, -\vec\sigma),
\lett
$$
and satisfy the anticommutators
$$
(\bar\sigma^{\mu})^{\dot\alpha\beta}
(\sigma^{\nu})_{\beta\dot\beta}
+
(\bar\sigma^{\nu})^{\dot\alpha\beta}
(\sigma^{\mu})_{\beta\dot\beta}
=
2 g^{\mu\nu} \delta_{\dot\beta}^{\dot\alpha},
\newlettlabel\sigmaanticommutator
$$
$$
(\sigma^{\mu})_{\alpha\dot\beta}
(\bar\sigma^{\nu})^{\dot\beta\beta}
+
(\sigma^{\nu})_{\alpha\dot\beta}
(\bar\sigma^{\mu})^{\dot\beta\beta}
=
2 g^{\mu\nu} \delta_{\alpha}^{\beta}.
\lett
$$

To each Lorentz 4-vector there corresponds a rank two multispinor,
formed from the contraction of the 4-vector with $\sigma^{\mu}$:
$$
\Wms_{\alpha\dot\beta} =
{1\over{\sqrt2}}
\sigma^{\mu}_{\alpha\dot\beta} W_{\mu},
\newlettlabel\msvector
$$
$$
{\overline\Wms}^{\dot\alpha\beta} =
{1\over{\sqrt2}}
\bar\sigma_{\mu}^{\dot\alpha\beta} W^{\mu}.
\lett
$$
For the purposes of normalization, it is convenient to use a
different convention when converting momenta:
$$
k_{\alpha\dot\beta} =
\sigma^{\mu}_{\alpha\dot\beta} k_{\mu},
\newlettlabel\msmomenta
$$
$$
{\bar k}^{\dot\alpha\beta} =
\bar\sigma_{\mu}^{\dot\alpha\beta} k^{\mu}.
\lett
$$
Useful consequences of (\msmomenta) and (\sigmaanticommutator) are
$$
\bar k^{\dot\alpha\beta} k_{\beta\dot\beta}
= k^2 \delta_{\dot\beta}^{\dot\alpha},
\newlettlabel\slashsquaredidentity
$$
$$
k_{\alpha\dot\beta} \bar k^{\dot\beta\beta}
= k^2 \delta_{\alpha}^{\beta}.
\lett
$$

The spinor indices may be raised and lowered using the
2-component antisymmetric tensor:
$$
u^{\alpha} = \vareps^{\alpha\beta}u_{\beta},
\newlettlabel\raiseandlower
$$
$$
\bar v^{\dot\alpha} = \vareps^{\dot\alpha\dot\beta}\bar v_{\dot\beta},
\lett
$$
$$
\vareps^{\alpha\beta}=\vareps_{\alpha\beta},
\lett
$$
$$
\vareps^{\dot\alpha\dot\beta}=\vareps_{\dot\alpha\dot\beta},
\lett
$$
$$
\vareps_{12}=\vareps_{\dot1\dot2}=1.
\lett
$$
Many useful relations may be easily proven from the Schouten identity
$$
\delta_{\gamma}^{\alpha} \delta_{\delta}^{\beta}
-\delta_{\delta}^{\alpha} \delta_{\gamma}^{\beta}
+\vareps^{\alpha\beta} \vareps_{\gamma\delta}=0,
\eqlabel\SCHOUTEN
$$
the generator of 2-component Fierz transformations.

We denote by $u(k)$ and $\bar u(k)$ the solutions to the
2-component Weyl equations:
$$
\bar k^{\dot\alpha\beta} u_{\beta}(k) = 0,
\newlettlabel\Weylequations
$$
$$
\bar u_{\dot\beta}(k) \bar k^{\dot\beta\alpha} = 0.
\lett
$$
These two spinors are related by complex conjugation
$$
\bar u_{\dot\alpha}(k) = \bigl[u_{\alpha}(k)\bigr]^{*},
\eqlabel\complexconj
$$
and have the normalization
$$
u_{\alpha}(k) \bar u_{\dot\alpha}(k) = k_{\alpha\dot\alpha}.
\eqlabel\spinornormalization
$$
It is useful to define a scalar product
$$
\braket{1}{2} \equiv u^{\alpha}(k_1) u_{\alpha}(k_2),
\eqlabel\scalarproduct
$$
which has two elementary properties
$$
\braket{1}{2} = - \braket{2}{1},
\newlettlabel\antisymmetry
$$
$$
\braket{1}{2} {\braket{1}{2}}^{*} =2k_1\cdot k_2.
\lett
$$
Contraction of $u_{\alpha}(k_1)u_{\beta}(k_2)u^{\gamma}(k_3)
u^{\delta}(k_4)$ into (\SCHOUTEN) produces the extremely
useful relation
$$
\braket{1}{2}\braket{3}{4}
+\braket{1}{3}\braket{4}{2}
+\braket{1}{4}\braket{2}{3}
=0.
\eqlabel\veryhelpful
$$
A second relation of great utility may be derived from (\veryhelpful):
$$
{
{ \braket{1}{2} }
\over
{ \braket{1}{P} \braket{P}{2} }
}
+
{
{ \braket{2}{3} }
\over
{ \braket{2}{P} \braket{P}{3} }
}
=
{
{ \braket{1}{3} }
\over
{ \braket{1}{P} \braket{P}{3} }
}.
\eqlabel\LINK
$$
Equation (\LINK) may be used to demonstrate that
$$
\sum_{j=\ell}^{m-1}
{
{ \braket{j}{j{+}1} }
\over
{ \braket{j}{P} \braket{P}{j{+}1} }
}
=
{
{ \braket{\ell}{m} }
\over
{ \braket{\ell}{P} \braket{P}{m} }
}.
\eqlabel\SUMLINK
$$

Helicities $\pm1$ for massless vector bosons may be described by
$$
\eps_{\alpha\dot\alpha}(k^{+}) \equiv
{
{ u_{\alpha}(q) \bar u_{\dot\alpha}(k) }
\over
{ \braket{k}{q} }
},
\newlettlabel\polarizations
$$
$$
\eps_{\alpha\dot\alpha}(k^{-}) \equiv
{
{ u_{\alpha}(k) \bar u_{\dot\alpha}(q) }
\over
{ {\braket{k}{q}}^{*} }
},
\lett
$$
where $q$ is any null-vector such that $k\cdot q\ne0$.  As the
choice of $q$ does not affect any physics result, we will refer
to $u(q)$ and $\bar u(q)$ as gauge spinors.  The  corresponding
polarization vectors $\eps^{\mu}(k)$ defined through (\msvector)
differ from the ``standard'' polarization vectors
$$
\vareps_{0}^{\mu}(k^{\pm}) =
\left(
0, \mp{1\over{\sqrt2}},{{-i}\over{\sqrt2}},0
\right),
\newlettlabel\standardpolarizations
$$
$$
k^{\mu}=(k,0,0,k),
\lett
$$
by a $q$-dependent phase and  gauge transformation  \cite\DY.

To save accounting for a large number of indices, an efficient
method is to initially write quantities in the usual formalism
and then convert to multispinor notation at a later stage using
the substitutions
$$
k\cdot k' =
{1\over2} \bar k^{\dot\alpha\alpha}{k'}_{\alpha\dot\alpha} =
{1\over2} k_{\alpha\dot\alpha} {\bar k}^{\prime\dot\alpha\alpha},
\newlettlabel\conversiondot
$$
$$
k\cdot\eps(k') =
{1\over{\sqrt2}} \bar k^{\dot\alpha\alpha}\eps_{\alpha\dot\alpha}(k') =
{1\over{\sqrt2}} k_{\alpha\dot\alpha}\bar\eps^{\dot\alpha\alpha}(k'),
\lett
$$
$$
\eps(k)\cdot\eps(k') =
\bar\eps^{\dot\alpha\alpha}(k)\eps_{\alpha\dot\alpha}(k') =
\eps_{\alpha\dot\alpha}(k)\bar\eps^{\dot\alpha\alpha}(k'),
\lett
$$
for Lorentz dot products and
$$
{1\over2}(1-\gamma_5)\psi \longrightarrow \psi_{\alpha},
\newlettlabel\conversionfermion
$$
$$
{1\over2}(1+\gamma_5)\psi \longrightarrow \psi^{\dot\alpha},
\lett
$$
$$
{1\over2}(1-\gamma_5)\thinspace{\not{\negthinspace\negthinspace W}}
{1\over2}(1+\gamma_5)
\longrightarrow \sqrt2 \thinspace {{\cal{W}}}_{\alpha\dot\alpha},
\lett
$$
$$
{1\over2}(1+\gamma_5)\thinspace{\not{\negthinspace\negthinspace W}}
{1\over2}(1-\gamma_5)
\longrightarrow \sqrt2
\thinspace {{\overline{{\cal{W}}}}}^{\dot\alpha\alpha},
\lett
$$
$$
{1\over2}(1-\gamma_5)\slash{k}{1\over2}(1+\gamma_5)
\longrightarrow k_{\alpha\dot\alpha},
\lett
$$
$$
{1\over2}(1+\gamma_5)\slash{k}{1\over2}(1-\gamma_5)
\longrightarrow \bar k^{\dot\alpha\alpha},
\lett
$$
in strings of Dirac matrices.  Note the unequal treatments of momenta
versus other 4-vectors caused by the conventions (\msvector)
and (\msmomenta).

\appendix{Proof of current conservation}
We begin with the permutation symmetric form (\WTrecursionperm)
of the transverse $W$ recursion relation.  Note that since we prove
current conservation inductively, it is permissible to use the
recursion relation even though current conservation was assumed true
in its derivation:  only $(n-1)$-photon and lower currents were
assumed conserved, the same assumption required for this proof.
For the zero-photon current, we have trivially
$$
P\cdot\Wnorm(P)=0.
\eq
$$
The one-photon current is not much harder.  Using (\WTrecursionnorm)
we obtain
$$
\eqalign{
(P+k_1)\cdot\Wnorm(P;1)&=
{e\over{(P+k_1)^2}}
[2k_1\cdot\Wnorm(P)\thinspace\eps(1)\cdot(P+k_1)
\cr&\qquad\qquad
-2P\cdot\eps(1)\thinspace\Wnorm(P)\cdot(P+k_1)
\cr&\qquad\qquad
-(P-k_1)\cdot(P+k_1)\thinspace\Wnorm(P)\cdot\eps(1)]
\cr&
=
{e\over{(P+k_1)^2}}
[(P^2-k_1^2)\thinspace\Wnorm(P)\cdot\eps(1)]
\cr&
=0,
}
\eq
$$
because of the on-shell conditions $P^2=k_1^2=0$.  For $n\geq 2$, we use
induction.
Then,
from (\WTrecursionperm) we obtain
$$
\eqalign{
[P+&\kappa(1,n)]\cdot\Wnorm\arglt{P}{1}{2}{n}=
\cr&
={-e\over{[P+\kappa(1,n)]^2}}
\Biggl\{
\permsum{1}{n}
{1\over{(n-1)!}} \thinspace [P+\kappa(1,n)]\cdot
\bigl[\eps(n),\Wnorm \arglt{P}{1}{2}{n{-}1}\bigr]
\cr&\quad
+e\negthinspace\negthinspace
\permsum{1}{n}
{1\over{2!\thinspace (n-2)!}} \thinspace[P+\kappa(1,n)]\cdot
\bigl\{ \eps(n{-}1),\Wnorm\arglt{P}{1}{2}{n{-}2},\eps(n)
\bigr\}
\Biggr\}.
}
\eq
$$
Assuming that the $m$-photon current is conserved, $m< n$, and inserting
the expressions (\sqbrak) and (\curlybrak) for the square and curly
brackets gives
$$
\eqalign{
[P&+\kappa(1,n)]\cdot\Wnorm\arglt{P}{1}{2}{n}=
\cr&
\negthinspace\negthinspace\negthinspace
={e\over{[P+\kappa(1,n)]^2}}
\Biggl\{
\permsum{1}{n}
{1\over{(n-1)!}} \thinspace
[P+\kappa(1,n{-}1)]^2
\cr & \qquad\qquad\qquad\qquad\qquad\qquad\times
\eps(n)\cdot\Wnorm\arglt{P}{1}{2}{n{-}1}
\cr&
\thinspace
-e\negthinspace\negthinspace
\permsum{1}{n}\negthinspace
{1\over{2!\thinspace (n-2)!}}
\Bigl[ 2(k_{n-1}+k_n) \cdot\Wnorm\arglt{P}{1}{2}{n{-}2}\thinspace
\eps(n{-}1)\cdot\eps(n)
\cr&\qquad\qquad\qquad
-\eps(n)\cdot[P+\kappa(1,n)]\thinspace
\eps(n{-}1)\cdot\Wnorm\arglt{P}{1}{2}{n{-}2}
\cr&\qquad\qquad\qquad
-\eps(n{-}1)\cdot[P+\kappa(1,n)]\thinspace
\eps(n)\cdot\Wnorm\arglt{P}{1}{2}{n{-}2}\Bigr]
\Biggr\},
}
\eqlabel\ccstart
$$
where the first two terms coming from the square brackets have
cancelled, and we have used the on-shell relation $k_n^2=0$.  We
insert the recursion relation for $\Wnorm\arglt{P}{1}{2}{n{-}1}$
into the first term of (\ccstart),
and use the permutation symmetry implied by the explicit permutation sum
to simplify the second piece.  This gives
$$
\eqalign{
[P&+\kappa(1,n)]\cdot\Wnorm\arglt{P}{1}{2}{n}=
\cr&
={-e^2\over{[P+\kappa(1,n)]^2}}
\Biggl\{
\permsum{1}{n}
{1\over{(n-2)!}} \thinspace
\eps(n)\cdot\bigl[\eps(n{-}1),\Wnorm \arglt{P}{1}{2}{n{-}2}\bigr]
\cr&\quad
+e\negthinspace\negthinspace
\permsum{1}{n}
{1\over{2! \thinspace (n-3)!}} \thinspace
\eps(n)\cdot
\bigl\{ \eps(n{-}2),\Wnorm\arglt{P}{1}{2}{n{-}3},\eps(n{-}1) \bigr\}
\cr&\quad
+e\negthinspace\negthinspace
\permsum{1}{n}
{1\over{(n-2)!}} \thinspace
\Bigl[ 2k_{n-1} \cdot\Wnorm\arglt{P}{1}{2}{n{-}2}\thinspace
\eps(n{-}1)\cdot\eps(n)
\cr&
\qquad\qquad\qquad
-\eps(n)\cdot[P+\kappa(1,n)]\thinspace
\eps(n{-}1)\cdot\Wnorm\arglt{P}{1}{2}{n{-}2}\Bigr]
\Biggr\}.
}
\eqlabel\ccend
$$
The first term in (\ccend) contains
$$
\eqalign{
\eps(n)\cdot&\bigl[\eps(n{-}1),\Wnorm \arglt{P}{1}{2}{n{-}2}\bigr]=
\cr & =
2[P+\kappa(1,n{-}2)]\cdot\eps(n{-}1)\thinspace
\eps(n)\cdot\Wnorm\arglt{P}{1}{2}{n{-}2}
\cr&\quad
-2k_{n-1} \cdot\Wnorm\arglt{P}{1}{2}{n{-}2}\thinspace
\eps(n)\cdot\eps(n{-}1)
\cr&\quad
+\eps(n)\cdot\bigl\{k_{n-1}-[P+\kappa(1,n{-}2)]\bigr\}\thinspace
\eps(n{-}1)\cdot\Wnorm\arglt{P}{1}{2}{n{-}2} .
}
\eqlabel\notthis
$$
Because (\notthis) appears inside a permutation sum, we may relabel
$n\leftrightarrow n{-}1$ in the first term.  If we also replace
$\kappa(1,n{-}2)$ with $\kappa(1,n)$ within the curly brackets appearing
in the third term, and compensate, recalling that $k_n \cdot \eps(n)=0$,
we obtain
$$
\eqalign{
\eps(n)\cdot&\bigl[\eps(n{-}1),\Wnorm \arglt{P}{1}{2}{n{-}2}\bigr]
=
\cr & =
2\eps(n)  \cdot[P+\kappa(1,n{-}2)]\thinspace
\eps(n{-}1)\cdot\Wnorm\arglt{P}{1}{2}{n{-}2}
\cr&\quad
-2k_{n-1} \cdot\Wnorm\arglt{P}{1}{2}{n{-}2}\thinspace
\eps(n{-}1)\cdot\eps(n)
\cr&\quad
+2\eps(n)\cdot(k_{n-1}+k_n) \thinspace
\eps(n{-}1)\cdot\Wnorm\arglt{P}{1}{2}{n{-}2}
\cr&\quad
-\eps(n)\cdot[P+\kappa(1,n)]\thinspace
\eps(n{-}1)\cdot\Wnorm\arglt{P}{1}{2}{n{-}2}.
}
\eqlabel\ugh
$$
The third term of (\ugh) may be used to extend the momentum sum
in the first term to $P+\kappa(1,n)$;  the result now matches
the last term, allowing us to write
$$
\eqalign{
\eps(n)\cdot&\bigl[\eps(n{-}1),\Wnorm \arglt{P}{1}{2}{n{-}2}\bigr]
=
\cr & =
-2k_{n-1}\cdot\Wnorm\arglt{P}{1}{2}{n{-}2}\thinspace
\eps(n{-}1)\cdot\eps(n)
\cr&\quad
+\eps(n)\cdot[P+\kappa(1,n)]\thinspace
\eps(n)\cdot\Wnorm\arglt{P}{1}{2}{n{-}2}.
}
\eqlabel\can
$$
When (\can) is put back into the first term of (\ccend), it cancels the
third term, leaving only the second term, which we now write as
$$
\eqalign{
[P&+\kappa(1,n)]\cdot\Wnorm\arglt{P}{1}{2}{n}=
\cr&=
{-e^3\over{[P+\kappa(1,n)]^2}} \permsum{1}{n}
{1\over{3! \thinspace (n-3)!}}
\cr & \qquad\qquad\times
\Bigl[
\eps(n)\cdot
\bigl\{ \eps(n{-}2),\Wnorm\arglt{P}{1}{2}{n{-}3},\eps(n{-}1) \bigr\}
\cr&\qquad\qquad\quad
+\eps(n{-}2)\cdot
\bigl\{ \eps(n{-}1),\Wnorm\arglt{P}{1}{2}{n{-}3},\eps(n) \bigr\}
\cr&\qquad\qquad\quad
+\eps(n{-}1)\cdot
\bigl\{ \eps(n),\Wnorm\arglt{P}{1}{2}{n{-}3},\eps(n{-}2) \bigr\}
\Bigr].
}
\eqlabel\ccgone
$$
The right hand side of  (\ccgone) vanishes because
$$
\eps(a)\cdot\bigl\{ \eps(b),\Wnorm,\eps(c) \bigr\} +
\eps(b)\cdot\bigl\{ \eps(c),\Wnorm,\eps(a) \bigr\} +
\eps(c)\cdot\bigl\{ \eps(a),\Wnorm,\eps(b) \bigr\} = 0,
\eqlabel\noplug
$$
which is trivially seen by explicitly writing out  the definition
(\curlybrak) of the curly brackets  appearing in (\noplug).
Thus, $\Wnorm\arglt{P}{1}{2}{n}$
is a conserved current.

\appendix{SOME PROPERTIES OF THE POLE FACTOR $\pole(P,1,\ldots,\jay)$}

In this appendix  we will prove two useful results involving
the ``universal'' pole factor
${\pole_{\alpha}}^{\beta}(P,1,\ldots,j)$ defined
in equation (\poledef).

The first identity that we will consider
involves the relationship between
$u^{\alpha}(g){\pole_{\alpha}}^{\beta}(P,1,\ldots,j)u_{\beta}(h)$
and the ``reversed'' form
$u^{\alpha}(h){\pole_{\alpha}}^{\beta}(P,1,\ldots,j)u_{\beta}(g)$.
{}From the definition (\poledef),  we have
$$
u^{\alpha}(g)
{{\pole}_{\alpha}}^{\beta}(P,1,2,\ldots,j)
u_{\beta}(h)=
{
{u^{\alpha}(g)(k_{j})_{\alpha\dot\alpha}
[\bar P + \bar \kappa(1,j)]^{\dot\alpha\beta}u_{\beta}(h)}
\over
{[P+\kappa(1,j{-}1)]^2
[P+\kappa(1,j)]^2}
}.
\eqlabel\sandwichstart
$$
We may rewrite the numerator in (\sandwichstart) using
$$
k_j= [P+\kappa(1,j)] - [P+\kappa(1,j-1)]
\eqlabel\jexpand
$$
to obtain
$$
\eqalign{
u^{\alpha}(g)
{{\pole}_{\alpha}}^{\beta}&(P,1,2,\ldots,j)
u_{\beta}(h)=
\cr & =
{
{u^{\alpha}(g)[\bar P+ \bar\kappa(1,j)]_{\alpha\dot\alpha}
[\bar P + \bar \kappa(1,j)]^{\dot\alpha\beta}u_{\beta}(h)}
\over
{[P+\kappa(1,j{-}1)]^2
[P+\kappa(1,j)]^2}
} \cr & \quad
-
{
{u^{\alpha}(g)[\bar P+\bar\kappa(1,j{-}1)]_{\alpha\dot\alpha}
[\bar P + \bar \kappa(1,j)]^{\dot\alpha\beta}u_{\beta}(h)}
\over
{[P+\kappa(1,j{-}1)]^2
[P+\kappa(1,j)]^2}
}
\cr & =
{
{\braket{g}{h}}
\over
{[P+\kappa(1,j{-}1)]^2}
}
-
{
{\braket{g}{h}}
\over
{[P+\kappa(1,j)]^2}
}
\cr & \quad
-
{
{u^{\alpha}(g)[\bar P+\bar\kappa(1,j{-}1)]_{\alpha\dot\alpha}
(k_j)^{\dot\alpha\beta}u_{\beta}(h)}
\over
{[P+\kappa(1,j{-}1)]^2
[P+\kappa(1,j)]^2}
}.
}
\eqlabel\almostreversed
$$
We have used (\slashsqr) in writing the last line of (\almostreversed).
Notice that we may use the Weyl equation to extend the remaining
$\kappa$-sum to $\kappa(1,j)$.  If we then apply the antisymmetry
of the spinor contractions to transpose the order of the matrix
multiplication, we see that the last term in (\almostreversed)
is $u^{\alpha}(h){\pole_{\alpha}}^{\beta}(P,1,\ldots,j)u_{\beta}(g)$.
Hence, we have the relation
$$
\eqalign{
u^{\alpha}(g)
{{\pole}_{\alpha}}^{\beta}&(P,1,2,\ldots,j)
u_{\beta}(h)=
\cr & =
u^{\alpha}(h){\pole_{\alpha}}^{\beta}(P,1,\ldots,j)u_{\beta}(g)
\cr & \quad
+
\braket{g}{h}
\biggl[
{
{1}
\over
{[P+\kappa(1,j{-}1)]^2}
}
-
{
{1}
\over
{[P+\kappa(1,j)]^2}
}
\biggr].
}
\eqlabel\REVERSE
$$
Note that the portion of (\REVERSE) appearing in square brackets
is easily summed over $j$ when that is necessary.

The second identity we consider involves
the contraction of $[P+\kappa(1,j)]$
into $\pole(P,1,\ldots,j)$.  From (\poledef) we have
$$
[\bar P + \bar\kappa(1,j)]^{\dot\beta\alpha}
{\pole_{\alpha}}^{\beta}(P,1,\ldots,j) =
{
{ [\bar P + \bar\kappa(1,j{-}1)]^{\dot\beta\alpha}
(k_j)_{\alpha\dot\alpha}
[\bar P + \bar\kappa(1,j)]^{\dot\alpha\beta} }
\over
{ [P+\kappa(1,j-1)]^2  [P+\kappa(1,j)]^2 }
}
\eqlabel\sumpolesstart
$$
where we have used
the Weyl equation to eliminate $k_j$ from one of the $\kappa$-sums.
Now apply (\jexpand) and (\slashsqr) to obtain
$$
[\bar P + \bar\kappa(1,j)]^{\dot\beta\alpha}
{\pole_{\alpha}}^{\beta}(P,1,\ldots,j)
=
{
{ [\bar P + \bar\kappa(1,j{-}1)]^{\dot\beta\beta} }
\over
{ [ P +  \kappa(1,j{-}1)]^2 }
} - {
{ [\bar P + \bar\kappa(1,j)]^{\dot\beta\beta} }
\over
{ [ P +  \kappa(1,j)]^2 }
}   .
\eqlabel\splitid
$$
Again, we have a combination which is easily summed over $j$ appearing
on the right-hand side of the equation.

\appendix{CONVERSION OF PRODUCT TO SUM}

In our study of the solutions to the recursion relations,
a single product structure kept reappearing:
$$
\Xi(j,n)\equiv \permsum{j}{n} \prod_{\ell=j}^{n}
{
{ {\bar u}_{\dot\beta}(k_{\ell})
[\bar P + \bar \kappa(1,\ell)]^{\dot\beta\beta} u_{\beta}(g) }
\over
{ \braket{\ell}{g} \varsp [P+\kappa(1,\ell)]^2 }
}.
\eqlabel\XIdef
$$
In this appendix, we build an inductive solution for
$\Xi(j,n)$.  If $n=j$, we have simply
$$
\eqalign{
\Xi(j,j) &=
{
{ {\bar u}_{\dot\beta}(k_{j})
[\bar P + \bar \kappa(1,j)]^{\dot\beta\beta} u_{\beta}(g) }
\over
{ \braket{j}{g} \varsp [P+\kappa(1,j)]^2 }
} \cr&
=
{
{ [P+\kappa(1,j{-}1)]^2 }
\over
{ \braket{g}{j} \braket{j}{g} }
}
{
{ u^{\alpha}(g)k_{j\alpha\dot\beta}
[\bar P + \bar \kappa(1,j)]^{\dot\beta\beta} u_{\beta}(g) }
\over
{ [P+\kappa(1,j{-}1)]^2 [P+\kappa(1,j)]^2 }
} .
}
\eqlabel\neqj
$$
If $n=j+1$, (\XIdef) gives us
$$
\eqalign{
\Xi(j,j{+}1) &=
{\sum_{{\cal{P}}(j \thinspace j{+}1)}}
{
{ {\bar u}_{\dot\beta}(k_{j})
[\bar P + \bar \kappa(1,j)]^{\dot\beta\beta} u_{\beta}(g) }
\over
{ \braket{j}{g} \varsp [P+\kappa(1,j)]^2 }
}
{
{ {\bar u}_{\dot\gamma}(k_{j+1})
[\bar P + \bar \kappa(1,j{+}1)]^{\dot\gamma\gamma} u_{\gamma}(g) }
\over
{ \braket{j{+}1}{g} \varsp [P+\kappa(1,j{+}1)]^2 }
} \cr &
= -{\sum_{{\cal{P}}(j \thinspace j{+}1)}}
{
{ u^{\beta}(g)[P+\kappa(1,j)]_{\beta\dot\beta}
\bar k_j^{\dot\beta\alpha}(k_{j+1})_{\alpha\dot\gamma}
[\bar P + \bar\kappa(1,j{+}1)]^{\dot\gamma\gamma}
u_{\gamma}(g) }
\over
{ \braket{g}{j} \braket{j}{j{+}1} \braket{j{+}1}{g}
\varsp [P+\kappa(1,j)]^2 [P+\kappa(1,j{+}1)]^2 }
},
}
\eqlabel\neqjplusonestart
$$
where in the second line we have introduced $\braket{j}{j{+}1}$ in
both numerator and denominator, and used the antisymmetry of the spinor
product.  Consider the numerator $\num$ of (\neqjplusonestart).
Apply the so-called Feynman identity,
$$
k_{j+1}= [P+\kappa(1,j{+}1)] - [P+\kappa(1,j)],
\eqlabel\feynman
$$
the Weyl equation,
and (\slashsqr) to obtain:
$$
\eqalign{
\num &=
-[P+\kappa(1,j{+}1)]^2 u^{\beta}(g)
[P+\kappa(1,j)]_{\beta\dot\beta} k_j^{\dot\beta\alpha}
u_{\alpha}(g) \cr&
+ u^{\beta}(g)
[P+\kappa(1,j{-}1)]_{\beta\dot\beta} k_j^{\dot\beta\alpha}
[P+\kappa(1,j)]_{\alpha\dot\gamma}
[\bar P + \bar\kappa(1,j{+}1)]^{\dot\gamma\gamma}
u_{\gamma}(g).
}
\eqlabel\stepone
$$
Transpose the matrix multiplication in the first term of (\stepone)
and apply (\feynman), written for $k_j$, in the second term:
$$
\eqalign{
\num &=
[P+\kappa(1,j{+}1)]^2
u^{\alpha}(g) k_{j\alpha\dot\beta}
[\bar P+\bar\kappa(1,j)]^{\dot\beta\beta} u_{\beta}(g) \cr&
+[P+\kappa(1,j)]^2
u^{\beta}(g)[P+\kappa(1,j{-}1)]_{\beta\dot\beta}
[\bar P+\kappa(1,j{+}1)]^{\dot\beta\gamma} u_{\gamma}(g) \cr&
-[P+\kappa(1,j{-}1)]^2
u^{\beta}(g)[P+\kappa(1,j)]_{\beta\dot\gamma}
[\bar P + \bar\kappa(1,j{+}1)]^{\dot\beta\gamma} u_{\gamma}(g).
}
\eq
$$
A small amount of additional rearrangement produces
$$
\eqalign{
\num &=
[P+\kappa(1,j{+}1)]^2
u^{\alpha}(g) k_{j\alpha\dot\beta}
[\bar P+\bar\kappa(1,j)]^{\dot\beta\beta} u_{\beta}(g) \cr&
+[P+\kappa(1,j{-}1)]^2
u^{\alpha}(g) (k_{j+1})_{\alpha\dot\beta}
[\bar P+\bar\kappa(1,j{+}1)]^{\dot\beta\beta} u_{\beta}(g) \cr&
+[P+\kappa(1,j)]^2
u^{\beta}(g)[P+\kappa(1,j{-}1)]_{\beta\dot\beta}
[\bar k_j + \bar k_{j+1} ]^{\dot\beta\gamma} u_{\gamma}(g).
}
\eqlabel\numdone
$$
Inserting (\numdone) into (\neqjplusonestart) yields:
$$
\eqalign{
\Xi&(j,j{+}1) =
{\sum_{{\cal{P}}(j \thinspace j{+}1)}}
{
{ [P+\kappa(1,j{-}1)]^2 }
\over
{ \braket{g}{j} \braket{j}{j{+}1}  \braket{j{+}1}{g} }
}
{
{ u^{\alpha}(g)k_{j\alpha\dot\beta}
[\bar P + \bar \kappa(1,j)]^{\dot\beta\beta} u_{\beta}(g) }
\over
{ [P+\kappa(1,j{-}1)]^2 [P+\kappa(1,j)]^2 }
} \cr & +
{\sum_{{\cal{P}}(j \thinspace j{+}1)}}
{
{ [P+\kappa(1,j{-}1)]^2 }
\over
{ \braket{g}{j} \braket{j}{j{+}1}  \braket{j{+}1}{g} }
}
{
{ u^{\alpha}(g)(k_{j+1})_{\alpha\dot\beta}
[\bar P + \bar \kappa(1,j{+}1)]^{\dot\beta\beta} u_{\beta}(g) }
\over
{ [P+\kappa(1,j)]^2 [P+\kappa(1,j{+}1)]^2 }
} \cr & +
{\sum_{{\cal{P}}(j \thinspace j{+}1)}}
{
{u^{\beta}(g)[P+\kappa(1,j{-}1)]_{\beta\dot\beta}
[\bar k_j + \bar k_{j+1} ]^{\dot\beta\gamma} u_{\gamma}(g)}
\over
{ \braket{g}{j} \braket{j}{j{+}1}  \braket{j{+}1}{g}
\thinspace[P+\kappa(1,j{+}1)]^2 }
}.
}
\eqlabel\neqjplusonemid
$$
Notice that the third term of (\neqjplusonemid) is antisymmetric
under the interchange $j \leftrightarrow j{+}1$.  Hence, it vanishes
when we perform the sum.  The first two terms are of the same form,
allowing us to write
$$
\Xi(j,j{+}1) = \negthinspace\negthinspace\negthinspace\negthinspace
{\sum_{{\cal{P}}(j \thinspace j{+}1)}}
{
{ [P+\kappa(1,j{-}1)]^2 }
\over
{ \braket{g}{j} \braket{j}{j{+}1}  \braket{j{+}1}{g} }
}
\sum_{\ell=j}^{j+1}
{
{ u^{\alpha}(g)k_{\ell\alpha\dot\beta}
[\bar P + \bar \kappa(1,\ell)]^{\dot\beta\beta} u_{\beta}(g) }
\over
{ [P+\kappa(1,\ell{-}1)]^2 [P+\kappa(1,\ell)]^2 }
}.
\eqlabel\neqjplusoneend
$$

The results (\neqj) and (\neqjplusoneend) suggest the following ansatz
for $\Xi(j,n)$:
$$
\Xi(j,n) =
\permsum{j}{n}
{
{[P+\kappa(1,j{-}1)]^2}
\over
{\bra{g} j,\ldots,n \ket{g}}
}
\sum_{\ell=j}^n
u^{\alpha}(g)
{{\pole}_{\alpha}}^{\beta}(P,1,2,\ldots,\ell)
u_{\beta}(g),
\eqlabel\XIidappend
$$
where ${{\pole}_{\alpha}}^{\beta}(P,1,2,\ldots,\ell)$ is
defined in (\poledef) and studied in Appendix \polepropappendix.

We now  prove (\XIidappend) by induction.  Assuming
this expression to be correct for
$\Xi(j,n{-}1)$, (\XIdef) yields
$$
\eqalign{
\Xi(j,n) = \permsum{j}{n} &
{
{ {\bar u}_{\dot\gamma}(k_{n})
[\bar P + \bar \kappa(1,n)]^{\dot\gamma\gamma} u_{\gamma}(g) }
\over
{ \braket{n}{g} \varsp [P+\kappa(1,n)]^2 }
}
{
{[P+\kappa(1,j{-}1)]^2}
\over
{\bra{g} j,\ldots,n{-}1 \ket{g}}
}
\cr & \times
\sum_{\ell=j}^{n-1}
u^{\alpha}(g)
{{\pole}_{\alpha}}^{\beta}(P,1,2,\ldots,\ell)
u_{\beta}(g).
}
\eqlabel\proofstart
$$
We rewrite this as
$$
\eqalign{
\Xi(j,n) = \permsum{j}{n} \sum_{\ell=j}^{n-1} &
{
{[P+\kappa(1,j{-}1)]^2}
\over
{\bra{g} j,\ldots,n \ket{g}}
}
{
{ {\bar u}_{\dot\gamma}(k_{n})
[\bar P + \bar \kappa(1,n)]^{\dot\gamma\gamma} u_{\gamma}(g) }
\over
{{\braket{n{-}1}{g}} [P+\kappa(1,n)]^2 }
}
\cr & \times
{
{
{\braket{n{-}1}{n}}\braket{g}{\ell}
\bar u_{\dot\beta}(\ell)
[\bar P + \bar\kappa(1,\ell)]^{\dot\beta\beta} u_{\beta}(g) }
\over
{  [P+\kappa(1,\ell{-}1)]^2 [P+\kappa(1,\ell)]^2 }
}
}
\eqlabel\uno
$$
by supplying a factor of $\braket{n{-}1}{n}$ to numerator and
denominator, and inserting the definition (\poledef) for
${{\pole}_{\alpha}}^{\beta}(P,1,2,\ldots,\ell)$.
Apply the identity (\fierz) in the form
$$
\braket{n{-}1}{n} \braket{g}{\ell} =
\braket{n{-}1}{g} \braket{n}{\ell}
- \braket{n}{g} \braket{n{-}1}{\ell}.
\eqlabel\dos
$$
This gives us the following expression for (\uno):
$$
\eqalign{
\Xi(j,n) &=
\permsum{j}{n} \sum_{\ell=j}^{n-1}
{
{[P+\kappa(1,j{-}1)]^2}
\over
{\bra{g} j,\ldots,n \ket{g}}
}
\cr & \quad\times
{
{ \braket{n}{\ell}
{\bar u}_{\dot\gamma}(k_{n})
[\bar P + \bar \kappa(1,n)]^{\dot\gamma\gamma} u_{\gamma}(g)
\bar u_{\dot\beta}(\ell)
[\bar P + \bar\kappa(1,\ell)]^{\dot\beta\beta} u_{\beta}(g) }
\over
{ [P+\kappa(1,n)]^2 [P+\kappa(1,\ell{-}1)]^2 [P+\kappa(1,\ell)]^2 }
} \cr &   -
\permsum{j}{n} \sum_{\ell=j}^{n-2}
{
{[P+\kappa(1,j{-}1)]^2}
\over
{\bra{g} j,\ldots,n{-}1 \ket{g}}
}
{ {\braket{n{-}1}{\ell}} \over {\braket{n{-}1}{n}} }
\cr & \quad\times
{
{ {\bar u}_{\dot\gamma}(k_{n})
[\bar P + \bar \kappa(1,n)]^{\dot\gamma\gamma} u_{\gamma}(g)
\bar u_{\dot\beta}(\ell)
[\bar P + \bar\kappa(1,\ell)]^{\dot\beta\beta} u_{\beta}(g) }
\over
{ [P+\kappa(1,n)]^2 [P+\kappa(1,\ell{-}1)]^2 [P+\kappa(1,\ell)]^2 }
}.
}
\eqlabel\tres
$$
We have removed $\ell=n-2$ from the sum appearing in the second term
since the factor $\braket{n{-}1}{\ell}$ is zero when $\ell=n-1$.

Let us examine the numerator in the first term of of (\tres).  It
may be written as
$$
{\num}_1 = u^{\gamma}(g) [P+\kappa(1,n)]_{\gamma\dot\gamma}
\bar k_n^{\dot\gamma\delta} k_{\ell\delta\dot\beta}
[\bar P+\bar\kappa(1,\ell)]^{\dot\beta\beta} u_{\beta}(g).
\eqlabel\cuatro
$$
We write $k_n=[P+\kappa(1,n)]-[P+\kappa(1,\ell)]-\kappa(\ell{+}1,n{-}1)$
to obtain
$$
\eqalign{
{\num}_1 &=
[P+\kappa(1,n)]^2 u^{\gamma}(g)k_{\ell\gamma\dot\beta}
[\bar P+\bar\kappa(1,\ell)]^{\dot\beta\beta} u_{\beta}(g)
\cr& -
u^{\gamma}(g) [P+\kappa(1,n)]_{\gamma\dot\gamma}
[\bar P + \bar\kappa(1,\ell)]^{\dot\gamma\delta}
k_{\ell\delta\dot\beta}
[\bar P + \bar\kappa(1,\ell{-}1)]^{\dot\beta\beta} u_{\beta}(g)
\cr& -
u^{\gamma}(g) [P+\kappa(1,n)]_{\gamma\dot\gamma}
{\bar\kappa}^{\dot\gamma\delta}(\ell{+}1,n{-}1)
k_{\ell\delta\dot\beta}
[\bar P + \bar\kappa(1,\ell)]^{\dot\beta\beta} u_{\beta}(g).
}
\eqlabel\cinco
$$
Application of
$k_{\ell}=[P+\kappa(1,\ell)]-[P+\kappa(1,\ell{-}1)]$ in the
second term of (\cinco) gives
$$
\eqalign{
{\num}_1 &=
[P+\kappa(1,n)]^2 u^{\gamma}(g)k_{\ell\gamma\dot\beta}
[\bar P+\bar\kappa(1,\ell)]^{\dot\beta\beta} u_{\beta}(g)
\cr& -
[P+\kappa(1,\ell)]^2 u^{\gamma}(g)
[P+\kappa(1,n)]_{\gamma\dot\gamma}
[\bar P + \bar\kappa(1,\ell{-}1)]^{\dot\gamma\beta} u_{\beta}(g)
\cr& +
[P+\kappa(1,\ell{-}1)]^2 u^{\gamma}(g)
[P+\kappa(1,n)]_{\gamma\dot\gamma}
[\bar P + \bar\kappa(1,\ell)]^{\dot\gamma\beta} u_{\beta}(g)
\cr& -
u^{\gamma}(g) [P+\kappa(1,n)]_{\gamma\dot\gamma}
{\bar\kappa}^{\dot\gamma\delta}(\ell{+}1,n{-}1)
k_{\ell\delta\dot\beta}
[\bar P + \bar\kappa(1,\ell)]^{\dot\beta\beta} u_{\beta}(g).
}
\eqlabel\seis
$$
Inserting (\seis) into (\tres) produces the
following formidable expression:
$$
\eqalign{
\Xi(j,n) &=
\permsum{j}{n} \sum_{\ell=j}^{n-1}
{
{[P+\kappa(1,j{-}1)]^2}
\over
{\bra{g} j,\ldots,n \ket{g}}
}
u^{\gamma}(g){{\pole}_{\gamma}}^{\beta}(P,1,2,\ldots,\ell) u_{\beta}(g)
\cr & -
\permsum{j}{n} \sum_{\ell=j}^{n-1}
{
{[P+\kappa(1,j{-}1)]^2}
\over
{\bra{g} j,\ldots,n \ket{g}}
}
{
{ u^{\gamma}(g) [P+\kappa(1,n)]_{\gamma\dot\gamma} }
\over
{ [P+\kappa(1,n)]^2 }
}  \cr& \qquad\qquad\times
\Biggl\{
{
{ [\bar P + \bar\kappa(1,\ell{-}1)]^{\dot\gamma\beta} u_{\beta}(g) }
\over
{ [P + \kappa(1,\ell{-}1)]^2 }
} - {
{ [\bar P + \bar\kappa(1,\ell)]^{\dot\gamma\beta} u_{\beta}(g) }
\over
{ [P + \kappa(1,\ell)]^2 }
}
\Biggr\}
\cr & -
\permsum{j}{n} \sum_{\ell=j}^{n-2}
{
{[P+\kappa(1,j{-}1)]^2}
\over
{\bra{g} j,\ldots,n \ket{g}}
}
{
{ u^{\gamma}(g) [P+\kappa(1,n)]_{\gamma\dot\gamma} }
\over
{ [P+\kappa(1,n)]^2 }
}
\cr & \qquad\qquad\times
{
{ {\bar\kappa}^{\dot\gamma\delta}(\ell{+}1,n{-}1)
k_{\ell\delta\dot\beta}
[\bar P + \bar\kappa(1,\ell)]^{\dot\beta\beta} u_{\beta}(g) }
\over
{ [P+\kappa(1,\ell{-}1)]^2 [P+\kappa(1,\ell)]^2 }
} \cr & -
\permsum{j}{n} \sum_{\ell=j}^{n-2}
{
{[P+\kappa(1,j{-}1)]^2}
\over
{\bra{g} j,\ldots,n{-}1 \ket{g}}
}
{ {\braket{n{-}1}{\ell}} \over {\braket{n{-}1}{n}} }
{
{ u^{\gamma}(g) [P+\kappa(1,n)]_{\gamma\dot\gamma} }
\over
{ [P+\kappa(1,n)]^2 }
}
\cr & \qquad\qquad\times
{
{ {\bar u}^{\dot\gamma}(k_{n})
\bar u_{\dot\beta}(\ell)
[\bar P + \bar\kappa(1,\ell)]^{\dot\beta\beta} u_{\beta}(g) }
\over
{  [P+\kappa(1,\ell{-}1)]^2 [P+\kappa(1,\ell)]^2 }
}.
}
\eqlabel\siete
$$
Let us denote the four terms appearing in (\siete) by ${\Xi}_1$,
${\Xi}_2$, ${\Xi}_3$ and ${\Xi}_4$.

The desired ansatz except for a missing $\ell=n$ term
is reproduced by ${\Xi}_1$.   We are able to do the
sum on $\ell$ in ${\Xi}_2$, with the result
$$
\eqalign{
{\Xi}_2 =& -
\permsum{j}{n}
{
{[P+\kappa(1,j{-}1)]^2}
\over
{\bra{g} j,\ldots,n \ket{g}}
}
\cr & \qquad\times
{
{ u^{\gamma}(g) [P+\kappa(1,n)]_{\gamma\dot\gamma}
[\bar P + \bar\kappa(1,j{-}1)]^{\dot\gamma\beta} u_{\beta}(g)  }
\over
{ [P+\kappa(1,n)]^2 [P + \kappa(1,j{-}1)]^2 }
}
\cr& +
\permsum{j}{n}
{
{[P+\kappa(1,j{-}1)]^2}
\over
{\bra{g} j,\ldots,n \ket{g}}
}
\cr & \qquad\times
{
{ u^{\gamma}(g) [P+\kappa(1,n)]_{\gamma\dot\gamma}
[\bar P + \bar\kappa(1,n{-}1)]^{\dot\gamma\beta} u_{\beta}(g) }
\over
{ [P+\kappa(1,n)]^2  [P + \kappa(1,n{-}1)]^2 }
} .
}
\eqlabel\XItwo
$$
Notice that most of the factors inside the permutation sum in the first
piece of (\XItwo) are invariant under ${{\cal{P}}(j\ldots n)}$.  These
may be moved outside the sum producing
$$
\eqalign{
{\Xi}_2 = &-
{
{ u^{\gamma}(g) [P+\kappa(1,n)]_{\gamma\dot\gamma}
[\bar P + \bar\kappa(1,j{-}1)]^{\dot\gamma\beta} u_{\beta}(g)  }
\over
{ [P+\kappa(1,n)]^2 }
}
\cr & \qquad\times
\permsum{j}{n}
{ 1\over{\bra{g} j,\ldots,n \ket{g}} }
\cr& +
\permsum{j}{n}
{
{[P+\kappa(1,j{-}1)]^2}
\over
{\bra{g} j,\ldots,n \ket{g}}
}
\cr & \qquad\times
{
{ u^{\gamma}(g) [P+\kappa(1,n)]_{\gamma\dot\gamma}
[\bar P + \bar\kappa(1,n{-}1)]^{\dot\gamma\beta} u_{\beta}(g) }
\over
{ [P+\kappa(1,n)]^2  [P + \kappa(1,n{-}1)]^2 }
} .
}
\eqlabel\killone
$$
Now recall (\permsumid),
$$
\permsum{j}{m}
{ 1\over {\bra{g} j,\ldots,m \ket{g'}} }
=
{
{ {\braket{g}{g'}}^{m-j} }
\over
{\prod\limits_{i=j}^m \bra{g} i \ket{g'} }
}.
\eqlabel\sumperm
$$
Even though we deduced this identity by comparing our solution
with the one obtained by Berends and Giele, we may use it at this
stage in the proof since we are proceeding inductively:
$n-j$ is always less than $n$, the value for which we {\it would}
be engaging in circular reasoning.  Thus, we
see that the first term of (\killone) vanishes, since $n>j+1$
and $\braket{g}{g}=0$.
The remaining term in (\killone) is just the ``missing'' $\ell=n$
piece for the sum in ${\Xi}_1$.
So, we expect the final two pieces to cancel each other.

Let us
concentrate on ${\Xi}_3$.  To make this cancellation explicit, we
use the freedom to relabel quantities inside the permutation sum.
Consider the following set of cyclic relabellings to be applied
to the successive terms of $\kappa(\ell{+}1,n{-}1) =
(k_{\ell+1}+\cdots+k_{n-1})$:
$$
\eqalign{
&
\{ \ell{+}1, \ell{+}2,\ldots,n \} \rightarrow
\{ n, \ell{+}1, \ell{+}2, \ldots, n{-}1 \} \cr &
\{ \ell{+}2, \ell{+}3,\ldots,n \} \rightarrow
\{ n, \ell{+}2, \ell{+}3, \ldots, n{-}1 \} \cr &
\qquad\vdots\cr&
\{ n{-}1, n \} \rightarrow \{n,n{-}1\}.
}
\eqlabel\relabelling
$$
The relabelling (\relabelling) has the following effect on a
general function of the
form $\kappa(\ell{+}1,n{-}1)f(k_1,k_2,\ldots,k_n)$:
$$
\eqalign{
(k_{\ell+1}+\cdots+k_{n-1})&
f(k_1,k_2,\ldots,k_n) \longrightarrow
\cr & \quad
k_n f(k_1,\ldots,k_{\ell},k_n,k_{\ell+1},\ldots,k_{n-1})
\cr & +
k_n f(k_1,\ldots,k_{\ell},k_{\ell+1},k_n,k_{\ell+2},\ldots,k_{n-1})
\cr & + \cdots +
\cr & +
k_n f(k_1,\ldots,k_{\ell},k_{\ell+1},\ldots,k_{n-2},k_n,k_{n-1}).
}
\eqlabel\effect
$$
This certainly simplifies the numerator, but at the expense of
introducing a whole series of denominators:
$$
\eqalign{
{\Xi}_3 &=
- \permsum{j}{n} \sum_{\ell=j}^{n{-}2}
[P+\kappa(1,j{-}1)]^2 \varsp
{
{ u^{\gamma}(g) [P+\kappa(1,n)]_{\gamma\dot\gamma} }
\over
{ [P+\kappa(1,n)]^2 }
}
\cr & \qquad\times
{
{ k_n^{\dot\gamma\delta} k_{\ell\delta\dot\beta}
[\bar P + \bar\kappa(1,\ell)]^{\dot\beta\beta} u_{\beta}(g) }
\over
{
[P+\kappa(1,\ell{-}1)]^2 \varsp
[P+\kappa(1,\ell)]^2 }
}
{ 1\over { \bra{g} j,\ldots,\ell{-}1 \ket{\ell} } }
\cr & \qquad\times
\Biggl\{
{1\over { \bra{\ell} n,\ell{+}1,\ldots,n{-}1 \ket{g} }}
+{1\over {\bra{\ell} \ell{+}1,n,\ell{+}2,\ldots,n{-}1 \ket{g} }}
\cr & \qquad\qquad
+ \cdots
+{1\over {\bra{\ell} \ell{+}1,\ldots,n{-}2,n,n{-}1 \ket{g} }} \Biggr\}.
}
\eqlabel\messy
$$
The curly brackets in (\messy) are easily simplified by making use of
(\linkidsummed):
$$
\eqalign{
&
{1\over { \bra{\ell} \ell{+}1,\ldots,n{-}1 \ket{g} }}
\Biggl[
{ {\braket{\ell}{\ell{+}1}}\over{\bra{\ell}n\ket{\ell{+}1}} } +
{ {\braket{\ell{+}1}{\ell{+}2}}\over{\bra{\ell{+}1}n\ket{\ell{+}2}} } +
\cdots +
{ {\braket{n{-}2}{n{-}1}}\over{\bra{n{-}2}n\ket{n{-}1}} }
\Biggr]
\cr &
=
{1\over { \bra{\ell} \ell{+}1,\ldots,n{-}1 \ket{g} }}
{ {\braket{\ell}{n{-}1}}\over{\bra{\ell}n\ket{n{-}1}} } .
}
\eqlabel\messytidyup
$$
If we combine (\messytidyup) with (\messy) we obtain
$$
\eqalign{
{\Xi}_3 =&
\permsum{j}{n} \sum_{\ell=j}^{n-2}
{
{[P+\kappa(1,j{-}1)]^2}
\over
{\bra{g} j,\ldots,n{-}1 \ket{g}}
}
{ {\braket{n{-}1}{\ell}} \over {\braket{n{-}1}{n}} }
{
{ u^{\gamma}(g) [P+\kappa(1,n)]_{\gamma\dot\gamma} }
\over
{ [P+\kappa(1,n)]^2 }
}
\cr & \qquad\quad\times
{
{ {\bar u}^{\dot\gamma}(k_{n})
\bar u_{\dot\beta}(\ell)
[\bar P + \bar\kappa(1,\ell)]^{\dot\beta\beta} u_{\beta}(g) }
\over
{  [P+\kappa(1,\ell{-}1)]^2 [P+\kappa(1,\ell)]^2 }
},
}
\eqlabel\proofend
$$
which indeed cancels ${\Xi}_4$, proving the ansatz.

\vfill\eject

\global \chap =1
\vfill \eject \vglue .2in
\centerline {\headingfont REFERENCES}
\vglue .5in\baselineskip =\single
\parindent =16pt \parskip =\single
\item {1.}C. Dunn and T.--M. Yan, Nucl. Phys.
{\fam \bffam \twelvbf B352}, 402 (1991). \hfill \par
\item {2.}S. L. Glashow, Nucl. Phys. {\fam \bffam \twelvbf 22},
579 (1961); S. Weinberg, Phys. Rev. Lett. {\fam \bffam \twelvbf 19},
1264 (1967); A. Salam in
{\fam \itfam \twelvit Proc. 8th Nobel Symposium,
Aspen{\accent "7F a}sgarden,}
edited by N. Svartholm, (Almqvist and Wiksell, Stockholm, 1968)
p. 367. \hfill \par
\item {3.}F. A. Berends and W. T. Giele, Nucl. Phys.
{\fam \bffam \twelvbf B306}, 759 (1988). \hfill \par
\item {4.}R. Kleiss and W.J. Stirling, Phys. Lett.
{\fam \bffam \twelvbf 179B}, 159 (1986). \hfill \par
\item {5.} P. De Causmaecker, R. Gastmans, W. Troost, and T.T. Wu,
Phys. Lett. {\fam \bffam \twelvbf 105B}, 215 (1981);
P. De Causmaecker, R. Gastmans, W. Troost, and T.T. Wu,
Nucl. Phys. {\fam \bffam \twelvbf B206}, 53 (1982);
F. A. Berends, R. Kleiss, P. De Causmaecker, R. Gastmans,
W. Troost, and T.T. Wu, Nucl. Phys. {\fam \bffam \twelvbf B206},
61 (1982); F.A. Berends, P. De Causmaecker, R. Gastmans, R. Kleiss,
W. Troost, and T.T. Wu, Nucl. Phys. {\fam \bffam \twelvbf B239},
382 (1984); {\fam \bffam \twelvbf B239}, 395 (1984);
{\fam \bffam \twelvbf B264}, 243 (1986); {\fam \bffam \twelvbf B264},
265 (1986).  \hfill \par
\item {6.}J. M. Cornwall, D. N. Levin, and G. Tiktopoulous,
Phys. Rev. {\fam \bffam \twelvbf D10}, 1145 (1974); B. W. Lee,
C. Quigg, and H. Thacker, Phys. Rev. {\fam \bffam \twelvbf D16},
1519 (1977); M. S. Chanowitz and M. K. Gaillard, Nucl. Phys.
{\fam \bffam \twelvbf B261}, 379 (1985); G. J. Gounaris,
R. Kogerler and H. Neufeld, Phys. Rev. {\fam \bffam \twelvbf D34},
3257 (1986). \hfill \par
\item {7.}J. Schwinger, {\fam \itfam \twelvit Particles,
Sources and Fields,}\ (Addison-Wesley, Redwood City, 1970), Vol. I;
Ann. Phys. {\fam \bffam \twelvbf 119}, 192 (1979). \hfill \par
\item {8.} The spinor technique was first introduced by the CALCUL
collaboration, in the context of massless Abelian gauge theory
(ref. [5]). By now, many papers have been published on the subject.
A partial list of references follows.
\vskip \smallskipamount \item { } P. De Causmaecker, thesis,
Leuven University, 1983; R. Farrar and F. Neri, Phys. Lett.
{\fam \bffam \twelvbf 130B}, 109 (1983); R. Kleiss, Nucl. Phys.
{\fam \bffam \twelvbf B241}, 61 (1984); Z. Xu, D.H. Zhang, and
Z. Chang, Tsingua University preprint TUTP-84/3, 84/4, and 84/5a (1984),
and Nucl. Phys. {\fam \bffam \twelvbf B291}, 392 (1984);
J.F. Gunion and Z. Kunszt, Phys. Lett. {\fam \bffam \twelvbf 161B},
333 (1985); F.A. Berends, P.H. Davereldt, and R. Kleiss,
Nucl. Phys. {\fam \bffam \twelvbf B253}, 441 (1985);
R. Kleiss and W.J. Stirling, Nucl. Phys. {\fam \bffam \twelvbf B262},
235 (1985); J.F. Gunion and Z. Kunszt, Phys. Lett.
{\fam \bffam \twelvbf 159B}, 167 (1985); {\fam \bffam \twelvbf 161B},
333 (1985); S.J. Parke and T.R. Taylor, Phys. Rev. Lett.
{\fam \bffam \twelvbf 56}, 2459 (1986); Z. Kunszt, Nucl. Phys.
{\fam \bffam \twelvbf B271}, 333 (1986); J.F. Gunion and J. Kalinowski,
Phys. Rev. {\fam \bffam \twelvbf D34}, 2119 (1986);
R. Kleiss and W.J. Stirling, Phys. Lett. {\fam \bffam \twelvbf 179B},
159 (1986); M. Mangano and S.J. Parke, Nucl. Phys.
{\fam \bffam \twelvbf B299}, 673 (1988); M. Mangano, S.J. Parke,
and Z. Xu, Nucl. Phys. {\fam \bffam \twelvbf B298}, 653 (1988);
D.A. Kosower, B.--H. Lee, and V.P. Nair, Phys. Lett.
{\fam \bffam \twelvbf 201B}, 85 (1988); M. Mangano and S.J. Parke,
Nucl. Phys. {\fam \bffam \twelvbf B299}, 673 (1988);
F.A. Berends and W.T. Giele, Nucl. Phys. {\fam \bffam \twelvbf B313},
595 (1989); M. Mangano, Nucl. Phys. {\fam \bffam \twelvbf B315},
391 (1989); D.A. Kosower, Nucl. Phys. {\fam \bffam \twelvbf B335},
23 (1990); Phys. Lett. {\fam \bffam \twelvbf B254}, 439 (1991);
Z. Bern and D.A. Kosower, Nucl. Phys. {\fam \bffam \twelvbf B379},
451 (1992); C.S. Lam, McGill preprint McGill/92-32, 1992. \hfill \par
\item {9.} Many of the results for processes containing six or
fewer particles are collected in R. Gastmans and T.T. Wu,
{\fam \itfam \twelvit The Ubiquitous Photon: Helicity Method
for QED and QCD} (Oxford University Press, New York, 1990). \hfill \par
\item {10.} The excellent review by Mangano and Parke provides
a guide to the various approaches to and extensive literature on
the subject: M. Mangano and S.J. Parke, Phys. Reports
{\fam \bffam \twelvbf 200}, 301 (1991). \hfill \par
\item {11.}M. Mangano, Nucl. Phys. {\fam \bffam \twelvbf B309},
461 (1988). \hfill \par
\item {12.} G. Mahlon, Cornell preprint CLNS 92/1154, 1992. \hfill \par
\item {13.}For a brief introduction to properties of two-component
Weyl-van der Waerden spinors, see, for example, M. F. Sohnius,
Phys. Reports {\fam \bffam \twelvbf 128}, 39 (1985). \hfill \par

\vfill\eject
\bye